\documentclass[prb,aps,amsmath,amssymb,showpacs,superscriptaddress,twocolumn]{revtex4-1}
\usepackage{bm,color,float,dcolumn,amsmath,amssymb,amsfonts,amsthm,nicefrac,graphicx,subfigure,dsfont}
\definecolor{darkred}{rgb}{0.8,0.0,0.0}
\definecolor{darkblue}{rgb}{0.0,0.0,0.8}
\usepackage[colorlinks,urlcolor=darkred,linkcolor=red,anchorcolor=darkred,citecolor=blue]{hyperref}

\begin{document}

\title{Vacancies in the Kitaev quantum spin liquids on the 3D hyper-honeycomb lattice}
\author{G. J. Sreejith}
\affiliation{Max-Planck-Institut f\"{u}r Physik komplexer Systeme, N\"{o}thnitzer Str. 38, 01187 Dresden, Germany}
\author{Subhro Bhattacharjee}
\affiliation{Max-Planck-Institut f\"{u}r Physik komplexer Systeme, N\"{o}thnitzer Str. 38, 01187 Dresden, Germany}
\affiliation{International Center for Theoretical Sciences, Tata Institute of Fundamental Research, Bangalore 560012, India}
\author{R. Moessner}
\affiliation{Max-Planck-Institut f\"{u}r Physik komplexer Systeme, N\"{o}thnitzer Str. 38, 01187 Dresden, Germany}
\date{\today}

\begin{abstract}
We study the effect of adding disorder to the exactly solvable Kitaev model on the hyper-honeycomb lattice, which hosts both gapped and gapless spin liquid phases with an emergent $\mathbb{Z}_2$ gauge field. The latter has an unusual gapless spectrum of Majorana fermion excitations, with a co-dimension-two Fermi ring. We thus address the question of the interplay of topological physics and disorder by considering the properties of isolated single and pair of vacancies. We show that near the vacancies, the local magnetic response to a field $h_z$ is parametrically enhanced in comparison to the pristine bulk. Unlike the previously studied case of the 2D honeycomb Kitaev model, the vacancies do not bind a flux of the $\mathbb{Z}_2$ gauge field. In the gapped phase, an isolated vacancy gives rise to effectively free spin-half moments with a non-universal coupling to an external field. In the gapless phase,  the low-field magnetization is suppressed parametrically, to $(-\ln h_z)^{-1/2}$ because of interactions with the surrounding spin-liquid. We also show that a pair of vacancies is subject to a sublattice-dependent interaction on account of coupling through the bulk spin liquid, which is spatially anisotropic even when all Kitaev couplings have equal strength. This coupling is thus exponentially suppressed with distance in the gapped phase. In the gapless phase,  two vacancies on the same (opposite) sublattice exhibit an enhanced (suppressed) low-field response, amounting to an effectively  (anti-)ferromagnetic interaction.
\end{abstract}
\maketitle

\section{Introduction}
The presence of various forms of disorder in condensed matter materials is both an inevitable reality and a rich source of new physics. While extensive disorder can give rise to qualitatively new phases through localization \cite{Anderson1958,Abrahams1979,Lee1985} and glass transitions,\cite{Binder1986,Ramirez1994} interesting physics can emerge even near individual defects, as in the case of Yu-Shiba-Rusinov \cite{Shiba1968,*yu1965bound,*rusinov1969theory} in-gap bound states in superconductors and Kondo \cite{Kondo1964} effect for magnetic impurities in a metal. Recent studies on disordered interacting quantum systems have revealed interesting effects on quantum entanglement in the many-body localized states. Entanglement entropy of high-energy states in such disordered systems show an area-law, as opposed to the volume-law seen in clean systems.\cite{Bauer2013,Kjall2014}

Connections between disorder and entanglement motivate a broader set of questions starting with the effect of various kinds of disorder on quantum phases that are themselves characterized by long-range entanglement. The issue is particularly interesting in the context of the growing number of magnetic materials which, at low temperatures, are believed to host a class of long-range entangled quantum paramagnetic phases called quantum spin liquids (QSLs). \cite{Anderson1973, Moessner2001, Wen2002, Balents2010, Lee2008}  The effective low-energy degrees of freedom in a QSL can carry fractions of quantum numbers of the constituent degrees of freedom, and interact with each other via interactions mediated by an emergent gauge field, similar to the fractional quantum Hall effect.\cite{ZHANG2012}

The above question has two major aspects. Firstly, the experimental search for candidate magnetic materials calls for a theoretical understanding of the impact of disorder on the properties of QSLs. Secondly, dilute disorder and isolated defects can be used to elucidate the unconventional nature of these exotic phases by making some of their properties accessible to experiments. \cite{Willans2010,*Willans2011,Kolezhuk2006,Sachdev1999,Vojta2000} This latter scenario, in a general sense, also includes the physics of vortices and boundaries in one dimensional chains,\cite{Jackiw1976,Su1979, Kitaev2001,Fendley2012} superconductors,\cite{Read2000,Matsuura2013} edges in FQH systems,\cite{Blok1990,*wenedge,*Chang2003} lattice-defects in spin-liquids,\cite{Willans2010,*Willans2011,Schaffer2015,Trousselet2011,Petrova2014} etc. These defects may carry low-energy modes described in terms of the fractional excitations that can be manipulated by external probes. An understanding of the formation of these defect modes starting from the Hilbert space of the original system, and their response to external probes can therefore be of theoretical interest and practical value. While such questions have received some attention for several systems in two spatial dimensions,\cite{Blok1990,*wenedge,*Chang2003,Read2000, Willans2010,*Willans2011,Petrova2014,Laumann2012, Das2015, Lahtinen2014,Mei2012,G.2012,Trousselet2011,Dhochak2010} interesting physics can also emerge in three dimensional systems.

The present work is concerned with understanding some of the above issues in the context of three dimensional QSLs. Due to the technical difficulties in studying the combination of disorder and emergent many body phenomena like fractionalization in a general setting, we formulate and address these questions in the context of a specific three dimensional {\it exactly solvable} QSL - the Kitaev model on a hyper-honeycomb\cite{Mandal2009} lattice that has been suggested to be relevant for the compound $\beta$-Li$_2$IrO$_3$.\cite{Takayama2014,Biffin2014,Jackeli2009,Lee2014,Kimchi2014,*Kimchi2014Analytis} However, our results should be valid in a regime where the perturbations to the pure Kitaev model are small compared to the field and interaction scales derived here. The Kitaev spin-model, though highly anisotropic, offers an exact solution in terms of  Majorana (fermionic) partons coupled to a static $\mathbb{Z}_2$ gauge field. The ground state is a $\mathbb{Z}_2$ QSL with fractionalized Majorana excitations that can be gapped or gapless (depending on the {coupling parameters of the Kitaev Hamiltonian, Eq-\ref{Spin-Hamiltonian}), and gapped $\mathbb{Z}_2$ flux excitations. There is no net magnetization at small magnetic fields and the spin-correlations are short-ranged (strictly nearest-neighbour).\cite{Smith2015}

Removal of a spin (vacancy) from the lattice in such a system creates new {\it local} low-energy degrees of freedom near the resulting vacancy that emerge as zero-energy Majorana modes.\cite{Willans2010,*Willans2011,Halasz2014} These modes carry finite magnetic moments contributing a non-zero magnetization at low magnetic fields. The magnetic susceptibility characteristically depends on the nature (gapless or gapped) of the spin liquid phase indicating that the lowest order contributions to the spin-spin correlations arise from the interaction between the vacancy-induced magnetic moments which in turn are governed by the nature of the QSL phase. In the gapped phase, an isolated vacancy-induced moment behaves like a free spin, polarizing under an arbitrarily small external field. In the gapless phase, the vacancy spin-moment interacts with the low-energy modes of the surrounding spin liquid, thereby suppressing the magnetization at low fields to $\frac{1}{\sqrt{\ln[1/h_z]}}$ for a magnetic field $h_z$ along $z$-direction.

Two vacancies that are a finite distance apart, interact with an anisotropic sublattice dependent interaction. In the gapped phase, this interaction is  non-zero only when the vacancies are on sub-lattices with opposite parity (Hyper-honeycomb is a bipartite lattice). This interaction is exponentially decaying with separation (falling off with a characteristic scale of the bulk excitation gap) and it suppresses the low-field magnetization, indicating an effective anti-ferromagnetic nature. In the gapless phase, however, the same interaction only decays with separation as a power law and are such that the magnetization is enhanced (suppressed) relative to isolated vacancies when the vacancies are on the same (opposite) sublattices. In comparison with the isolated vacancies, when two vacancies are in unit-cells separated along the direction of the strongest interaction, the low-field magnetization increase to a constant (decrease to $m\sim h_z$) for two vacancies are on the same (opposite) sublattices.

Compared to vacancies in the two dimensional Kitaev model on a honeycomb lattice, the three dimensional hyper-honeycomb lattice exhibits key differences. Most importantly, vacancies in the honeycomb lattice carry a low-energy $\mathbb{Z}_2$ flux through the ``defective" plaquette associated with  the vacancy, \cite{Willans2010,*Willans2011}. By contrast,  such fluxes are absent in the three dimensional case as we show here. This is due to the fact that $\mathbb{Z}_2$ fluxes, in the three spatial dimensions form closed loops and threading the ``defective" plaquette with $\mathbb{Z}_2$ flux is impossible without penalizing ``healthy" plaquettes by also threading them with flux. This makes flux binding to the vacancy energetically expensive in the hyper-honeycomb. This difference in terms of flux binding between two and three dimensional Kitaev models can lead to important differences between the properties of the two systems. Also, the hyper-honeycomb lattice is inherently anisotropic (only two of the three nearest-neighbour bonds being equivalent). As a result, the interaction between the vacancies are anisotropic even for an `isotropic' choice of parameters $J_{x,y,z}=1$ (Fig-\ref{unitcell}).

More broadly, the analysis of vacancies presented in Ref-[\onlinecite{Willans2010,*Willans2011},\onlinecite{Halasz2014}] and this work, suggests a general strategy to understand disorder in the class of Kitaev spin liquids, consisting of first identifying the number and nature of the low-energy modes nucleated by the vacancies; and then to consider their coupling to an external field, as well as their mutual interactions, with the latter depending primarily on some gross features of the model, such as spatial dimensionality and the low-energy Majorana spectrum and its co-dimension, distinguishing between gapped, point- or line-like Fermi surfaces.

The rest of the paper is organized as follows. In Sec-\ref{sec:cleanKM}, we review the Kitaev Model on the hyper-honeycomb lattice and introduce notation. In Sec-\ref{sec:vacancydof}, we discuss the changes and redistribution of overall and low-energy degrees of freedom of the system upon introducing a vacancy. The low-energy vacancy-induced local spin degrees of freedom are explicitly identified.  In Sec-\ref{sec:absflux}, we show that the ground state of the system does not contain any flux-loops. General arguments are supported with numerical results in finite size systems.
Section \ref{sec:gapped-vac} discusses the properties of the vacancy in the gapped phase. We identify the zero-energy Majorana modes around the vacancy. Magnetic response is shown to result from the hybridization of these modes. We calculate the magnetization of an isolated vacancy as well as a pair of interacting vacancies. In Sec-\ref{sec:gapless-vacancy}, a similar analysis of the magnetization is performed for the case of gapless system at the point $J_{x,y,z}=1$. A considerable amount of technical material for some of the more invovled calculations has been relegated to a set of appendices.


\section{Review of Kitaev Model on a hyper-honeycomb lattice}
\label{sec:cleanKM}

\begin{figure}
\includegraphics[width=\columnwidth]{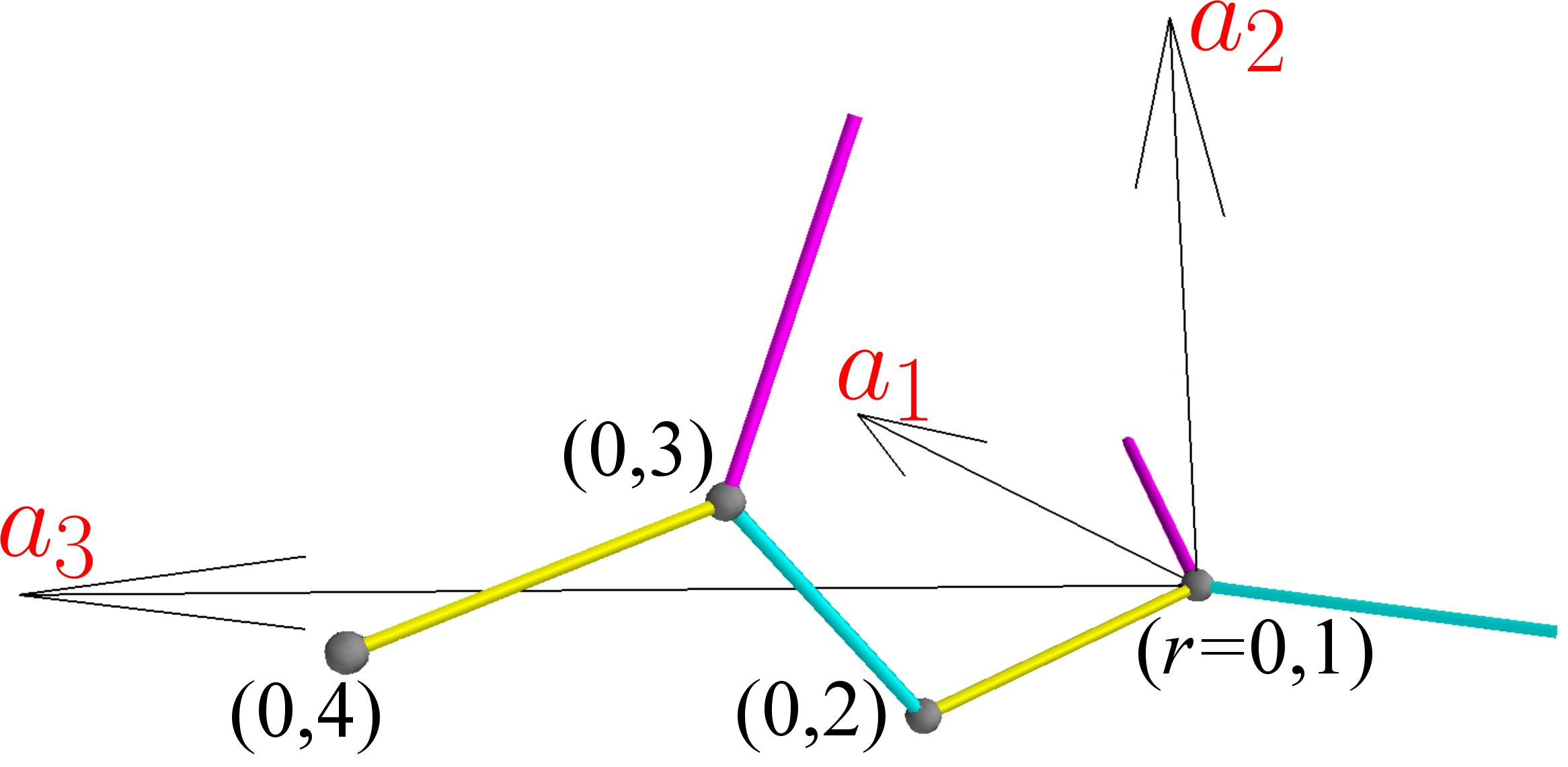}
\caption{ A unit-cell at location $r=0$ of the hyper-honeycomb lattice, showing its four sites (grey spheres) indexed by sublattice index $i=1,2,3,4$. Translations along the three lattice vectors $a_1=(0,-2\sin{\frac{\pi}{3}},0)$, $a_2=(0,0,2\sin{\frac{\pi}{3}})$ and $a_3=(3,-\sin{\frac{\pi}{3}},\sin{\frac{\pi}{3}})$ produce the whole lattice. Blue, magenta and yellow bonds connect the spins that interact through $J_x\sigma^x\sigma^x$, $J_y\sigma^y\sigma^y$ and $J_z\sigma^z\sigma^z$ interactions.}\label{unitcell}
\end{figure}

We begin with a brief reminder to the the physics of the Kitaev Model on a hyper-honeycomb lattice\cite{Mandal2009} in order to introduce notation and provide essential background for the analysis of vacancies in the model.

The 3D Kitaev Model consists of spin-$\frac{1}{2}$'s on the sites of a hyper-honeycomb lattice.\cite{Lee2014, Jackeli2009, Takayama2014} The hyper-honeycomb lattice is tri-coordinated and has a four-site unit-cell (Fig-\ref{unitcell}) arranged on an orthorhombic Bravais lattice $\mathcal{T}$ with basis sites given by $r = n_1 a_1 + n_2 a_2 + n_3 a_3$ where $a_i$ are the Bravais lattice vectors and $n_i\in \mathbb{Z}$. We denote the set of all sites by $\mathcal{S}$ and label each site by a pair $(r,i)$ where unit-cell location $r\in \mathcal{T}$ and sublattice index $i\in\{1,2,3,4\}$. It is useful to define odd and even sublattices as the sites with odd and even values of sublattice indices. We shall use bold font lower case roman alphabets  to label sites; and normal font to label unit-cell locations and sublattice indices. To define the Kitaev model, we denote the bonds connecting each spin to its three nearest-neighbours as $x$ (blue), $y$ (magenta) and $z$ (yellow) bonds as shown in Fig-\ref{unitcell}.

\begin{figure}
\includegraphics[width=\columnwidth]{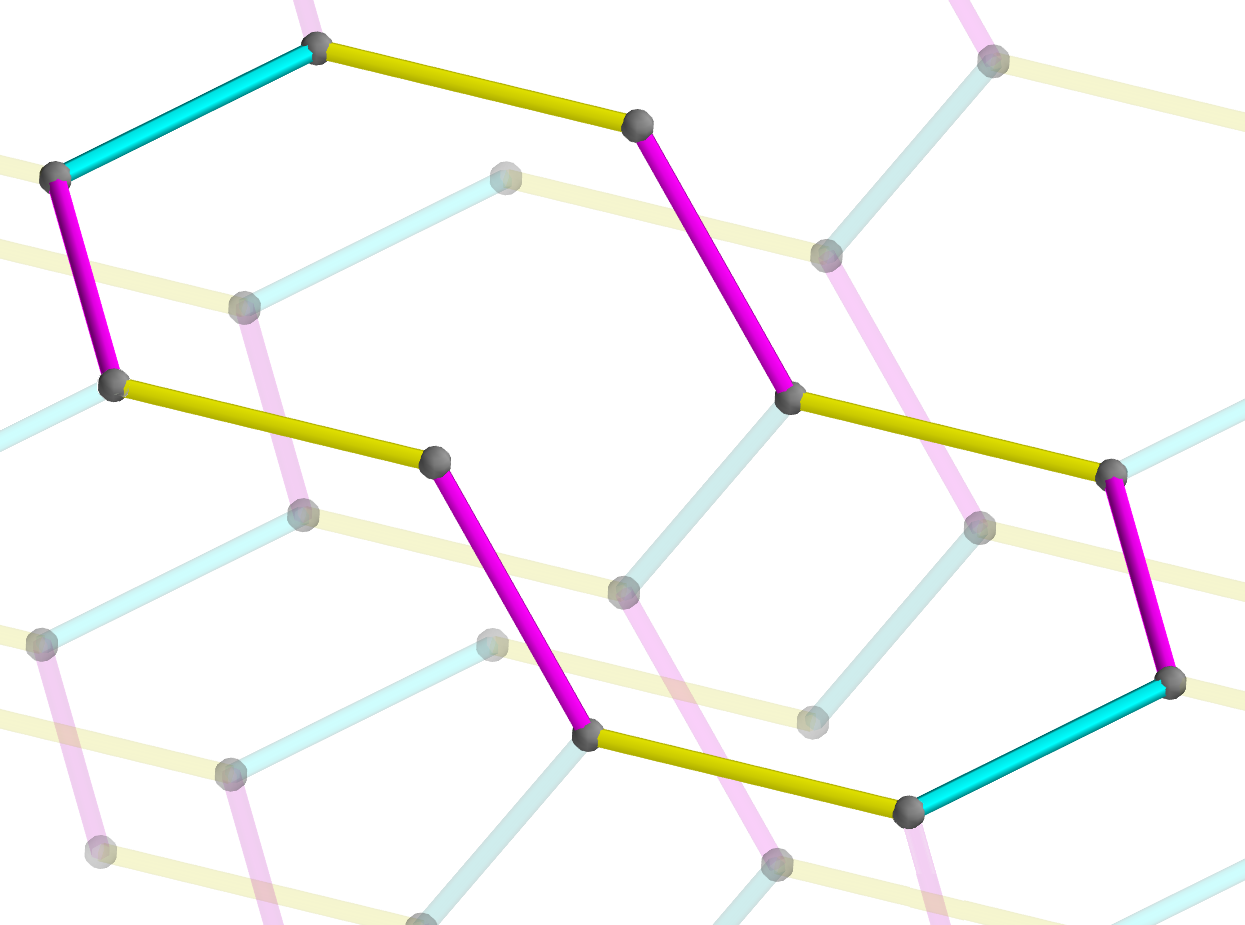}
\caption{A plaquette in a hyper-honeycomb lattice. Loop-operator for any loop that does not wind around the 3-torus can be constructed by combining such plaquettes. }\label{elementary-plaquette}
\end{figure}

The Kitaev spin-Hamiltonian is given by\cite{Mandal2009,Lee2014} 
\begin{align}
H=\sum_{\langle \mathbf{r},\mathbf{r}'\rangle} J_\alpha \sigma^\alpha_\mathbf{r}\sigma^\alpha_{\mathbf{r}'}
\label{Spin-Hamiltonian}
\end{align}
where $\sigma^{\alpha}_\mathbf{r}$ are the spin-$\frac{1}{2}$ operators at site $\mathbf{r}$. Sum is over all pairs $\mathbf{r},\mathbf{r}'$ of nearest-neighbour sites and $\alpha$ ($=x,y$ or $z$) is the type of bond connecting each pair. The Hamiltonian couples every odd sublattice spin to three nearest-neighbour even sublattice spins and vice-versa.

The Hamiltonian, as in the honeycomb model, has an extensive number of conserved loop-operators or $\mathbb{Z}_2$ fluxes, one associated with every closed loop on the lattice.\cite{Wegner1971,Kitaev06} For a loop $g$ formed by a sequence of $L$ nearest-neighbour sites  $\{\mathbf{s}_a\}_{1\leq a\leq L}$, the loop-operator, $W(g)$, is given by 
\begin{equation}
W(g) = \prod_{a=1}^L \sigma_{\mathbf{s}_a}^{\alpha_a}
\end{equation}
where the $\alpha_a$ is the type of bond connected to the site $\mathbf{s}_a$, but which is not in the loop $g$. The eigenvalues of $W(g)$ are $\pm1$ and a state $|\phi\rangle$ of the system is said to have a flux through a loop $g$ if $W(g)|\phi\rangle=-|\phi\rangle$. The spin-algebra implies that the loop-operators form a discrete Abelian group isomorphic to the cycle-space\cite{gross2005graph} of the lattice (when treated as a graph). However, not all such loop-operators are independent. 

A periodic system with $N$ unit-cells has $4N$ sites (vertices) and $6N$ bonds (edges). The number of independent generators in the group is given by the circuit-rank \footnote{\label{circuitrank}The circuit rank of a connected graph is given by ${\rm Edges-Vertices+1}$. It is the number of independent generators of the cycle space $H_1(\mathcal{S},\mathbb{Z}_2)$} $2N+1$ and represents the number of independent loop-operators/flux degrees of freedom.

\begin{figure} 
\includegraphics[width=\columnwidth]{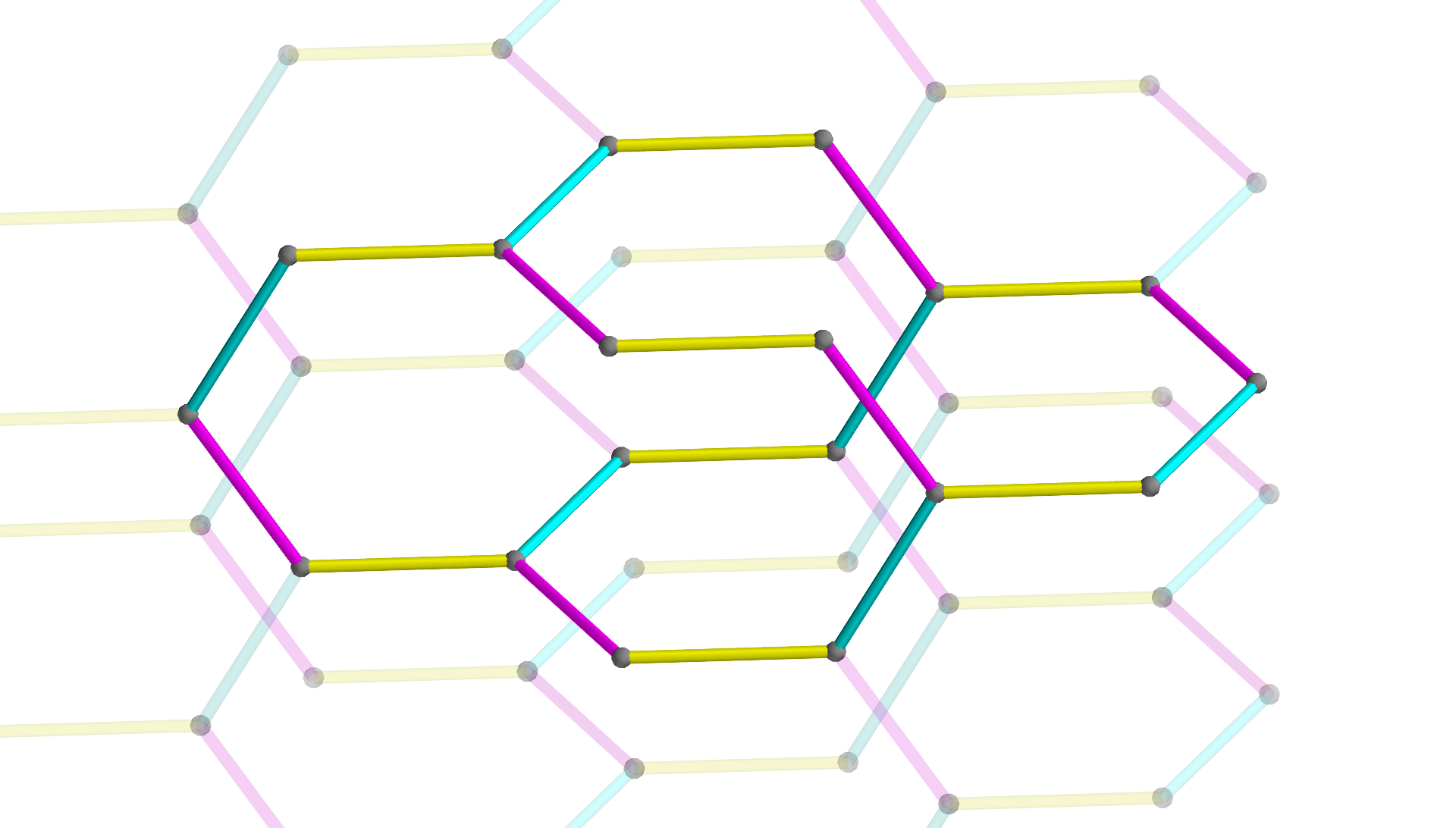}
\caption{The highlighted set of bonds contains four plaquettes, which together define a volume. The product of the corresponding four loop-operators equals $+1$, constraining the values that these operators can take. Centers of these four plaquettes form a tetrahedron of the pyrochlore lattice formed by the centers of all the plaquettes.}
\label{tetrahedra}
\end{figure}

Generators can be chosen to be the following $\mathbb{Z}_2$  (since $W(g)^2=1$) elements: (1) loop-operators on the $4N$ elementary plaquettes (Fig-\ref{elementary-plaquette}) which generate the operators on all loops that do not wind around the torus, and (2) loop-operators on the three loops that wind around the three periodic directions. The loop-operators satisfy the following constraints: (1) {\it Local constraints}: Centers of four adjacent plaquettes of the kind shown in Fig-\ref{tetrahedra} form the corners of a tetrahedron. The product of the corresponding loop-operators is  identity, i.e. $W(g_1)W(g_2)W(g_3)W(g_4)=1$. There are $2N$ tetrahedra and associated constraints of which $2N-1$ are independent. (2) {\it Global constraints}: Product of loop-operators on plaquettes tiling three planes that cut across the periodic directions of the three-torus result in $3$ constraints.

These generators and constraints account for the $2N+1$ independent $\mathbb{Z}_2$ fluxes. The local constraints imply that an even number of plaquettes in each tetrahedron have fluxes (i.e. $W=-1$), thus these fluxes are constrained to form closed flux-loops formed by joining the mid-points of the plaquettes threaded with flux.

Conservation of fluxes imply that the Hamiltonian governing the $4N$ spins is $2^{2N+1}$-fold block diagonal, with each block labeled by the eigenvalues of the $2N+1$ independent $\mathbb{Z}_2$-valued loop-operators. Each block couples the remaining $2N-1$ dynamic degrees of freedom. The above picture, following Kitaev, is conveniently described in terms of Majorana fermions acting on an extended Hilbert space as follows.

At each site  $\mathbf{s}$, define four Majorana operators $b^x_\mathbf{s}$, $b^y_\mathbf{s}$, $b^z_\mathbf{s}$, $c_\mathbf{s}$ acting on an extended four dimensional Hilbert space at each site. These operators satisfy the algebra
\begin{gather}
\{ b_\mathbf{s}^\alpha, b_\mathbf{s'}^\beta \} = 2\delta_\mathbf{ss'}\delta_{\alpha\beta};\;
\{ c_\mathbf{s}, c_\mathbf{s'} \} = 2\delta_\mathbf{ss'};\;
\{ c_\mathbf{s},b^{\alpha}_\mathbf{s'} \}=0\nonumber
\label{MajoranaOperatorDefinitions}
\end{gather}
for all $\mathbf{s},\mathbf{s'}\in \mathcal{S}$. The extensions of the spin operators are identified to be $\tilde\sigma^\alpha_{\mathbf{s}}=\imath b^\alpha_{\mathbf{s}} c_{\mathbf{s}}$. Extension $\tilde{H}$ of the Hamiltonian (Eq-\ref{Spin-Hamiltonian}) is obtained by replacing the spin operators in the Hamiltonian by their extensions.

The physical spin Hilbert space is a two dimensional subspace of the extended one in which extensions of the spin operators satisfy the spin-algebra $1+\imath \tilde{\sigma}^x \tilde{\sigma}^y \tilde{\sigma}^z=0$. The projection into the physical subspace at site $\mathbf{s}$ is achieved by the operator $P_\mathbf{s}=\frac{1+D_\mathbf{s}}{2}$ where $D_\mathbf{s}\equiv-\imath \tilde{\sigma}^x_\mathbf{s} \tilde{\sigma}^y_\mathbf{s} \tilde{\sigma}^z_\mathbf{s} = b^x_\mathbf{s}b^y_\mathbf{s}b^z_\mathbf{s}c_\mathbf{s}$. $D_\mathbf{s}$ is Hermitian, with eigenvalues $\pm 1$; as a result $P_\mathbf{s}$ orthogonally projects any state into the physical space (where $D_\mathbf{s}=1$). $P_\mathbf{s}$ at different sites commute and therefore the physical projection of a state from the combined Hilbert space of all sites, is achieved by $\prod_{\mathbf{s}\in \mathcal{S}} P_\mathbf{s}$. It can be seen that two states in the extended Hilbert space have identical projections iff the two states are related by the action of $D_\mathbf{s}$ on a suitable subset of sites. The discrete Abelian group $\mathfrak{D}$ generated by $\{D_\mathbf{r}\}_{\mathbf{r}\in \mathcal{S}}$ maps between states with equivalent projections. Thus physical states are $\mathfrak{D}$-orbits in the extended Hilbert space. 

The extended Hilbert space $\mathcal{V}$ can be constructed as the Fock space of complex fermions. To this end, we define `bond' fermions on each bond, and `matter' fermions on each $z$-bond. For a bond of type $\alpha$ ($=x$, $y$ or $z$) connecting odd and even sublattice sites $\mathbf{r}$ and $\mathbf{s}$ respectively, the bond-fermion is defined as $\chi_\mathbf{r}^\alpha=\frac{1}{2}(b^\alpha_\mathbf{r}+\imath b^\alpha_\mathbf{s})$. On $z$-bonds connecting sublattices $1$ with $2$ and $3$ with $4$ (Fig-\ref{unitcell}) of a unit-cell $r\in \mathcal{T}$, matter-fermions $f^A_{r}$ and $f^B_{r}$ can be defined as $f^A_{r}=\frac{1}{2}(c_{(r,1)}+\imath c_{(r,2)})$ and $f^B_{r}=\frac{1}{2}(c_{(r,3)}+\imath c_{(r,4)})$. The extended Hilbert space can then be identified with the $2^{8N}$ dimensional Fock space of the $6N$ bond and $2N$ matter-fermions: $\mathcal{V}=\mathcal{V}_{\rm bond}\otimes \mathcal{V}_{\rm matter}$.

Projection $P$ can be expressed as
\begin{equation}
\sum_{X\subseteq \mathcal{S}}\frac{\prod_{\mathbf{r}\in X}D_{\mathbf{r}}+\prod_{\mathbf{r}\in \mathcal{S}-X}D_{\mathbf{r}} }{2^{4N+1}}= \frac{1+\mathcal{D}}{2^{4N+1}} \sum_{X\subseteq\mathcal{S}}\prod_{\mathbf{r}\in X}D_{\mathbf{r}},\label{D=1constraint}
\end{equation}
where $\mathcal{D}=\prod_{\mathbf{r}\in \mathcal{S}}D_\mathbf{r}$ is a $\mathbb{Z}_2$ operator. As shown in Appendix-\ref{App:FermionParity}, $\mathcal{D}$ is the parity of the total number of fermions (both bond and matter). Thus, a basis-state $|\phi\rangle$ with fixed fermion-number has a physical projection iff it has even fermion parity, i.e. only even fermion parity orbits are physical ($\mathfrak{D}$ preserves total fermion-parity).

The action of $D_\mathbf{r}$ on a basis-state $|\phi\rangle\in\mathcal{V}$ with fixed fermion-number is to flip the occupancy of the bond and matter-fermions on the bonds connected to $\mathbf{r}$. Considering that $\mathcal{D}|\phi\rangle=\pm |\phi\rangle$, it can be inferred that the orbit $\mathfrak{D}\phi$ has $2^{4N-1}$ elements (all independent); and $\mathcal{V}/\mathfrak{D}$ is $2^{4N+1}$ dimensional. The even fermion-parity orbit-space has dimension $2^{4N}$ which matches that of the physical space. Thus physical space is the space of even parity $\mathfrak{D}$-orbits in $\mathcal{V}$.

It is useful to define the quantity $u_{\mathbf{rs}}=\imath b^\alpha_\mathbf{r} b^\alpha_\mathbf{s}$ called the bond-operator for any bond of type $\alpha$ ($=x,y$ or $z$) connecting nearest-neighbour odd and even sublattice sites $\mathbf{r}$ and $\mathbf{s}$. This is simply the parity of the bond-fermion for the particular bond. Extensions of the loop-operators introduced earlier can be chosen to be
\begin{equation}
\tilde{W} = -\prod_{a=1}^{10} u_{\mathbf{r}_a,\mathbf{r}_{a+1}}
\label{gauge2flux}
\end{equation}
for a loop like the one in Fig-\ref{elementary-plaquette} made of sites $\{\mathbf{r}_a\}_{1\leq a\leq 10}$, ($a=11$ identified with $a=1$). Note that $\tilde{W}$ is invariant under $\mathfrak{D}$

The spin-Hamiltonian $H$ has $2^{2N+1}$ blocks, each block corresponding to a fixed flux sector (an eigenspace of loop-operators). The extended Hamiltonian has $2^{6N}$ blocks, each block corresponding to a an eigenspace of the $6N$ bond-operators. While Eq-\ref{gauge2flux} determines all the loop-operators for a given choice of all bond-operators, the converse is not true. Picking a bond-operator sector corresponds to a gauge-choice, equivalent to identifying a representative element in each $\mathfrak{D}$-orbit; gauge-transformations being $\mathfrak{D}$. From the fact that $\tilde{H}$ commutes with all the bond-operators, loop-operators and gauge-transformations $\mathfrak{D}$, it can be inferred that the spectrum of $H$ in a specific flux-sector can be obtained as the spectrum of $\tilde{H}$ in a chosen gauge-sector.

For a given gauge-choice within a flux sector, the Hamiltonian of the system reduces to a tight binding Hamiltonian for the Majorana operators which has the form:
\begin{equation}
\mathcal{H}=-\imath\sum_{\langle\mathbf{rs}\rangle} J_{\alpha}u_{\mathbf{rs}} c_\mathbf{r} c_\mathbf{s}
\label{tightbindingHamiltonian}
\end{equation}
where $\alpha$ is the type of bond connecting the sites $\mathbf{r},\mathbf{s}$.
From numerical studies in finite systems, it is known that the lowest energy states occur in a sector where the loop-operators are all $+1$.\cite{Mandal2009} Any gauge-choice in which the six bond-operators of a unit-cell (Fig-\ref{unitcell}) have identical values in every unit-cell give $W=+1$ (i.e. $u_{(r,i)(s,j)}$ are independent of $r,s$). Therefore, for the flux sector containing the ground state, the bond-operators $u_\mathbf{rs}$ in the above equation can be chosen to be $-1$s when $\mathbf{r}$  and $\mathbf{s}$ are on odd and even sublattices. In this gauge, Eq-\ref{tightbindingHamiltonian} can be explicitly written as:
\begin{align}
\mathcal{H}=\frac{1}{2}\sum_{r_1,r_2\in \mathcal{T}} \left[\begin{array}{c}
c_{\left(r_{1},1\right)}\\
c_{\left(r_{1},3\right)}\\
c_{\left(r_{1},2\right)}\\
c_{\left(r_{1},4\right)}
\end{array}\right]^T \mathfrak{H}(r_1,r_2) \left[\begin{array}{c}
c_{\left(r_{2},1\right)}\\
c_{\left(r_{2},3\right)}\\
c_{\left(r_{2},2\right)}\\
c_{\left(r_{2},4\right)}
\end{array}\right]
 \label{cleanmatrixham}
\end{align}
with
\begin{align}
\mathfrak{H}(r_1,r_2)=\left[\begin{array}{cc}
0_{2\times 2} & M(r_1,r_2)\\
M(r_2,r_1)^{\dagger} & 0_{2\times 2}\\
\end{array}\right] 
\end{align} 
where $M(r,s)$ for $r,s\in\mathcal{T}$ is 
\begin{equation}
\imath\left[\begin{array}{cc}
J_{z}\delta_{r,s} & J_{x}\delta_{r,s+a_{3}}+J_{y}\delta_{r,s-a_{1}+a_{3}}\\
J_{x}\delta_{r,s}+J_{y}\delta_{r-a_{2},s} & J_{z}\delta_{r,s}
\end{array}\right]\nonumber
\label{eq_ttham}
\end{equation}

The Hamiltonian $\mathcal{H}$ can be written as $\sum E {d_E^\dagger d_E}$ where $E$ and $d_E$ are single-mode energies (eigenvalues of $\mathfrak{H}$) and annihilation operators. Symmetries of the Hamiltonian imply that the spectrum is symmetric about zero and $d^\dagger_E=d_{-E}$. The spectrum is gapped if $(J_x,J_y,J_z)$ does not satisfy triangle inequalities, and has a line-node in the three dimensional Brillouin zone otherwise.\cite{Mandal2009} 


\section{Kitaev model on hyper-honeycomb lattice with vacancies}

Having introduced the Kitaev model on the hyper-honeycomb lattice,  we now proceed to formulate the problem of static non-magnetic impurities (vacancies) in it, created by replacing magnetic atoms by non-magnetic ones. The spin Hamiltonian $H^V$ of this system is the same as that of the clean system (Eq-\ref{Spin-Hamiltonian}) but with the spin-interaction terms involving the vacancy site absent.  

The Hamiltonian $H^V$, even in the presence of vacancies, commutes with all loop-operators and hence the problem is exactly solvable. As before, the group formed by the loop-operators is isomorphic to the cycle-space of the lattice treated as a graph. A lattice with $n$ isolated vacancies has $4N-n$ sites (vertices) and $6N-3n$ bonds (edges). The number of independent loop-operators, given by the circuit rank,\cite{Note1} is now $2N-2n+1$. Thus two flux degrees of freedom are removed from the vicinity of each vacancy as explained below
\begin{enumerate}
\item Each site in the clean lattice is a part of ten plaquettes (Fig-\ref{elementary-plaquette}). Loop-operators of these ten plaquettes are removed from the set of generators and are replaced by two loop-operators on the two `defect' plaquettes as shown in Fig-\ref{newplaquettes}.
\item Each site in the clean lattice is a part of nine constraint-generating tetrahedral volumes (Fig-\ref{tetrahedra}). These are removed and replaced by three new constraints involving the loop-operators on plaquettes surrounding the three volumes in Fig-\ref{newvolumes}.
\end{enumerate}
Other generators and constraints on the fluxes are unaltered by the vacancies.
\begin{figure}
\includegraphics[width=\columnwidth]{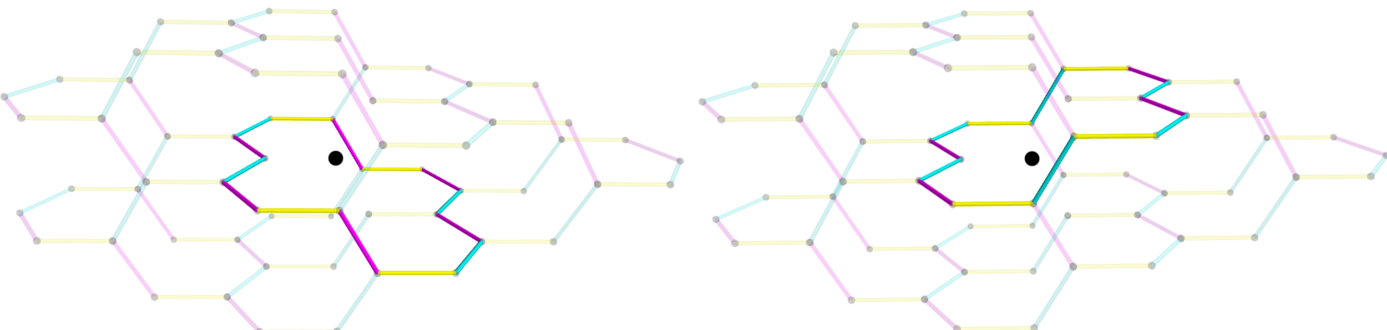}
\caption{(Colour online) Volume around the vacancy (dot). All plaquettes and constraints outside this volume are unaffected by the vacancy. The two highlighted loops are the new (merged) elementary plaquettes that are formed due to the vacancy.}
\label{newplaquettes}
\end{figure}

\begin{figure}
\includegraphics[width=\columnwidth]{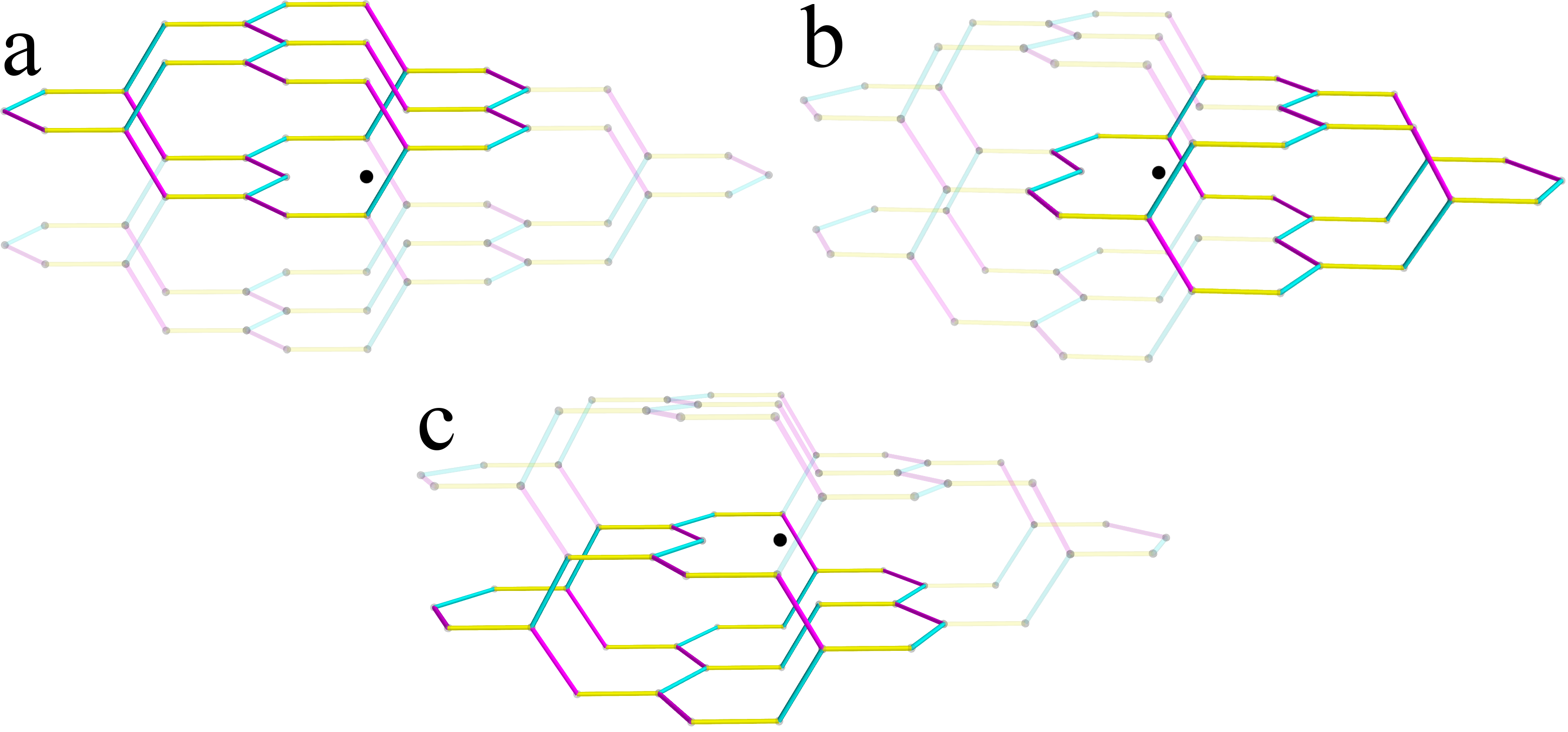}
\caption{(colour online) Highlighted set of bonds in each of the three panels contain seven (a,c) or six (b) plaquettes, each set enclosing a volume. The product of the corresponding loop-operators in each case equals identity. Thus these three panels represent three constraints on the loop-operators.}
\label{newvolumes}
\end{figure}

The physics of the vacancies essentially arises from the sites surrounding the vacancy. It is therefore useful to set up a notation for these sites. The vicinity of a vacancy is shown in Fig-\ref{localdof}. The sites that were connected to the vacancy site through $x,y$ and $z$-bonds are labeled $\mathbf{1,2}$ and $\mathbf{3}$ respectively. In terms of the operators in the extended Hilbert space, these three sites host three Majorana $b$ operators that do not appear in the Hamiltonian, thus contributing three Majorana zero-modes of the system.

There are $c$ Majoranas at these three sites which couple only to two other sites as opposed to generally three couplings. Careful analysis in the gapped phase shows that this results in another Majorana zero-mode. In the next several sections we show that these modes hybridize to produce a low-energy spin-moment and respond to external magnetic fields. 

\begin{figure}
\includegraphics[width=.8\columnwidth]{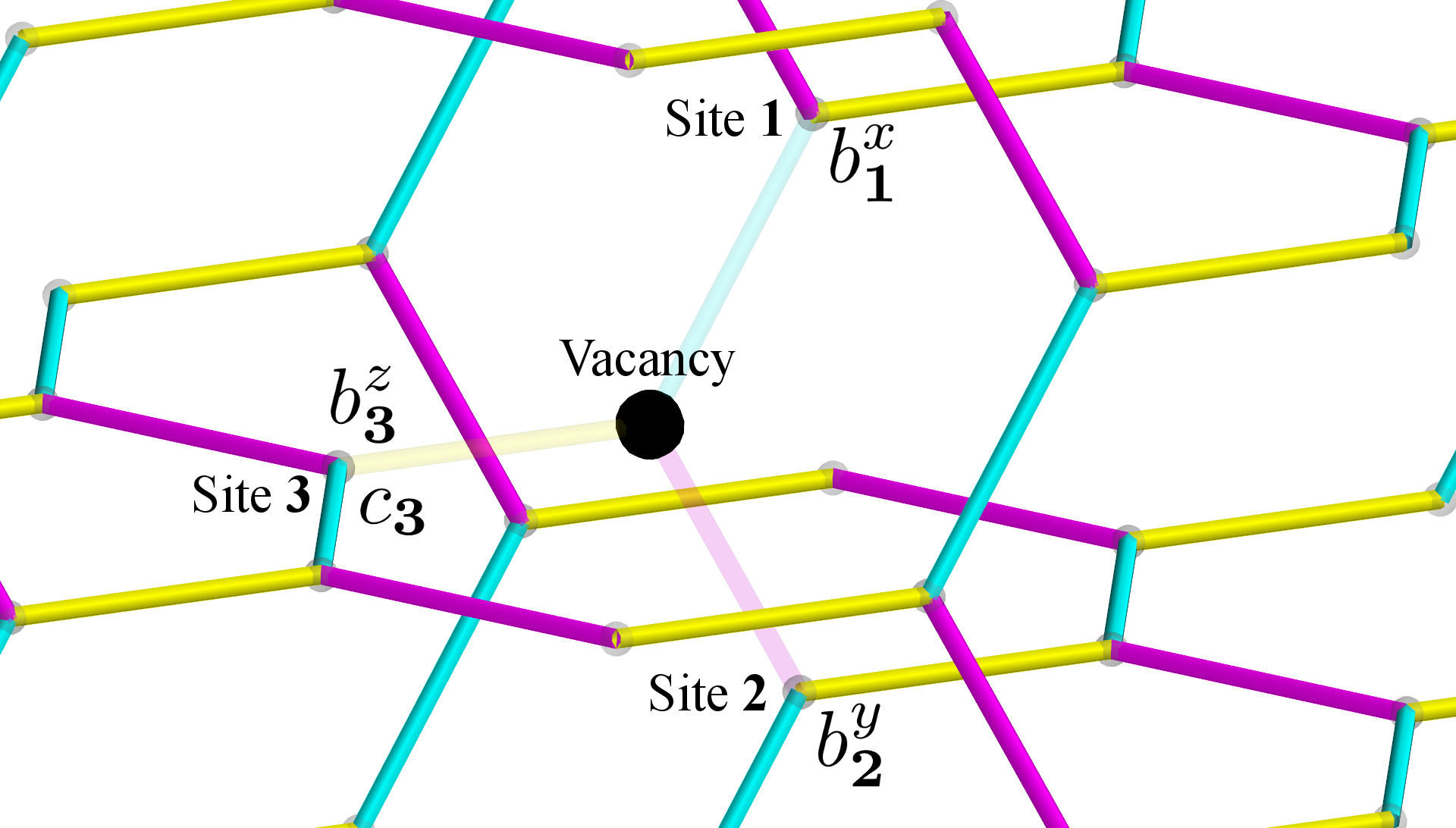}
\caption{(Colour online) Removal of a spin eliminates its interactions (represented by blurred bonds) with sites $\mathbf{1}$, $\mathbf{2}$ and $\mathbf{3}$ surrounding it. As a result, three Majorana operators - $b_{\mathbf{1,2,3}}^{x,y,z}$ do not appear in the Hamiltonian. A fourth zero-mode of the Hamiltonian arises due to a $c_3$ Majorana operator that is weakly coupled to the rest of the system.}
\label{localdof}
\end{figure}

The extended Hilbert space of the clean Kitaev model was constructed as the Fock space of bond-fermions $\chi$ and matter-fermions $f$. In the presence of the vacancy, the corresponding Hilbert space can be written as 
\begin{equation}
\mathcal{V} = \mathcal{V}_{\rm bond} \otimes \mathcal{V}_{\rm matter} \otimes \mathcal{V}_{b^z_\mathbf{3},c_\mathbf{3}} \otimes \mathcal{V}_{b^x_\mathbf{1},b^y_\mathbf{2}}\label{Hilbertspace-vac}
\end{equation}
The bond-fermions on the $6N-3$ bonds and matter-fermions on the $2N-1$ $z$-bonds are defined as before. $\mathcal{V}_{b^z_\mathbf{3},c_\mathbf{3}}$ and $\mathcal{V}_{b^x_\mathbf{1},b^y_\mathbf{2}}$ are simply the Fock spaces for the fermionic operators $\frac{1}{2}(b^z_\mathbf{3}+\imath c_\mathbf{3})$ and $\frac{1}{2}(b^x_\mathbf{1}+\imath b^y_\mathbf{2})$ respectively. The group of gauge-transformations $\mathfrak{D}$ is generated by the ${4N-1}$ operators $D_\mathbf{r}$. The physical space corresponds to the $2^{4N-1}$ dimensional even fermion-parity subspace of the orbit-space $\mathcal{V}/\mathfrak{D}$. (The fermion-parity constraint is discussed in Appendix-\ref{App:FermionParity}) The extended Hamiltonian commutes with loop-operators, bond-operators and gauge-transformations even in the presence of the vacancy. As before,\cite{Willans2010,*Willans2011} this allows us to obtain the spectrum within a flux sector by considering a gauge fixed, tight-binding Hamiltonian.
\subsection{Absence of Flux through the defect-plaquettes near a vacancy}
\label{sec:absflux}

 \begin{figure}
\includegraphics[width=\columnwidth]{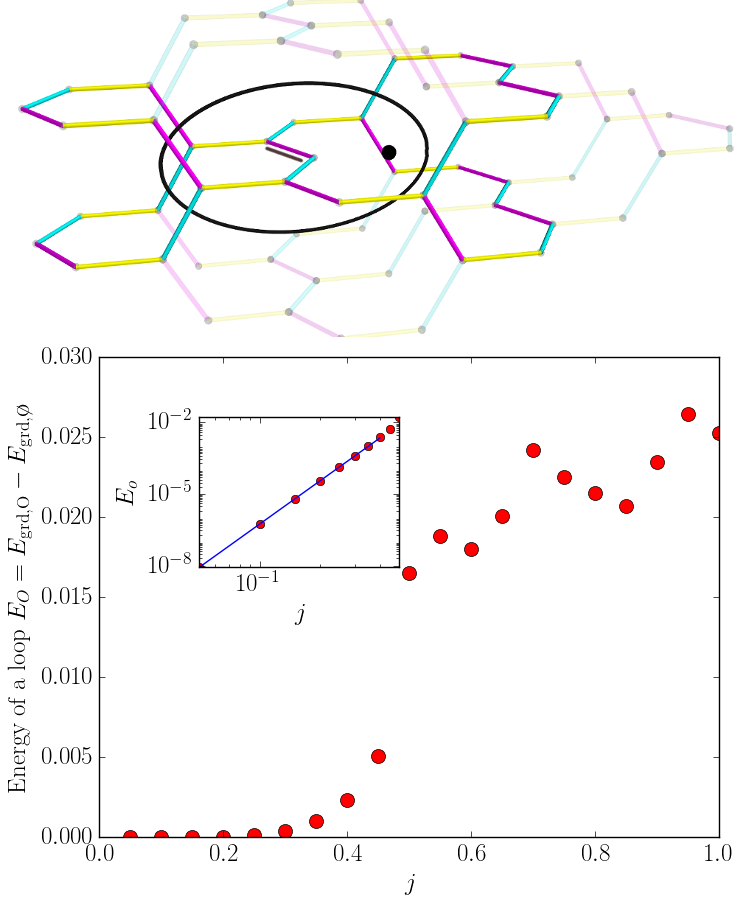}
\caption{(Colour online) (top) Shortest flux-loop that passes through the defect-plaquettes. The corresponding gauge-sector is obtained by flipping the bond-operator on the $y$ bond marked in the figure. (bottom) Energy $E_O$ of the flux-loop calculated as the difference between the ground state energies in the matter-fermion sector with ($E_{\rm grd,O}$) and without ($E_{\rm grd,\not{O}}$) the flux-loop. Inset shows  $E_{O}$ as a function of $j$ in the gapped phase. The solid line shows the energy estimate from the perturbative calculations on the two $10$-sided plaquettes. The energy estimates in the gapless phase show considerable finite size effects but are always positive.}
\label{fluxloopenergy}
\end{figure}

In the two-dimensional honeycomb lattice, upon introducing a vacancy, three six-sided plaquettes around the vacancy are removed and replaced by a single twelve-sided plaquette. Perturbative calculations for the gapped phase as well as numerical simulations showed that it is energetically favourable to have a flux ($W_{\rm defect}=-1$) through the defect plaquette.\cite{Willans2010,*Willans2011} In systems with periodic boundary conditions, however, a global constraint that these fluxes should be even in number allows their presence only if there are two or more vacancies (If the number of vacancies is odd, fluxes can be present only on an even number of them).

A vacancy in hyper-honeycomb, as described at the beginning of this section, results in removal of ten ten-sided plaquettes (Fig-\ref{elementary-plaquette}) and formation of two fourteen-sided defect-plaquettes (Fig-\ref{newplaquettes}). Perturbative calculations for the gapped phase $J_z>J_x+J_y$ tells us that at the lowest order of perturbation, the energy cost of a flux through such a plaquette is positive and hence a flux through the vacancy is energetically unfavourable. A more important energetic restriction on this flux can be seen by recalling that local constraints in three dimensions force these fluxes to form closed loops. Thus any flux-loop that passes through the defect plaquette has to pass through the healthy plaquettes away from the vacancy. The energy cost of such an extended flux-loop prevents its presence in the ground state. This is an important difference between vacancy problems in two and three dimensional Kitaev models and one of the central results of the present work.

The above arguments are valid only in a perturbative sense in which the energy cost of a flux arises only from the plaquettes it passes through. In order to test the absence of flux binding to a vacancy, beyond this perturbative limit, we numerically study the ground state energy for a system with a flux-loop shown in Fig-\ref{fluxloopenergy} (top). This flux-loop passes through two fourteen-sided and two ten-sided plaquettes. The ground state energy is calculated in a finite system with $10^3$ unit-cells and periodic boundaries by diagonalizing the tight-binding model obtained in a suitable gauge. Calculations were done for parameters $j=J_x=J_y$ and $J_z=1$. In the gapped phase, the energy cost of the flux-loop arises primarily from the ten-sided plaquettes. At lowest order of perturbation, this energy cost is $2\times\frac{35}{128}j^6$, which agrees well with the numerical estimates as shown in Fig-\ref{fluxloopenergy} (inset). Energy cost in the gapless phase ($j>0.5$) show strong finite-size effects but is always positive. We have studied only the energy cost of the shortest loop possible in the system. It is unlikely that a longer flux-loop will result in energy gains as this would entail passing the flux through even more number of healthy plaquettes.

The energy cost of such a flux-loop is very small ($\sim j^6$) in the gapped phase. In presence of other perturbations like an external magnetic field, these fluxes become dynamical. However for sufficiently weak fields ($\ll j^6$), this effect can be ignored and the fluxes can be treated as static. We postpone the discussion of the response to such external fields to later sections and concentrate on understanding the physics of a single vacancy, particularly the low-energy degrees of freedom associated with it.

\subsection{Degrees of freedom associated with a single vacancy}
\label{sec:vacancydof}

Introduction of vacancies leads to formation of new low-energy degrees of freedom that are absent in the clean system. As discussed earlier, introduction of $n$ vacancies reduces the number of flux degrees of freedom from $2N+1$ in the clean system to $2N-2n+1$. Since there are $4N-n$ spin-$\frac{1}{2}$ degrees of freedom in such a system, the number of non-flux degrees of freedom is $2N+n-1$. $2N-n$ of these arise from the matter-fermions on the $2N-n$ $z$-bonds. Deep in the gapped phase, where $J_z\gg J_x, J_y$, it is energetically expensive (energy cost $\sim O(J_z)$) to change the number of such fermions on any bond. This leaves $2n-1$ low-energy degrees of freedom. We show in this subsection that $n$ out of these are modes bound to each vacancy. In the next section, we show that each of these vacancy-modes couples to an external field i.e. has a spin-moment, and thus contributes to the low-field magnetization. The remaining $n-1$ degrees of freedom do not couple to the external field. We defer the discussion of these to the section on a pair of vacancies.

In terms of the Majorana fermion representation, there are three operators $b_{\mathbf{1,2,3}}^{x,y,z}$ (Fig-\ref{localdof}) on the sites around the vacancy that do not enter into the Hamiltonian, and therefore commutes with the Hamiltonian. Consider the following local, gauge invariant operators 
\begin{equation}
\tau_z = \imath b_\mathbf{1}^x S_{\mathbf{12}} b_\mathbf{2}^y;\; \tau_x = \imath b_\mathbf{2}^y S_{\mathbf{23}} b_\mathbf{3}^z;\; \tau_y = \imath b_\mathbf{1}^x S_{\mathbf{31}} b_\mathbf{3}^z
\label{vacancy-spin}
\end{equation}
$S_{\mathbf{ij}}=S_{\mathbf{ji}}$ is the product of an even-length string of bond-operators along a path between sites $\mathbf{i}$ and $\mathbf{j}$. It is convenient to choose  $S_{\mathbf{31}}=S_{\mathbf{12}}S_{\mathbf{23}}$. Different choices of the string $S_{\mathbf{ij}}$ and $S_{\mathbf{ij}}'$ result in different realizations of operators $\tau$ and $\tau'$. These are related to each other by a loop-operator $\tau=\tau' W$ with $W=S_{\mathbf{ij}}S_{\mathbf{ij}}'$. 

It is easy to check that the $\tau$ operators can be written as products of an odd number of spin operators and hence are gauge invariant and odd under time-reversal. The $\tau$ operators satisfy the spin-algebra and commute with the Hamiltonian, loop and bond-operators. As a result of the algebra, the $\tau_{x,y}$ flips the $\tau_z$ eigenvalue without changing the energies of states. Thus they represent a zero-energy mode near the vacancy. 

Note that after a gauge-choice, the operators $S_{\mathbf{ij}}$ become $c$-numbers and $\tau$ become parity operators for fermions defined using pairs from $b_{\mathbf{1,2,3}}^{x,y,z}$.
\subsection{Effect of external magnetic field on a vacancy}
Having shown the existence of a free spin-$\frac{1}{2}$ like degree of freedom $\tau$ near the vacancy, we now aim to understand its response to an external field. Note that the $\tau$ variables are not the same as any of the underlying spins $\sigma$ and therefore their coupling to an external magnetic field is not obvious. However, we show that in an effective low-energy description, Zeeman coupling of the underlying spins $\sigma$ appears as a coupling of the vacancy spin-moments $\tau$ to the external field. We show this explicitly in the gapped phase $J_z>J_x+J_y$.  In  the gapless phase, we simply calculate the contribution to magnetization arising from these vacancy-induced moments. 

The underlying spins $\sigma_\mathbf{r}$, couple to an external magnetic field through a Zeeman term in the Hamiltonian
\begin{equation}
H_{Z}=\sum_{\alpha=x,y,z} \;\sum_{\mathbf{r}\in \mathcal{S}-V}h^\alpha \sigma^\alpha_\mathbf{r}\label{Z-man}
\end{equation}
where $\mathcal{S}-V$ are the non-vacancy sites. Since the loop-operators do not commute with $H_Z$, fluxes are not conserved and the system is no longer exactly solvable. However, for fields small enough that the Zeeman energy scale is much smaller than the flux-gap, the low-energy behavior can be understood by projecting the Hamiltonian to a flux-free sector, while treating the Zeeman correction perturbatively. In a clean system, the corrections that are first order in $h$ vanish as they necessarily couple different flux sectors and the leading order corrections are of order $h^3$. This leads to next-nearest-neighbour hopping for the $c_i$ Majorana fermions.\cite{Kitaev06,Hermanns2015} 

In the presence of a vacancy, adjacent pairs of plaquettes around the vacancy merge together, voiding the above argument. The result is that leading order perturbative corrections arising from Zeeman terms on the sites surrounding the vacancies are linear:
\begin{align}
H_{Z}^{eff}=(h^x\sigma^x_\mathbf{1}+h^y\sigma^y_\mathbf{2}+h^z\sigma^z_\mathbf{3})
\label{eq_eff_zeeman}
\end{align}
where the three spins denoted by $\sigma^x_\mathbf{1},\sigma^y_\mathbf{2},\sigma^z_\mathbf{3}$ are the three nearest-neighbour spins of the vacancy (Fig-\ref{localdof}).

The vacancy spins $\tau^\alpha$ do not commute with $H_Z^{eff}$:
\begin{align}
[\tau^\alpha, H^{eff}_Z]=2\tau^\alpha(H^{eff}_Z-H^{\alpha}_Z)
\label{eq_eff_tau_comm}
\end{align}
where  $H^x_Z=h^x\sigma_\mathbf{1}^x, H^y_Z=h^y\sigma_\mathbf{2}^y$ and $H^z_Z=h^z\sigma^z_\mathbf{3}$. As a result, the $\tau$-mode splits, leading to a field-dependent ground state energy. This causes the finite magnetic response originating from the vacancy.

In the following sections, we explore this low-field response in greater detail. We organize our discussion in the following way. We first discuss the precise nature (such as the wavefunctions) of the low-energy modes surrounding a vacancy in the gapped phase $J_z>J_x+J_y$. We use this to analyse the response of a  vacancy-spin to an external field. The free-spin like behavior of isolated vacancy is analysed first. This is followed by the discussion of a pair of impurities, where we need to take into account the effect of interactions between vacancies at finite separations.

The discussion of the gapless phase follows that of the gapped phase. We start with the analysis of an isolated vacancy, whose response to external field is reduced by the interaction with the finite density of low-energy states in the surrounding spin liquid. Analysis of a pair of impurities demonstrates a longer range, anisotropic, sublattice dependent interaction.

For simplicity we focus mainly on the calculation of the response to a $z$-directed external field.

\section{Vacancy in gapped phase}
\label{sec:gapped-vac}

In this section, we analyse the effect of a vacancy in the gapped phase of the clean system. There are three gapped phases corresponding to $J_x>J_y+J_z$, $J_y>J_x+J_z$ and $J_z>J_x+J_y$. In the two-dimensional model on honeycomb lattice, the three gapped phases are equivalent due to a $C_3$ lattice symmetry. In case of the hyper-honeycomb lattice, the $x$ and $y$ bonds are related by a $C_2$ symmetry, but the $z$-bonds are distinct from them.\cite{Lee2014} In this paper, we will focus on the gapped phase $J_z>J_x+J_y$ only.
\subsection{Single vacancy}
The tight binding Hamiltonian $\mathcal{H}$ (Eq. \ref{cleanmatrixham}) for a clean system can be written as
\begin{align}
\mathcal{H} =
\frac{1}{2}\left[\begin{array}{cc}
\mathbf{c_{o}} & \mathbf{c_{e}}\end{array}\right]\mathfrak{H}\left[\begin{array}{c}
\mathbf{c_{o}}^T\\
\mathbf{c_{e}}^T
\end{array}\right]\nonumber\\
 \mathfrak{\mathfrak{H}=\left[\begin{array}{cc}
0 & \mathfrak{H}_{oe}\\
\mathfrak{H}_{oe}^{\dagger} & 0
\end{array}\right]}
\label{eq_ttham2}
\end{align}
$\mathfrak{H}_{oe}$ is a $2N\times 2N$ matrix that couples odd and even sublattices. $\mathbf{c_{o(e)}}$ is a vector of size $2N$ containing all the $c$ operators on odd (even) sublattice. 

If there are $n$ vacancies, all on the odd sublattice sites $V=\{\mathbf{v}_i\}_{i=1\to n}$, the Hamiltonian  $\mathcal{H}^V$ for the system in a fixed gauge has a similar tight-binding form but $\mathbf{c_{o}}$, $\mathbf{c_{e}}$ and $\mathfrak{H}_{oe}$ now have sizes $2N-n$, $2N$ and $(2N-n)\times(2N)$ respectively. Rank-deficiency of $\mathfrak{H}_{oe}$ implies that $\mathfrak{H}$ has $n$ orthogonal null vectors $\{\psi_\alpha\}_{\alpha=1\to n}$
\begin{equation}
\psi_{\alpha}\left(\mathbf{r}\right)=\begin{cases}
0 & \mathbf{r}\in\text{odd sublattice}\\
\phi_{\alpha}\left(\mathbf{r}\right) & \mathbf{r}\in\text{even sublattice}
\end{cases}
\end{equation}
where $\{\phi_{\alpha}\}_{\alpha=1\to n}$ are $n$ non-zero, orthonormal null vectors of $\mathfrak{H}_{oe}$, existence of which is guaranteed by rank-deficiency of $\mathfrak{H}_{oe}$. These null vectors are associated with $n$ Majorana zero-modes 
\begin{equation}
C_{\alpha} =\sum_{\mathbf{r}\in \mathcal{S}-V}\psi_\alpha(\mathbf{r})c_{\mathbf{r}}.
\label{eq_majzero}
\end{equation}

In particular, for a single vacancy at $\mathbf{v}$ on the odd sublattice there is one zero-mode which can be obtained as
\begin{equation}
\psi(\mathbf{r})\propto [\mathfrak{H}^{-1}]_{\mathbf{r},\mathbf{v}}
\end{equation}
where $\mathfrak{H}^{-1}$ is the inverse of the single-mode Hamiltonian of the clean system (See Appendix-\ref{app:zeromode} for an explanation). 
The single-mode Hamiltonian can be inverted as shown in Appendix-\ref{appendix-invert}. For a vacancy at the location $\mathbf{v}\equiv(r_v=0,i_v=1)$, the wavefunction $\psi$ is given by 
\begin{equation}
\psi(r) \equiv a_0\left[0,0,F(-r),F_{q}\left(-r\right)\right]
\label{zeromodewf}
\end{equation}
where the row-vector on the RHS gives the amplitudes on the sublattices $1,3,2$ and $4$ in the unit-cell $r$.  $F_q(-r)$ and $F(-r)$ are functions that exponentially decay with distance, shown in Eq-\ref{FqFp} and \ref{F}, and $a_0$ is the normalization. $\psi(\mathbf{r})$ has support only on the even sublattices falling inside a pyramid shaped zone on one side of vacancy (Fig-\ref{zeromode}).

\begin{figure}
\includegraphics[width=\columnwidth]{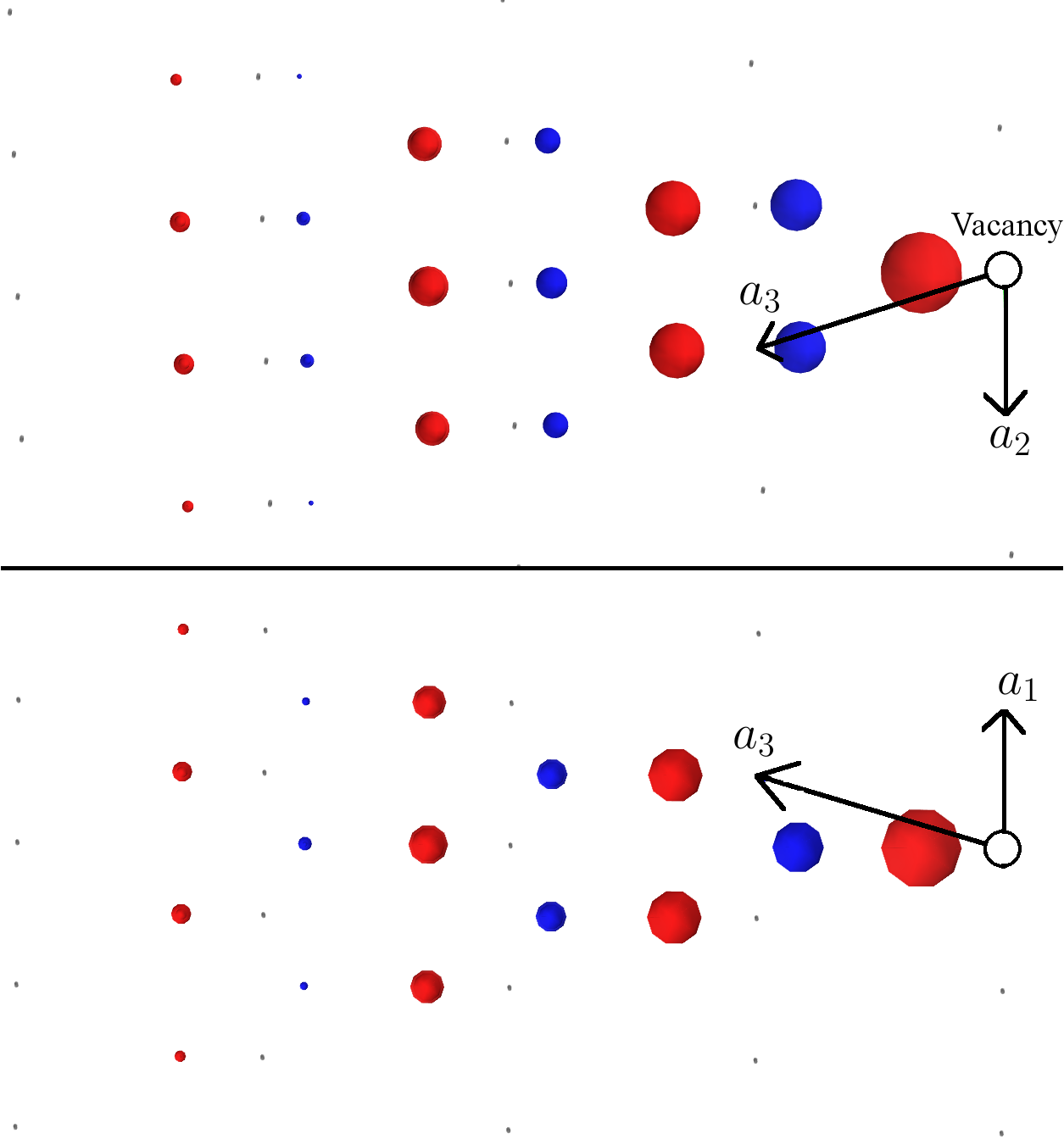}
\caption{Wavefunction of a zero-mode in the matter-fermion sector due to a vacancy on the sublattice $1$ for a gapped system $J_x=J_y=jJ_z$. Top and bottom figures show the view from $a_1$ and $a_2$ directions. The radius of the circles show the magnitude of the wavefunction on a log-scale. The wavefunction has a  phase difference of $\pi$ between sublattices $2$ (red circles) and $4$ (blue circle) . Black dots in the background represent the location of sublattice 1 sites of every unit-cell.}
\label{zeromode}
\end{figure}

For simplicity of analysis, we consider the case $J_x=J_y=jJ_z$ with $j<.5$. The wavefunction decays the slowest along the direction of $z$-bonds with a length scale  $\frac{1}{\ln[4j^2]}$. The normalization $a_0$ is
\begin{eqnarray}
\left[\frac{2\pi j^{2}}{(1+4j^{2})K(16j^{4})-E(16j^{4})}\right]^\frac{1}{2}
 \sim 
\frac{1}{\sqrt{1+2j^2}}
\label{eq_norm}
\end{eqnarray}
where $K,E$ are the elliptic integrals.

This Majorana zero-mode, arising out of the matter-fermions, is in addition to the three other Majorana zero-modes $b^{x,y,z}_{\mathbf{1,2,3}}$ surrounding the vacancy.

\subsubsection*{Coupling to the magnetic field}

Having identified the zero-modes, we now show that these zero-modes couple to an external magnetic field. For  the gapped phase obtained by making the $z$-bonds strong, the two spins sharing a $z$-bond are parallel (for $J_z<0$) or antiparallel (for $J_z<0$) with the quantization axis being along $\sigma^z$. For small magnetic fields, staying within the ground state flux sector, the low-energy effective Zeeman Hamiltonian is given by Eq-\ref{eq_eff_zeeman}. For strong $z$-bonds, this further reduces to
\begin{align}
\mathcal{H}_{Z}^{eff}=h_z\sigma^z_3
\label{eq_heffz}
\end{align}
as $h_x$ and $h_y$ terms are suppressed by the energy cost to flip the spins at sites $\mathbf{1}$ and $\mathbf{2}$ (Fig-\ref{localdof}) as these form parts of $z$-bonds. In such a low-energy space, the commutation relations given by Eq-\ref{eq_eff_tau_comm} become:
\begin{align}
&[\tau^x,\mathcal{H}_Z^{eff}]=2\tau^x\mathcal{H}^{eff}_Z;~~[\tau^y,\mathcal{H}_Z^{eff}]=2\tau^y\mathcal{H}^{eff}_Z;\nonumber\\
&[\tau^z,\mathcal{H}_Z^{eff}]=0
\end{align}
Considering that $\tau$ is odd under time reversal, this tells us that the low-energy Hamiltonian has the form
\begin{align}
\mathcal{H}_{\rm low-energy}=gh_z \tau_z
\end{align}
where $g$ is a coupling constant whose magnitude will be determined below using a more microscopic calculation.

\begin{figure}
\includegraphics[width=\columnwidth]{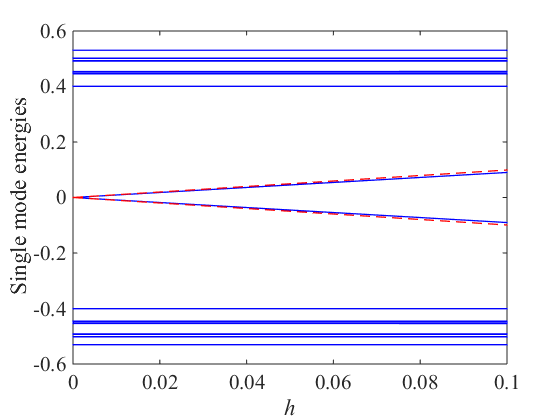}
\caption{Single-mode spectrum around $0$ for a system with $8^3$ unit-cells as a function of $h_z$ calculated by exact diagonalization of Eq-\ref{HamiltonianwithVacAndH} ($\frac{J_x}{J_z}=\frac{J_y}{J_z}=0.3$). Blue lines indicate the mode energies as a function of $h_z$. The splitting of mid-gap modes increases with $h_z$, while other modes are unaffected. The red line shows the predicted energies of the mid gap modes given in Eq-\ref{predictedmidgapenergies}.}
\label{singlemodesvsh}
\end{figure}

To gain more insight into the nature of the splitting, we recall from the discussion below Eq-\ref{vacancy-spin}, that within a gauge-sector, $\tau$ operators can be thought of as parity operators. In particular  $\tau_z\sim\imath b_\mathbf{1}^x b_\mathbf{2}^y$. For $\tau_z=+1$, lowest energy state of the system is obtained by occupying all modes below zero energy. When $\tau_z$ is flipped to $-1$, the system gains an additional fermion. In order to satisfy the fermion parity constraint (Appendix-\ref{App:FermionParity}), one less fermionic mode needs to be occupied in the remaining Hilbert-space $\mathcal{V}'=\mathcal{V}/\mathcal{V}_{b^x_{\mathbf{1}},b^y_{\mathbf{2}}}$ (See Eq-\ref{Hilbertspace-vac}). The Hamiltonian in the presence of the vacancy acts only on $\mathcal{V}'$.  Therefore, at lowest energy, the fermionic parity constraint is met by un-occupying the mode closest to zero energy. The energy change upon flipping the vacancy-spin is just the energy of this mode. This is schematically shown in Fig-\ref{schematic-spectrum}.

\begin{figure}
\includegraphics[width=\columnwidth]{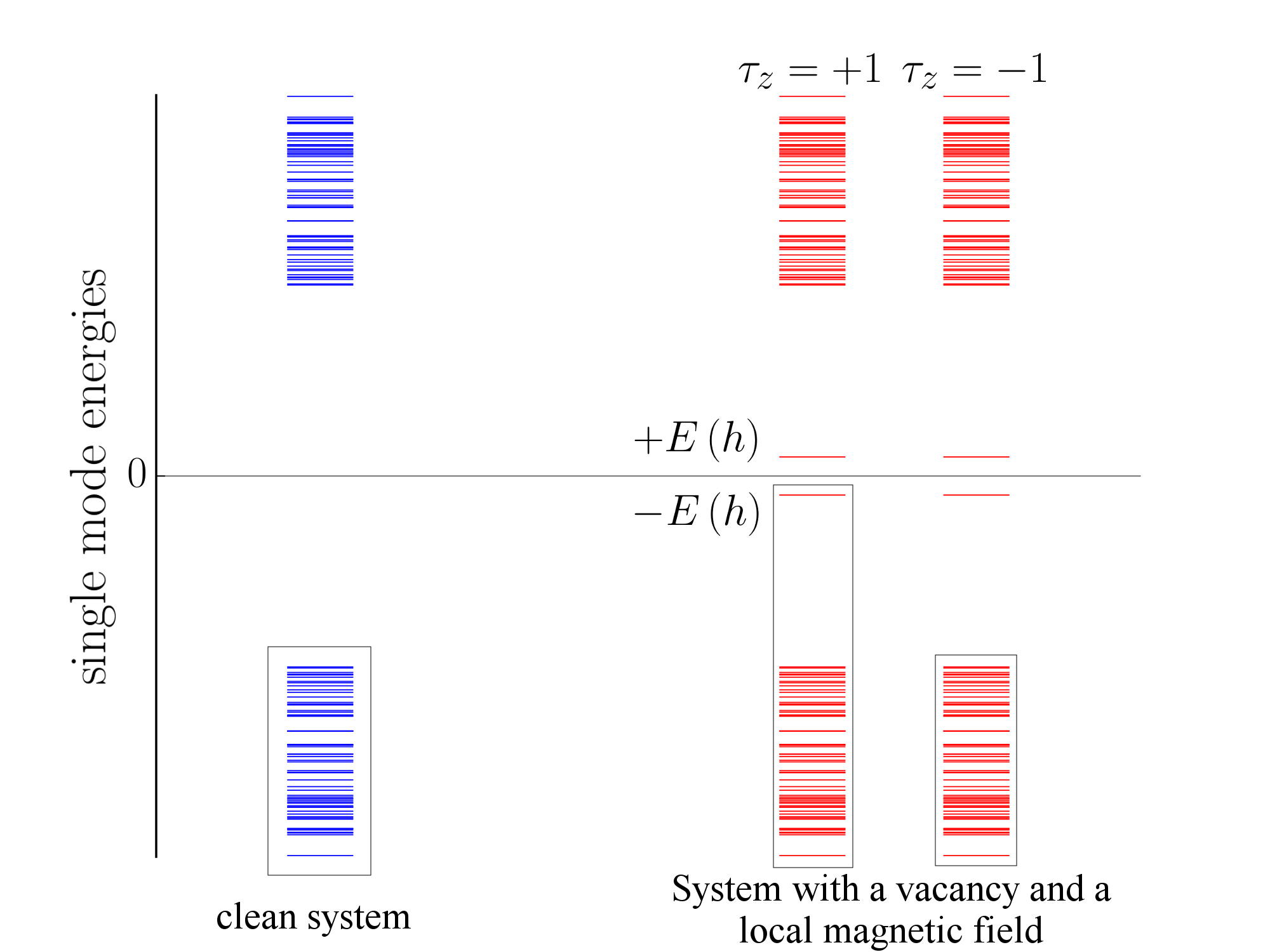}
\caption{Cartoon of the single-mode spectrum of the clean system (LHS), and of a system with a vacancy in a field (RHS) for the cases $\tau_z=\pm 1$. Rectangular boxes enclose the modes occupied in each case. In case of $\tau_z=-1$, one mode of energy $-E(h_z)$ is unfilled in order to satisfy fermion-parity, increasing the total energy by $E(h_z)$. The energy of the other modes are only weakly dependent on the magnetic field due to the large fermionic gap.}
\label{schematic-spectrum}
\end{figure}
A more controlled microscopic description of the above ideas is obtained by recalling that for a very weak field, we can work in the zero flux sector, and obtain the tight binding Hamiltonian as
\begin{equation}
\mathcal{H}^V_{h_z} =\mathcal{H}^V+\mathcal{H}^{eff}_Z= \mathcal{H}^V + \imath h_z b_\mathbf{3}^z c_\mathbf{3}\label{tightbindinggappedv1hz}
\end{equation}
where $\mathcal{H}^{eff}_Z$ is given by Eq-\ref{eq_heffz} , $b_\mathbf{3}^z$ and $c_\mathbf{3}$ are operators at site $\mathbf{3}$ in Fig-\ref{localdof}. $\mathcal{H}^V$ is the Hamiltonian with a single vacancy. $\mathcal{H}^V_{h_z}$ has the following block form:
\begin{eqnarray}
\mathcal{H}^V_{h_z} &=\frac{1}{2} \left[b^{z}_\mathbf{3},\mathbf{c_{o}},\mathbf{c_{e}}\right] \mathfrak{H}_{h_z}^V \left[\begin{array}{c}
b^{z}_\mathbf{3}\\
\mathbf{c_{o}}^T\\
\mathbf{c_{e}}^T
\end{array}\right]  \label{HamiltonianwithVacAndH} \\ 
&\mathfrak{H}_{h_z}^V = \left[\begin{array}{ccc}
0 & 0 & U\\
0 & 0 & \mathfrak{H}_{oe}\\
U^{\dagger} & \mathfrak{H}_{oe}^{\dagger} & 0
\end{array}\right]\nonumber\\
& U_j =  \imath h_z \delta_{j R_\mathbf{3}}\nonumber
\end{eqnarray}
where the $U$ and $\mathfrak{H}_{oe}$ have sizes $2N$ and $2N-1\times 2N$. Row index $R_\mathbf{3}$ corresponds to the $c_\mathbf{3}$ mode. Fig-\ref{singlemodesvsh} shows a part of the single-mode-spectrum of the above Hamiltonian around zero energy, as a function of $h_z$. The spectrum is symmetric and has a gap around zero, with two mid-gap modes of field-dependent energy $\pm E(h_z)$. Energies of other modes have much weaker field-dependence.

The leading order effect of $H^{eff}_Z$ is to hybridize the zero-modes $b^z_\mathbf{3}$  and $\psi$ discussed in the previous subsection. The energies of the hybridized mid-gap modes can be obtained by projecting into the space of these two modes. 
\begin{multline}
\mathfrak{H}_{\rm mid-gap} = \left[\begin{array}{cc}
\left\langle 1|\mathfrak{H}_{h_z}^V|1\right\rangle  & \left\langle \psi|\mathfrak{H}_{h_z}^V|1\right\rangle \\
\left\langle 1|\mathfrak{H}_{h_z}^V|\psi\right\rangle  & \left\langle \psi|\mathfrak{H}_{h_z}^V|\psi\right\rangle 
\end{array}\right]=a_0 h_{z} \left[\begin{array}{cc}
0 & -\imath\\
\imath & 0
\end{array}\right]
\label{predictedmidgapenergies}
\end{multline}
where $\left|1 \right \rangle$ and $\left| \psi \right \rangle$ represent zero-modes $[1,0,0,0,\dots]$ and $\psi$ (Eq-\ref{zeromodewf}) respectively. This gives the energies $E(h)=\pm a_0 h_z$ and a constant magnetization of $a_0$ for the vacancy spin. This estimate is compared with the numerical derivative of the ground state energy of the Hamiltonian (Eq-\ref{HamiltonianwithVacAndH}) in a finite system in Fig-\ref{singlevacmagnetization}. 

Thus, the vacancy-induced degrees of freedom carry a magnetic moment which responds to the external magnetic field and thus contributes to the low-field magnetization. In the next sub-section, we shall show that in case of more than one vacancies, these moments interact with each other and this interaction leads to, among other things, a non-trivial dependence of the magnetization on the external Zeeman field.
\begin{figure}
\includegraphics[width=\columnwidth]{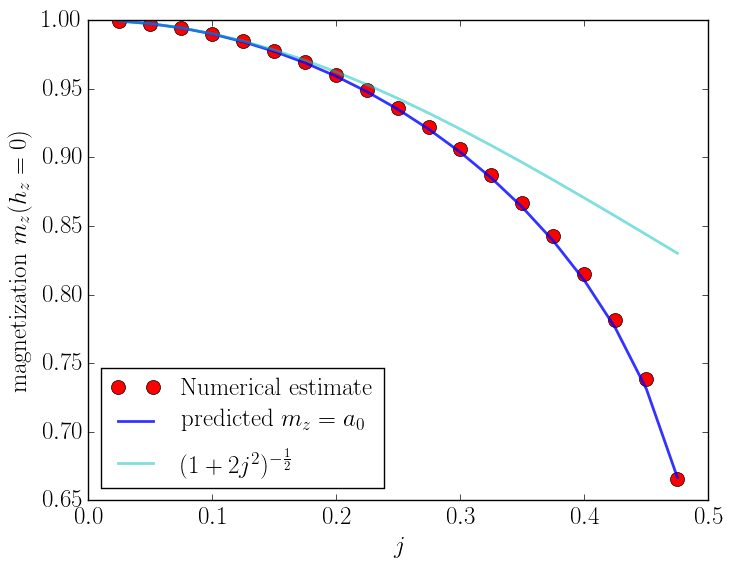}
\caption{Magnetization due to a single vacancy in a vanishing magnetic field. The numerical estimates were obtained in a system of $8^3$ unit-cells and periodic boundary conditions, by calculating the numerical derivative of the ground state energy of Eq-\ref{HamiltonianwithVacAndH}.}
\label{singlevacmagnetization}
\end{figure}
\subsection{Two Vacancies}

To understand the nature of the interaction between two vacancy-induced moments, we consider a system with two vacancies located at sites $\mathbf{v},\mathbf{v}'$. First, we study the scenario where these sites are infinitely far away from each other, such that any interaction between the local degrees of freedom can be neglected. In this case, the magnetization from the vacancy-induced degrees of freedom is a sum of the individual contributions.

After fixing the gauge, the remaining degrees of freedom of the system form the Hilbert space 
\begin{equation}
\mathcal{V}_{\rm matter} \otimes \mathcal{V}_{b^z_\mathbf{3},c_\mathbf{3}} \otimes \mathcal{V}_{b^y_\mathbf{2},b^x_\mathbf{1}} \otimes \mathcal{V}_{b^z_\mathbf{3'},c_\mathbf{3'}} \otimes \mathcal{V}_{b^y_\mathbf{2'},b^x_\mathbf{1'}}
\label{2vacHilbertSpace}
\end{equation}
of dimension $2^{2N-2}\times 2 \times 2 \times 2 \times 2 =2^{2N+2}$. Considering the fermion-parity constraint also, the physical subspace has dimension $2^{2N+1}$. A straightforward generalization of the single vacancy arguments of the last sub-section show that the tight-binding Hamiltonian $\mathcal{H}^V$ in the presence of the field couples only the components $\mathcal{V}_{\rm matter} \otimes \mathcal{V}_{b^z_\mathbf{3},c_\mathbf{3}}\otimes \mathcal{V}_{b^z_\mathbf{3'},c_\mathbf{3'}}$. The remaining two components $\mathcal{V}_{b^y_\mathbf{2'},b^x_\mathbf{1'}}$ and $\mathcal{V}_{b^y_\mathbf{2},b^x_\mathbf{1}}$ can be identified to be the $\tau_z$ and $\tau'_z$ degrees of freedom discussed previously.

As discussed in the case of a single vacancy in the previous section, the spectrum of the single-mode Hamiltonian has a gap with two mid-gap eigenvalues $\pm E(h)$ (close to $0$) for each vacancy. These four Majorana modes form two low-energy fermionic modes localized around the vacancies. Low-energy subspace of  Eq-\ref{2vacHilbertSpace} has the following decomposition.
\begin{equation}
\mathcal{V}_v\otimes \mathcal{V}_{\tau_z} \otimes \mathcal{V}_{v'}  \otimes \mathcal{V}_{\tau_z'}
\end{equation}
where $\mathcal{V}_v,\mathcal{V}_v'$ are the space of the two mid-gap modes of the Hamiltonian. The basis states of this low-energy space can be written as $\left |0,0,0,0 \right \rangle$, $\left |0,0,0,1 \right \rangle$, $\left |0,0,1,0 \right \rangle$ $\dots$ $\left |1,1,1,1 \right \rangle$ by labeling the  occupancy of each fermionic mode. Physical states have even number of fermions. 

Now we describe the local spin degrees of freedom in this picture. Flipping of one of the vacancy-bound spin degree of freedom is represented by simultaneous flipping of the parities of the two fermions bound to that vacancy. For example, starting from $\left |0,0,0,0 \right \rangle$, flipping the spin at $\mathbf{v}$ takes the state to $\left |1,1,0,0 \right \rangle$ and flipping the spin at $\mathbf{v}'$ changes the state to $\left |0,0,1,1 \right \rangle$. These processes are associated with an energy $E(h)$. In addition to these, there is one non-local degree of freedom which corresponds to simultaneous flipping of the $\tau_z$ and $\tau_z'$ eigenvalues. For example $\left |0,0,0,0 \right \rangle\to \left |0,1,0,1 \right \rangle$. These are not associated with any energy cost.

Such a description of the low-energy space can be generalized to the case of $n$ vacancies, where there are $n$ local spin-flips and $n-1$ degrees of freedom corresponding to pairwise flipping of $\tau_z$ eigenvalues.

Now we consider the case of vacancies that are a finite distance apart. This, as we show below, creates the possibility of the interaction between the vacancy moments, leading to a change in their total magnetization. The form of such an interaction depends crucially on the sublattice index of the sites in which the vacancies reside.

\subsubsection*{Vacancies on the same sublattice}

In the presence of a pair of impurities, both on the odd or both on the even sublattice and without a magnetic field, the mid-gap modes are at  zero energy. For a sufficiently distant pair of impurities, the normalization $a_0$ and the amplitudes of the wavefunctions near the vacancies are same as that of an individual vacancy. As a result, the arguments presented in the case of single vacancy can be applied to each vacancy independently and the total magnetization is again a constant, similar to the case of isolated vacancies.

\subsubsection*{Vacancies on opposite sublattices}

For two vacancies on opposite sublattices, the vacancy zero-modes hybridize, resulting in a small finite energy splitting (even in the absence of an external magnetic field) that decreases with their separation. The tight binding Hamiltonian has the form schematically shown in Fig-\ref{pictureofHamiltonian}.

\begin{figure}
\includegraphics[width=\columnwidth]{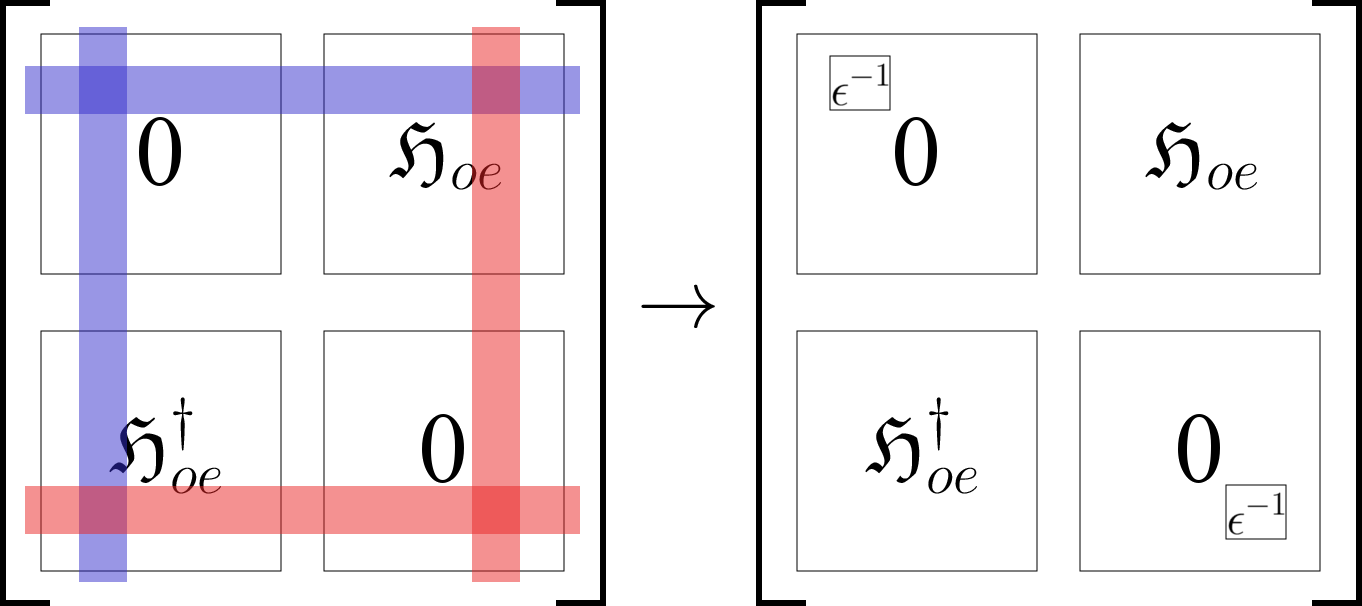}
\caption{Single-mode Hamiltonian in the presence of 2 vacancies on opposite sublattices is obtained by removing the rows and columns corresponding to those vacancy locations, shown here schematically as blue and red masks on the single-mode Hamiltonian of the clean system. Equivalently one can insert infinite on-site potentials $\epsilon^{-1}$ on the two vacancy sites.}
\label{pictureofHamiltonian}
\end{figure}

In order to obtain this splitting,  we consider a single-mode Hamiltonian $\mathfrak{H}^V$ with a diverging potential at the vacancy sites $\mathbf{v}$ and $\mathbf{v}'$, as shown in Fig- \ref{pictureofHamiltonian}. This has the same effect as removing two lattice sites.\cite{Willans2010,*Willans2011}

The new energy levels can then be inferred from the Green's function using t-matrix methods. Green's function at zero magnetic field, $G_{h=0}$, for the single-mode Hamiltonian $\mathfrak{H}^V$ can be written in terms of the Green's function $g$ of the clean system as $G_{h=0} = g + g T g$ where the t-matrix $T$ is given by (as described in Appendix-\ref{binomialInverse})
\begin{equation}
\left[\begin{array}{cc}
T_{\mathbf{v},\mathbf{v}} & T_{\mathbf{v},\mathbf{v}'}\\
T_{\mathbf{v}',\mathbf{v}} & T_{\mathbf{v}',\mathbf{v}'}
\end{array}\right]=\left[\begin{array}{cc}
\epsilon-g_{\mathbf{v},\mathbf{v}} & -g_{\mathbf{v},\mathbf{v}'}\\
-g_{\mathbf{v}',\mathbf{v}} & \epsilon-g_{\mathbf{v}',\mathbf{v}'}
\end{array}\right]^{-1}
\end{equation}
It can be shown that the energies of the hybridized mid-gap states occur at the poles of the t-matrix. In the limit, of $\epsilon$ going to zero, the poles occur at the solutions of $|g(\omega,\mathbf{v},\mathbf{v})|= |g(\omega,\mathbf{v},\mathbf{v}')|$. To linear order in $\omega$, $g(\omega)$ can be evaluated as $h^{-1}+\omega h^{-1}h^{-1}$ (this approximation is valid for $\omega$ inside the gap in the single-mode spectrum). This gives, for the energy of the hybridized states,
\begin{equation}
E=\pm a_0|\psi_\mathbf{v}(\mathbf{v}')|=\pm a_0|\psi_\mathbf{v}'(\mathbf{v})|
\label{energysplitting}
\end{equation}
where $\psi_{\mathbf{v}}(\mathbf{v}')$ is the magnitude at $\mathbf{v}'$ of the wavefunction of the zero-mode associated with a single impurity at $\mathbf{v}$. Thus to leading order, the vacancies do not interact if one vacancy does not sit on the support of the zero-mode of the other vacancy.

Thus, placing the vacancy $\mathbf{v}$ on the sublattice $1$, the modes hybridize if $\mathbf{v}'$ is on sublattices $2$ or $4$. The hybridized states are approximately given by $(\psi_\mathbf{v}\pm \psi_{\mathbf{v}'})/\sqrt{2}$ where the wavefunctions have energies $\pm E$ if $\mathbf{v}'$ is on sublattice $2$ and $\mp E$ if  it is on the sublattice $4$.

In the presence of an external field $h$, as for the case of the single vacancies, the leading contribution to the Hamiltonian arising from Zeeman terms is $H_Z^{eff} = h_z\sigma_{\mathbf{3}}^z+ h_z\sigma_{\mathbf{3}'}^z$.

With the knowledge of the wavefunctions, we can project the single-mode Hamiltonian in the presence of a magnetic field $h_z$, into the space of the mid-gap modes $(\psi_\mathbf{v}\pm \psi_{\mathbf{v}'})/\sqrt{2}$  and the modes associated with $b^z_\mathbf{3}$ and $b^z_{\mathbf{3}'}$ near the two vacancies to get 
\begin{equation}
h_{\text{mid-gap}} = \left[\begin{array}{cccc}
0 & 0 & \imath h_{z}\frac{a_{0}}{\sqrt{2}} & \imath h_{z}\frac{a_{0}}{\sqrt{2}}\\
0 & 0 & \imath h_{z}\frac{a_{0}}{\sqrt{2}} & -\imath h_{z}\frac{a_{0}}{\sqrt{2}}\\
-\imath h_{z}\frac{a_{0}}{\sqrt{2}} & -\imath h_{z}\frac{a_{0}}{\sqrt{2}} & \eta E & 0\\
-\imath h_{z}\frac{a_{0}}{\sqrt{2}} & \imath h_{z}\frac{a_{0}}{\sqrt{2}} & 0 & -\eta E
\end{array}\right] 
\end{equation}
$\eta$ is $+1$ and $-1$ for $v'$ on sublattices $2$ and $4$. This can be diagonalized to obtain the ground state energies  and magnetization as a function of the magnetic field $h_z$. The result is identical to that of the two dimensional system \cite{Willans2010,*Willans2011}
\begin{gather}
E_{\rm grd}(h_z)\approx-\sqrt{E^2+4a_0^2h^2_z}\nonumber\\
m(h_z)=\frac{4a_0^2 h_z}{\sqrt{E^2+4a^2_0 h_z^2}}
\end{gather}
where the energies $E$ are given by Eq-\ref{energysplitting}. For weak magnetic fields below a threshold $E/2a_0$, the magnetization is linear in $h_z$, and less than the sum of the magnetizations of isolated vacancies. This suppression arises from the interaction between the impurities, and is a signature of the hybridization discussed above. For fields larger than the threshold, the interaction is overcome by the external field, and the vacancy-induced moments behave like two independent polarized spins  leading to saturated magnetization. The crossover scale separating the low-field and high-field behavior decays exponentially with the distance between the vacancies.
\section{Vacancy in the gapless phase}
\label{sec:gapless-vacancy}

\begin{figure}
\includegraphics[width=\columnwidth]{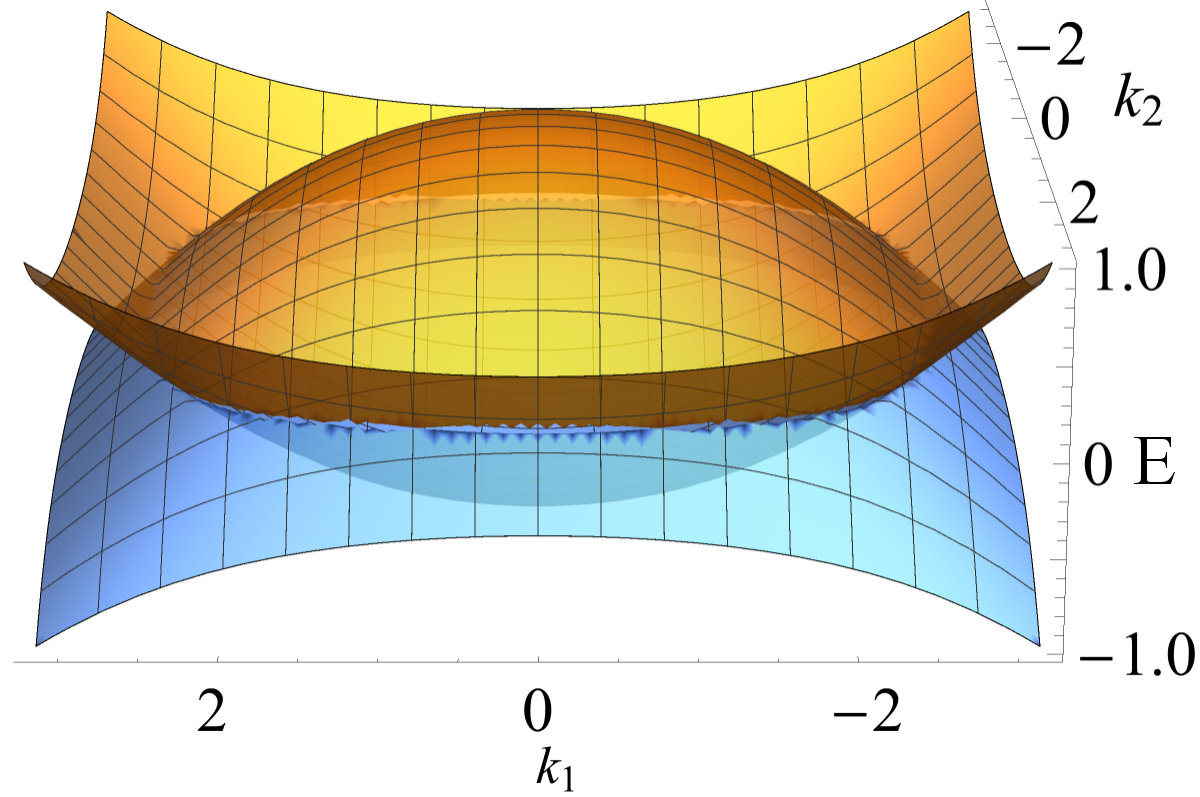}
\includegraphics[width=\columnwidth]{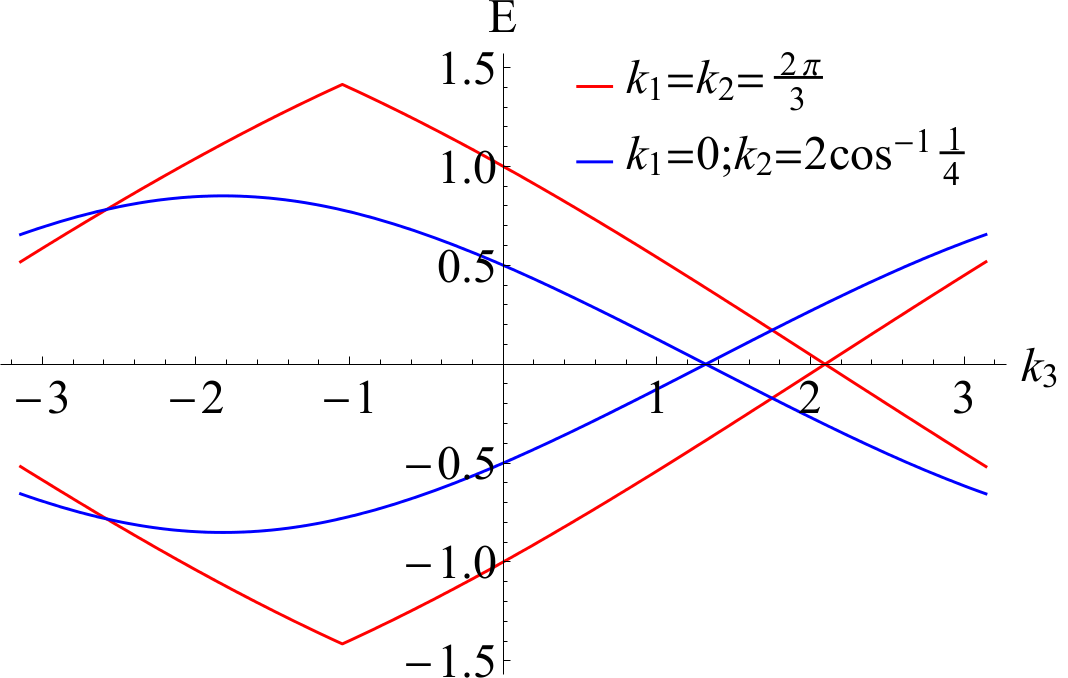}
\caption{(Top) Two out of the four bands in the single-mode spectrum touch at $E=0$ along a loop in the $k_1+k_2-2k_3=0$ plane. Figure shows these two bands as a function of $k_{1,2}$ in this plane. (Bottom) Spectrum as a function of $k_3$ along the lines $k_1=k_2=\frac{2\pi}{3}$ (red) and $k_1=0,k_2=2\cos^{-1}\frac{1}{4}$(blue). The line-node cuts this line at $k_3=\frac{k_1+k_2}{2}$. Near such points, the spectrum is linear along two directions normal to the line-node.}
\label{gaplessspectrum}
\end{figure}

Having discussed the physics of vacancies in the gapped phase, we now focus on the gapless $\mathbb{Z}_2$ spin liquid phase, i.e., when the parameters $J_{x,y,z}$ satisfy the triangle inequalities. In this phase, two out of the four bands touch at $E=0$ along a loop-like nodal line in the three dimensional Brillouin zone. For simplicity, here we study the system at parameter values $J_x=J_y=J_z=1$ (Fig-\ref{gaplessspectrum}) and analyse its response to a $z$-directed field. The nodal line occurs at the intersection of surfaces $k_1+k_2-2k_3=0$ and $4\cos \frac{k_1}{2} \cos \frac{k_2}{2}=1$ (for the gauge we have chosen).

It can be shown that the wavefunctions of the mid-gap modes derived for the gapped phase become non-normalizable on approaching the gapless phase. This is expected since, while the $\tau$s are still well-defined local degrees of freedom, because there is a finite density of modes near the zero energy (even for a clean system) the vacancy degrees of freedom can strongly hybridize with them invalidating the `local-mode' picture described above for the gapped phase. Therefore, instead of working with the explicit wavefunctions of various local-modes, we infer the response of the gapless system from the Green's functions in the presence of the vacancy and an external field. Magnetization can be estimated from the field-dependence of the ground state energy, which can be calculated from the first moment of the density of single-modes below zero energy. While the details differ, the methods used here are based on the ones introduced in Ref-\onlinecite{Willans2010,Willans2011}.

In a clean system, a weak external magnetic field opens up a gap which is (at the most) proportional to $h^3$ in the zero flux sector.\cite{Hermanns2015}  However, extrapolation of the results from the gapped phase in the previous section tells us that the field induced splitting of the vacancy modes increase faster - as $h^1$. Therefore, the vacancy modes can still hybridize with the modes of the surrounding spin-liquid, and the gap can be ignored. We will show that such a hybridization with the surrounding spin liquid reduces the magnetization of the isolated vacancy.

\subsection{Single vacancy response to magnetic field}

The leading contribution to magnetization of the system arises from the sites near the vacancy. In particular, for a $z$-directed field, the leading Zeeman correction arises from the action of the field on the site $\mathbf{3}$ (Fig-\ref{localdof}). In the presence of this contribution, namely  $H_Z^{eff}=h_z\sigma^z_\mathbf{3}$, the density of single-modes and thus the ground state energy are field-dependent. For sufficiently weak fields, we can work in the flux-free sector. The ground state energy can be calculated from the tight-binding Hamiltonian (Eq-\ref{HamiltonianwithVacAndH}) obtained after gauge fixing.

The vacancy contribution to the magnetization in the $z$-directed field can be calculated as as:
\begin{align}
m_z=-\partial_{h_z}({ E_{\rm grd}(h_z)-E_{\rm grd}(0)})
\end{align}
where the $E_{\rm grd}(h_z)$ is the ground state energy of the system with a vacancy and in the presence of a field-dependent contribution $H_Z^{eff}$.

The ground state energy is the first moment of the density of single-modes, integrated up to $0$ from $-\infty$.
\begin{gather}
\delta E_{\rm grd}(h_z)=E_{\rm grd}(h_z)-E_{\rm grd}(0) = \int_{-\infty}^{0} \omega \rho(\omega) d\omega\nonumber\\
\rho(\omega,h_z) = -\frac{1}{\pi} {\rm Tr}[{\rm Im} [G_{h_z}(\omega+\imath 0)-G_0(\omega+\imath 0)]]
\label{densitychange}
\end{gather}
where $\rho(\omega)$ is the change in the density of modes. $G_{h_z}$ is the Green's function matrix for the single-mode Hamiltonian $\mathfrak{H}_{h_z}^V$ for the system with the vacancy and a magnetic field, i.e. $G_{h_z}(\omega)=[\omega-\mathfrak{H}^V_{h_z}]^{-1}$. Green's function $G_{h_z=0}$ at zero field is denoted by $G_0$.

The single-mode Hamiltonian $\mathfrak{H}_{h_z}^V$ (Eq-\ref{HamiltonianwithVacAndH}) is a $4N\times 4N$ matrix with $4N-1$ rows corresponding to the $c$ modes on all the sites of the lattice and one row, with index $\mathbf{r_z}$, corresponding to the mode $b^z_\mathbf{3}$ of Fig-\ref{localdof} (In Eq-\ref{HamiltonianwithVacAndH}, $\mathbf{r_z}$ was $1$). We use the symbol $\mathbf{R_3}$ to represent the row corresponding to the mode $c_\mathbf{3}$. The leading Zeeman contribution $H_Z^{eff}\equiv \imath h_z b^z_\mathbf{3} c_\mathbf{3}$,  couples the modes at rows $\bf R_3$ and $\bf r_z$.

As shown in Appendix-\ref{appen_gapless_single}, the change in density of states upon introduction of a  field $h_z$ is
\begin{eqnarray}
\rho\left(\omega,h_z\right)= -\frac{1}{\pi}{\rm Im} [\partial_{\omega}\ln D(\omega+\imath 0^+)]
\label{singlevacdensity}
\end{eqnarray}
where
\begin{eqnarray}
D&=&1-\frac{h_z^{2}}{\omega}G_{0} (\omega,\mathbf{R_3},\mathbf{R_3})\\
G_0(\omega,\mathbf{R_3},\mathbf{R_3}) &=& g(\omega,\mathbf{R_3},\mathbf{R_3})-\frac{g(\omega,\mathbf{R_3},\mathbf{v})g(\omega,\mathbf{v},\mathbf{R_3})}{g(\omega,\mathbf{v},\mathbf{v})}. \nonumber
\end{eqnarray} 
$g$ is the Green's function for the clean lattice. We assume, without loss of generality, that the vacancy is located on sublattice $1$. For small ${\rm Re}[\omega]$ close to the real axis, $g(\omega,\mathbf{r},\mathbf{r})\approx A\omega \ln (-B\omega^2) $ and $g(\omega,\mathbf{R_3},\mathbf{v})\approx C$, where $A\sim 0.5$, $B\sim 0.45$ and $C\sim -0.4$ (Appendix-\ref{app:GFinGaplessPhase}). It can be seen that  imaginary part $D_I$ of $D$ is a smooth negative valued function. The main contribution to $\rho(\omega, h)$ arises from the point where the real part $D_R$ of $D$ changes sign, at which ${\rm Im } [\ln D]$ discontinuously changes from $0$ to $-\pi$.  More precisely, the contribution to $\delta E_{\rm grd}(h_z)$ from around this point $\omega_0$ can be obtained as 
\begin{multline}
\delta E_{\rm grd}(h_z)=\int_{-\infty}^0 \omega \rho(\omega) d\omega \\
\approx  \frac{1}{\pi}\int^{\infty}_{-\infty} \frac{\omega_0 D_R'(\omega_0)D_I(\omega_0)}{D_R'(\omega_0)^2 x^2+D_I(\omega_0)^2} dx \nonumber\\= {\rm sgn}[D_I(\omega_0)D_R'(\omega_0)]\omega_0
\end{multline}

The solution to $D_R (\omega_0)=0$, corresponding to the point where $D_R$ changes sign, can be obtained as follows. For small $|{\rm Re }[\omega]|$, $D_R$ can be approximated as 
\begin{equation}
D_R (\omega) \approx 1 + \frac{C^2h_z^2}{A\omega^2\ln[B\omega^2]}\nonumber
\end{equation}
which gives (using results in Chapter-$4.13$ of Ref-\onlinecite{NIST:DLMF}), 
\begin{equation}
\omega_0 =
-\frac{|C|h_z}{\sqrt{-2A W_{-1}
(-\frac{BC^2h_z^2}{A}) }}
\approx-\frac{|C|h_z}{\sqrt{2A\ln[1/h_z]}}
\label{omega0}
\end{equation}
$W_{-1}$ is the $-1$ branch of the Lambert W function. Magnetization at vanishing fields $h_z$ can be calculated to be
\begin{equation}
m(h_z) = -\partial_{h_z} \delta E_{\rm grd}(h_z) \approx \frac{|C|}{\sqrt{2A\ln[1/h_z]}}
\end{equation}

Thus, unlike the gapped phase, where the vacancy spin was free (such that it was polarized completely by an infinitesimal magnetic field), we see here that the hybridization between the vacancy-induced zero-modes and the gapless bulk modes of the spin-liquid suppresses the low-field magnetization of an isolated vacancy. Magnetization rises rapidly from $0$ as $m(h_z)\sim \frac{1}{\sqrt{-\ln h_z}}$.
\subsection{Two vacancies}
In this subsection, we study a pair of vacancies at a finite distance from each other with the aim of understanding their interactions. As we show below, the spin-moment associated with two such vacancies interact with a strength that depends on their separation and relative sublattices. We calculate the magnetization of the pair and show that the interaction modifies the magnetization at low fields. At high fields the pair behaves like two isolated vacancies.

As before, vacancies are introduced effectively by adding (to the tight-binding model) infinite on-site potentials at the vacancy sites $\mathbf{v}$ and $\mathbf{v}'$. \cite{Willans2010,*Willans2011} The single-mode Green's function $G_{h_z}$ in the presence of two vacancies and field is a $4N\times4N$ matrix with $4N-2$ row-indices corresponding to the $c$ modes on all the sites and two row-indices $\mathbf{r_z}$ and $\mathbf{r_z}'$ corresponding to the modes of the type $b^z_\mathbf{3}$ next to these two vacancies (Fig-\ref{localdof}). The indices corresponding to the modes of the type $c_\mathbf{3}$ next to the vacancies are labeled $\mathbf{R_3}$ and $\mathbf{R_3}'$.

The Green's function $G_0$ in the presence of the vacancies and at zero field can be written in terms of the Green's function $g$ for the clean lattice as
\begin{equation}
G_{0}(\omega,\mathbf{r,s})=\begin{cases}
g_{\mathbf{r},\mathbf{s}}+\left[gTg\right]_{\mathbf{r},\mathbf{s}} & \mathbf{r},\mathbf{s}\not\in\left\{ \mathbf{r_{z}},\mathbf{r_{z}}'\right\} \\
\frac{1}{\omega} & \mathbf{r}=\mathbf{s}\in\left\{ \mathbf{r_{z}},\mathbf{r_{z}}'\right\} \\
0 & \text{otherwise}
\end{cases}
\label{2vacGaplessGFnofield}
\end{equation}
where the non-zero entries of the t-matrix $T$ are (Appendix-\ref{binomialInverse})
\begin{gather}
\left[\begin{array}{cc}
T_{\mathbf{v,v}} & T_{\mathbf{\mathbf{v,v}'}}\\
T_{\mathbf{v',v}} & T_{\mathbf{v',v'}}
\end{array}\right]=-\left[\begin{array}{cc}
g_{\mathbf{v},\mathbf{v}} & g_{\mathbf{v},\mathbf{v}'}\\
g_{\mathbf{v}',\mathbf{v}} & g_{\mathbf{v}',\mathbf{v}'}
\end{array}\right]^{-1}
\label{impurityTmatrx}
\end{gather}
The leading contributions from the Zeeman correction due to a $z$-directed field, now arise from the two sites of the type $\mathbf{3}$ (Fig-\ref{localdof}) next to each vacancy. These contributions correspond to a  perturbation $Q$ of the single-mode Hamiltonian $\mathfrak{H}^V$. The perturbation has four non-zero matrix elements : $Q_{\mathbf{R_3},\mathbf{r_z}}=Q_{\mathbf{R_3}',\mathbf{r_z}'}=Q_{\mathbf{r_z},\mathbf{R_3}}=Q_{\mathbf{r_z}',\mathbf{R_3}'}=h_z$. 

The Green's function $G_{h_z}$ in the presence of the perturbation $Q$ is given by $G_{h_z}=G_0 + G_0TG_0$, where the non-zero entries of the t-matrix $T$ are given by
\begin{gather}
T\equiv\left[\begin{array}{cc}
-\mathbb{I}_{2}/\omega & \mathbb{I}_{2}/h_z\\
\mathbb{I}_{2}/h_z & -\mathfrak{G}
\end{array}\right]^{-1}\\
\mathfrak{G}=\left[\begin{array}{cc}
G_{0}\left(\mathbf{R_3},\mathbf{R_3}\right) & G_{0}\left(\mathbf{R_3},\mathbf{R_3}'\right)\\
G_{0}\left(\mathbf{R_3}',\mathbf{R_3}\right) & G_{0}\left(\mathbf{R_3}',\mathbf{R_3}'\right)
\end{array}\right]\nonumber
\end{gather}
The rows and columns correspond to the indices in the order $(\mathbf{r_z,r_z',R_3,R_3'})$.

Using the identity Eq-\ref{greenfuncidentity}, the field-induced change in density of states can be calculated to be
\begin{gather}
\rho (\omega) = -\frac{1}{\pi} {\rm Im}[\partial_\omega \ln D(\omega+\imath 0^+)] \nonumber \\
D = 1-\frac{h_z^2}{\omega}X_1 +  \frac{h_z^4}{\omega^2}X_2
\label{defineDfor2vac}
\end{gather}
where 
\begin{gather}
X_1=G_0(\mathbf{R_3},\mathbf{R_3})+G_0(\mathbf{R_3}',\mathbf{R_3}')\nonumber\\
X_2=G_0(\mathbf{R_3},\mathbf{R_3})G_0(\mathbf{R_3}',\mathbf{R_3}')-G_0(\mathbf{R_3},\mathbf{R_3}')G_0(\mathbf{R_3}',\mathbf{R_3})\label{X1X2}
\end{gather}
The functions $X_1$ and $X_2$ can be explicitly evaluated in terms of the Green's functions $g$ of the clean system by substituting the expressions for $G_0$ given in Eq-\ref{2vacGaplessGFnofield} and Eq-\ref{impurityTmatrx}. For vacancies at finite separations, and at small $|{\rm Re}[\omega]|$, the functions $X_{1}$ and $X_2$ can be approximated as (Appendix-\ref{app:X1X2})
\begin{align}
X_1 \approx -\frac{2g_{\bf vR_3}^2g_{\bf vv}}{g_{\bf vv}^2-g_{\bf vv'}^2}\\
X_2 \approx \frac{g_{\bf vR_3}^4}{g_{\bf vv}^2-g_{\bf vv'}^2}\nonumber
\end{align}
Upon applying these approximations to the definition of $D$ in Eq-\ref{defineDfor2vac}, it factorizes as follows:
\begin{gather}
D = D_1D_2\label{DequalsD1D2}\\
D_1=1+\frac{h_z^2 }{\omega} \frac{g_{\mathbf{v},\mathbf{R_3}}^2}{g_{\mathbf{v},\mathbf{v}}-g_{\mathbf{v},\mathbf{v}'}};\;D_2=1+\frac{h_z^2 }{\omega} \frac{g_{\mathbf{v},\mathbf{R_3}}^2}{g_{\mathbf{v},\mathbf{v}}+g_{\mathbf{v},\mathbf{v}'}}\nonumber
\end{gather}
The change in density of single-mode splits into a sum of two terms as follows:
\begin{equation}
\rho(\omega) = -\frac{1}{\pi}{\rm Im}\left[\partial_\omega \ln D_1\right]  -\frac{1}{\pi}{\rm Im}\left[\partial_\omega \ln D_2\right]
\label{RhoequalsRho1plusRho2}
\end{equation}
Each term has a structure similar to that for a single vacancy discussed in the previous subsection. Using the same arguments presented there, we can see that the contribution from $D_i$ to the field-dependent change $\delta E_{\rm grd}$ of the ground state energy is $\omega_i$, which is the point where the real part of $D_i(\omega)$ changes sign.

In the next subsections, we use these results to calculate the magnetization of a pair of at vacancies in the gapless system for parameters $J_{x,y,z}=1$. Lattice Green's functions for the hyper-honeycomb lattice is not available, and therefore the analysis of arbitrarily placed pair of vacancies is not feasible. However, we are able to estimate the qualitative behavior for two  vacancies separated along certain directions.

Green's function $g_{\bf a,b}$ is easiest to calculate if $\mathbf{a}$ and $\mathbf{b}$ are separated along the high symmetry direction, which is perpendicular to the plane of the line-node. This corresponds to the direction of the $z$-bonds, which, in terms of the lattice vectors, is $\mathcal{A}=a_3-a_1/2-a_2/2$. These matrix elements of the Green's function are estimated in Appendix-\ref{app:GFinGaplessPhase}. In subsection-\ref{sec:gapless-samesublattice} we use these Green's functions to calculate the magnetization of two vacancies separated along the direction $\mathcal{A}$, both on sublattice $1$. Qualitative behavior of the Green's functions $g_{\bf a,b}$ when $\bf a$ and $\bf b$ are separated along a direction away from $\mathcal{A}$, can be obtained assuming a simpler shape for the line-node in the same plane ($2k_3=k_1+k_2$) as described in Appendix-\ref{app:circle-approx}. Using this assumption, we can infer the qualitative behavior of the magnetization for vacancies separated along directions away from $\mathcal{A}$. Lastly, magnetization of two vacancies on sublattices $1$ and $3$ (or equivalently $2$ and $4$) are shown to be qualitatively similar to that of two vacancies on the sublattice $1$.

In the subsection-\ref{sec:gapless-oppsublattice}, a similar analysis is presented for the case of  two vacancies on opposite sublattices.

\subsubsection{Vacancies on the same sublattice}
\label{sec:gapless-samesublattice}

We begin by analyzing the case of two vacancies, both on the sublattice $1$, separated by $\mathbf{r}=n\mathcal{A}$. In this case, the sites $\mathbf{R_3}$ and $\mathbf{R_3}'$ are on the sublattice $2$. The Green's functions appearing in the expression for $D$ in Eq-\ref{DequalsD1D2} are (from Appendix-\ref{app:GFinGaplessPhase}).
\begin{gather}
g_{\mathbf{v}\mathbf{v}} = A \omega \ln \left (-B\omega^2 \right );\; g_{\mathbf{v},\mathbf{R}_3} = C \nonumber\\
g_{\mathbf{v}\mathbf{v}'} = A\pi \left[ \omega Y_0(\sqrt{\Delta_0} |n\omega|) \mp \imath |\omega| J_0 (\sqrt{\Delta_0} |n\omega|)\right]
\label{samesublatticeGFs}
\end{gather}
where the $A,B,C,\Delta_0$ are constants.

When these are substituted into Eq-\ref{DequalsD1D2}, we find that the functions $D_1$ and $D_2$  change sign once - at some $\omega_1$ and $\omega_2$ on the negative real axis. $\delta E_{\rm grd}$ and $m$ are given by $\omega_1+\omega_2$ and $-\partial_{h_z}(\omega_1+\omega_2)$ respectively (See discussion below Eq-\ref{RhoequalsRho1plusRho2}).

At large fields $h_z$, $\omega_{1}$ and $\omega_{2}$ are in a regime where $g_{\mathbf{v}\mathbf{v}'}\ll g_{\mathbf{v}\mathbf{v}}$, and $g_{\mathbf{v}\mathbf{v}'}$ in the denominators of $D_1$ and $D_2$ can be ignored. In this case, $D_1$ and $D_2$ are identical to $D$ for an isolated vacancy. $\omega_{1}$ and $\omega_2$ are given by Eq-\ref{omega0}, and therefore $\delta E_{\rm grd}$ and the magnetization are the same as that of two isolated vacancies.

As the field is lowered, $\omega_i$ approaches $0$. In this regime,
\begin{equation}
{\rm Re}\,g_{\bf v,v'}=\pi A\omega Y_0(|n\omega|)\to  A\omega \ln [n^2\omega^2]
\label{eq_logapprox}
\end{equation}
and $g_{\bf vv'}$ cancels the $\omega\ln \omega^2$ behavior of $g_{\bf vv}$ in the denominator of $D_1$. The result of this exact cancellation is that $\omega_1$ behaves as
\begin{equation}
\omega_1\approx-\frac{h_z|C|}{\sqrt{2A\ln n}}.
\label{omega1lf}
\end{equation}
The behavior of $\omega_2$ is similar to that of a single vacancy. 

Thus the vacancy contribution to magnetization is
\begin{gather}
m_{\rm low} (h_z) \approx  \frac{|C|}{\sqrt{2A\ln n}} + \frac{|C|}{\sqrt{-4A\ln [h_z]}}\nonumber\\
m_{\rm high} (h_z) \approx  \frac{2|C|}{\sqrt{-2A\ln [h_z]}}
\label{samesublatticeAresults}
\end{gather}
The cross-over energy scale $\Omega$ can be interpreted as the intersection point of the low-field and the high-field behavior of $\omega_1$ (Eq-\ref{omega1lf} and Eq-\ref{omega0}), giving $\Omega\sim \frac{1}{n\sqrt{\ln n}}$.

\begin{figure}
\includegraphics[width=\columnwidth]{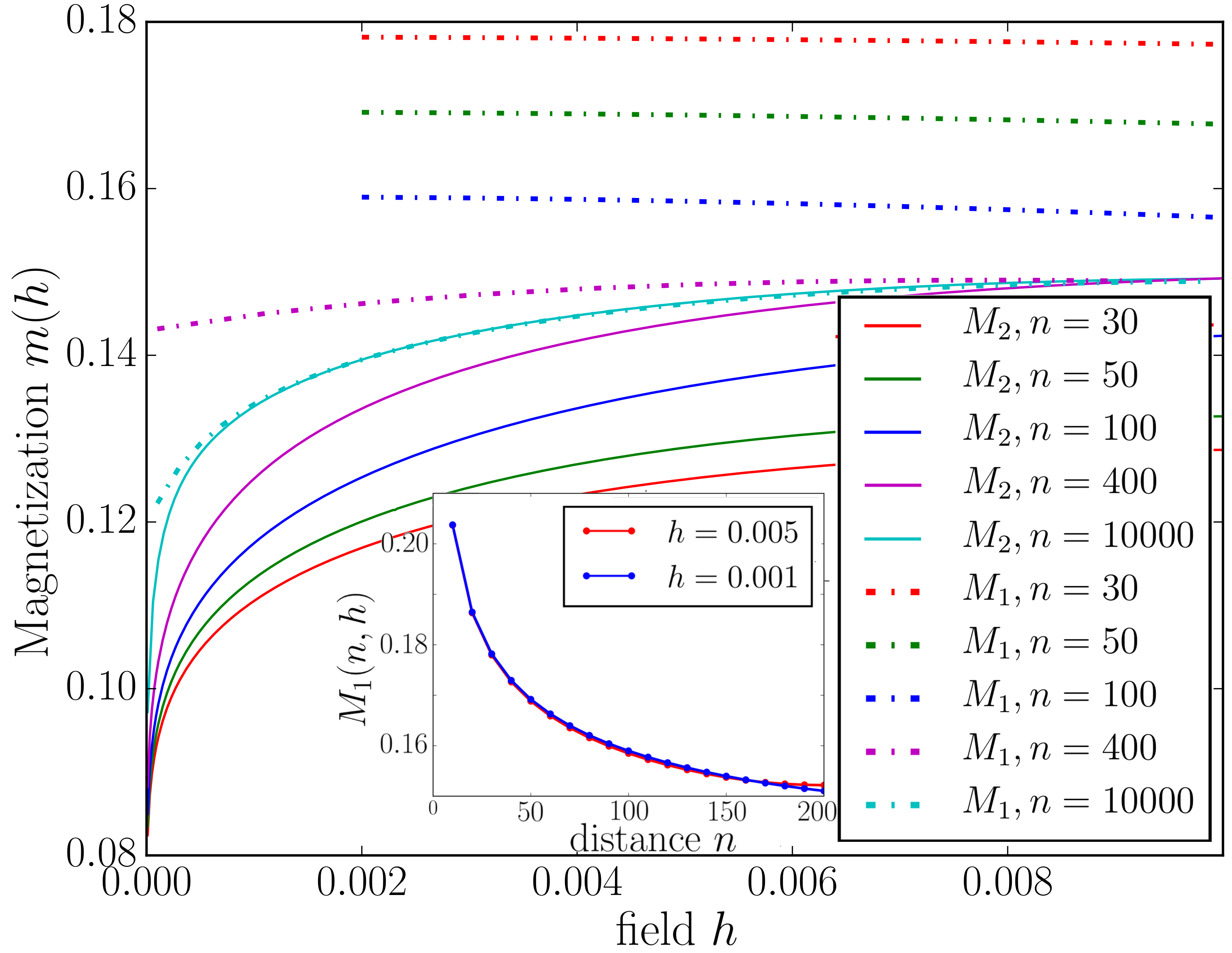}
\caption{Contributions $M_1$ and $M_2$ to magnetization of two vacancies (separated in the $z$-bond direction, on sublattice $1$), arising from the two terms in Eq-\ref{DequalsD1D2}. Different lines indicate different distances $n$. For small distances, the contribution from first term is a constant while the contribution from the second term is $\propto \frac{1}{\sqrt{-\ln [h_z]}}$. The contribution $M_1$ decreases with distance (inset) in such a way that at large distances $M_1$ matches the magnetization of an isolated vacancy.}
\label{magnetizationsamesublattice}
\end{figure}

Magnetization can be estimated by numerical integration in the negative $\omega$ axis as follows
\begin{gather}
m=M_1+M_2\nonumber\\
M_i=-\frac{1}{\pi}{\rm Im}\int d\omega \;\omega \left[\frac{\partial_h \partial_\omega D_i}{D_i} - \frac{\partial_h{D_i} \partial_\omega{D_i}}{D_i^2} 
\right],
\end{gather}
where $M_1$ and $M_2$ are the contributions to magnetization from $D_1$ and $D_2$ respectively. Fig-\ref{magnetizationsamesublattice} shows the two contributions as a function of the field. Numerical integrations were performed using the forms of the Green's functions given in Eq-\ref{samesublatticeGFs}. At distance $n=10^4$, both $M_1$ and $M_2$ behave in the same manner as the magnetization of an isolated vacancy. The behavior described in Eq-\ref{samesublatticeAresults} is best illustrated by the $n=400$ case, for which we see that $M_1$ crosses over from a constant at low fields to a $1/\sqrt{-\ln[h_z]}$ behavior at large fields.

We find that the low-field magnetization of two vacancies on the same sublattice is higher than that of two isolated vacancies. This indicates an effectively ferromagnetic interaction between these vacancy spin moments. The interaction strength decreases with distance between the two vacancies, resulting in a lowering of the magnetization with distance (Fig-\ref{magnetizationsamesublattice} inset).

In order to understand the behavior of the magnetization of two vacancies separated along directions away from the $z$-bond direction $\mathcal{A}$, we resort to an approximation in which the line-node is assumed to be along a simpler contour (such that its projection on the $k_1$-$k_2$ plane is a circle of radius $R$) in the Brillouin zone. The Green's functions calculated using this approximation are described in Appendix-\ref{app:circle-approx}. For two vacancies separated in a direction away from $\mathcal{A}$, say $\mathbf{r}=n\mathcal{A}+\delta_1 a_1$, the forms of the Green's functions in Eq-\ref{samesublatticeGFs} are replaced by
\begin{gather}
g_{\bf vv} = A \omega \ln \left[ -\frac{\Delta_0\omega^2}{\Lambda^2} \right];\; g_{\bf v,R_3} = C\\
g_{\bf vv'}=A J_{0}(R\delta_{1}) \pi \left[\omega Y_{0}(\sqrt{\Delta_{0}}\left|n\omega\right|)\mp\imath\left|\omega\right|J_{0}(\sqrt{\Delta_{0}}\left|n\omega\right|)\right]\nonumber
\end{gather}
While the nature of the $\omega$-dependence (for small $|{\rm Re}[\omega]|$) of $g_{\bf vv'}$ is the same as that of $g_{\bf vv}$, the prefactor $J_0(R\delta_1)$ prevents the exact cancellation of the $\omega \ln \omega^2$ term in the denominator $g_{\bf vv}-g_{\bf vv'}$ of $D_1$. Instead, it only reduces the coefficient of $\omega \ln \omega^2$. Note that an exact cancellation of this leading term is what resulted in the qualitatively different behavior (Eq-\ref{omega1lf}) of $\omega_1$ at low fields in the previous case. The result is that the finite magnetization seen at low fields, is now replaced by a rapidly increasing one:
\begin{align}
&m_{\rm low}(h_z) \approx \frac{|C|}{\sqrt{(1-J_0(R\delta_1)) 2A\ln[1/h_z]}}+ \dots \nonumber \\ & \qquad\qquad\qquad \dots + \frac{|C|}{\sqrt{(1+J_0(R\delta_1))2A\ln[1/h_z]}}\nonumber\\
&m_{\rm high}(h_z) \approx \frac{2|C|}{\sqrt{2A\ln[1/h_z]}}
\label{arbitAnglesamesublattice}
\end{align}
Magnetization scales as $\frac{1}{\sqrt{-\ln h_z}}$ in both regimes, the difference being in the prefactors only. Thus, for two vacancies separated along a direction at a finite angle from $\mathcal{A}$, only the $\frac{1}{\sqrt{-\ln h_z}}$ behavior is seen.

Now we address the case of two vacancies located on sublattices $1$ and $3$. From Appendix-\ref{app:GFinGaplessPhase}, it can be seen that the nature of $\omega$-dependence of $g_{\bf vv'}$ here is the same as that in the case of two vacancies on the sublattice $1$. Since the prefactors prevent exact cancellation of the leading $\omega\ln\omega^2$ term, only the $\frac{1}{\sqrt{-\ln h_z}}$ behavior is observed in the magnetization.

\subsubsection{Vacancies on opposite sublattice}
\label{sec:gapless-oppsublattice}

We first consider the case of two vacancies on sublattices $1$ and $2$  located in unit-cells that are separated in the direction $r=n\mathcal{A}$. The sites $\mathbf{R_3}$ and $\mathbf{R_3}'$ are now on sublattices $2$ and $1$ respectively.

As in the previous subsection, we infer the magnetization by analyzing the contributions from the two terms $D_1$ and $D_2$ of Eq-\ref{DequalsD1D2}. The relevant Green's functions are 
(Appendix-\ref{app:GFinGaplessPhase})
\begin{align}
&g_{\mathbf{v}\mathbf{v}} = A \omega \ln \left (-B\omega^2 \right );\; g_{\mathbf{v},\mathbf{R}_3} = C; \label{OppsublatticeGFs}\\
&g_{\mathbf{v}\mathbf{v}'} = -\frac{\pi\alpha\sqrt{\Delta_0}{\rm sign}(n) }{2} \times \dots \nonumber \\& \qquad \dots \times \left[ |\omega| Y_1(|n\omega|\sqrt{\Delta_0})\mp \imath \omega J_1(|n\omega|\sqrt{\Delta_0})\right]\nonumber
\end{align}
where $\alpha=-\frac{4K(-15)}{\pi^2}$, $A\approx 0.5$, $B\approx0.45$, $C\approx -0.4$ and $\Delta_0\approx 7.5$. 

At large fields, the main contributions to the magnetization come from larger $|{\rm Re}[\omega]|$ where $g_{\mathbf{vv}}\gg g_{\mathbf{vv}'}$ such that $g_{\mathbf{vv}'}$ can be ignored in the denominators of $D_1$ and $D_2$. The resulting field-dependent contributions $\omega_1$ and $\omega_2$ to $\delta E_{\rm grd}$ in this regime are identical to the corresponding contribution $\omega_0$ (Eq-\ref{omega0}) in the single vacancy case. Therefore at large fields, the two vacancies behave like two isolated ones.

As the field is lowered, the main contributions to magnetization come from smaller $|{\rm Re}[\omega]|$ where $g_{\mathbf{vv}'}$ is significant. In this regime, ${\rm Re}[g_{\mathbf{vv}'}(\omega)]$ is a negative constant  with respect to $\omega$ given by $\frac{\alpha}{n}$ (we assume that $n$ is positive), whereas ${\rm Re}[g_{\mathbf{vv}}(\omega)]$ is $A\omega\ln B\omega^2$ and positive. 

As a result, for small fields, $g_{\mathbf{vv}}(\omega)$ can be neglected from the denominator $D_1$ of Eq-\ref{DequalsD1D2}. This results in a contribution $\omega_1$ to $\delta E_{\rm grd}$ of the form $\frac{nh_z^2C^2}{|\alpha|}$, and a magnetization of $M_1\approx \frac{2nh_zC^2}{|\alpha|}$.

The contribution to $\delta E_{\rm grd}$ from $D_2$ arises from two points $\omega_{2a}$ and $\omega_{2b}$, which are where the real part of $D_2$ changes sign. The first contribution $\omega_{2a}$ occurs where the change of sign happens across a divergence arising from vanishing of the denominator $g_{\mathbf{v}\mathbf{v}'}+g_{\mathbf{v}\mathbf{v}}$. This can be estimated to be at $\omega_{2a} \approx \frac{\alpha}{2An\ln n}$. Since this is independent of $h_z$, it does not contribute to the magnetization.  The second contribution $\omega_{2b}$ occurs when ${\rm Re}[D_2]=0$. 
The denominator $\omega(g_{\bf vv}+g_{\bf vv'})$ of $D_2$ can be approximated by a Taylor expansion around $\omega_{2a}$, as $\frac{|\alpha|}{n}(\omega-\omega_{2a})$. Using this, $\omega_{2b}$ can be estimated to be $\omega_{2a}-\frac{nh_z^2C^2}{|\alpha|}$. Magnetization $M_2$ resulting from this is identical to $M_1$. At very small $h_z$, $\omega_{2b}$ approaches $\omega_{2a}$, and it is not clear that a separate analysis of contributions that we have performed is valid. However, numerical estimates of contribution to magnetization from $D_2$ agrees with our estimate (Fig-\ref{oppsublatticeM1}).
\begin{figure}
\includegraphics[width=\columnwidth]{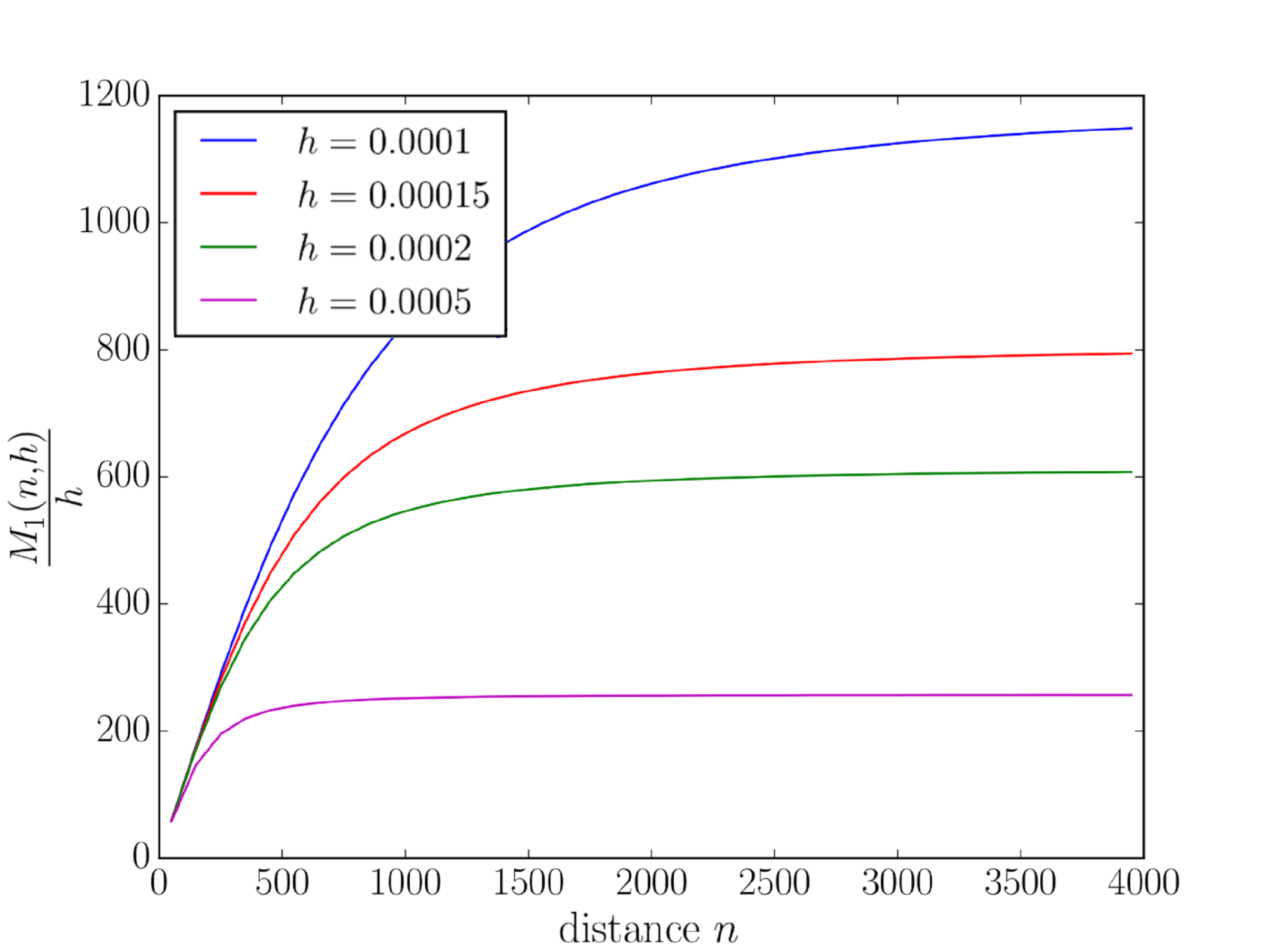}
\includegraphics[width=\columnwidth]{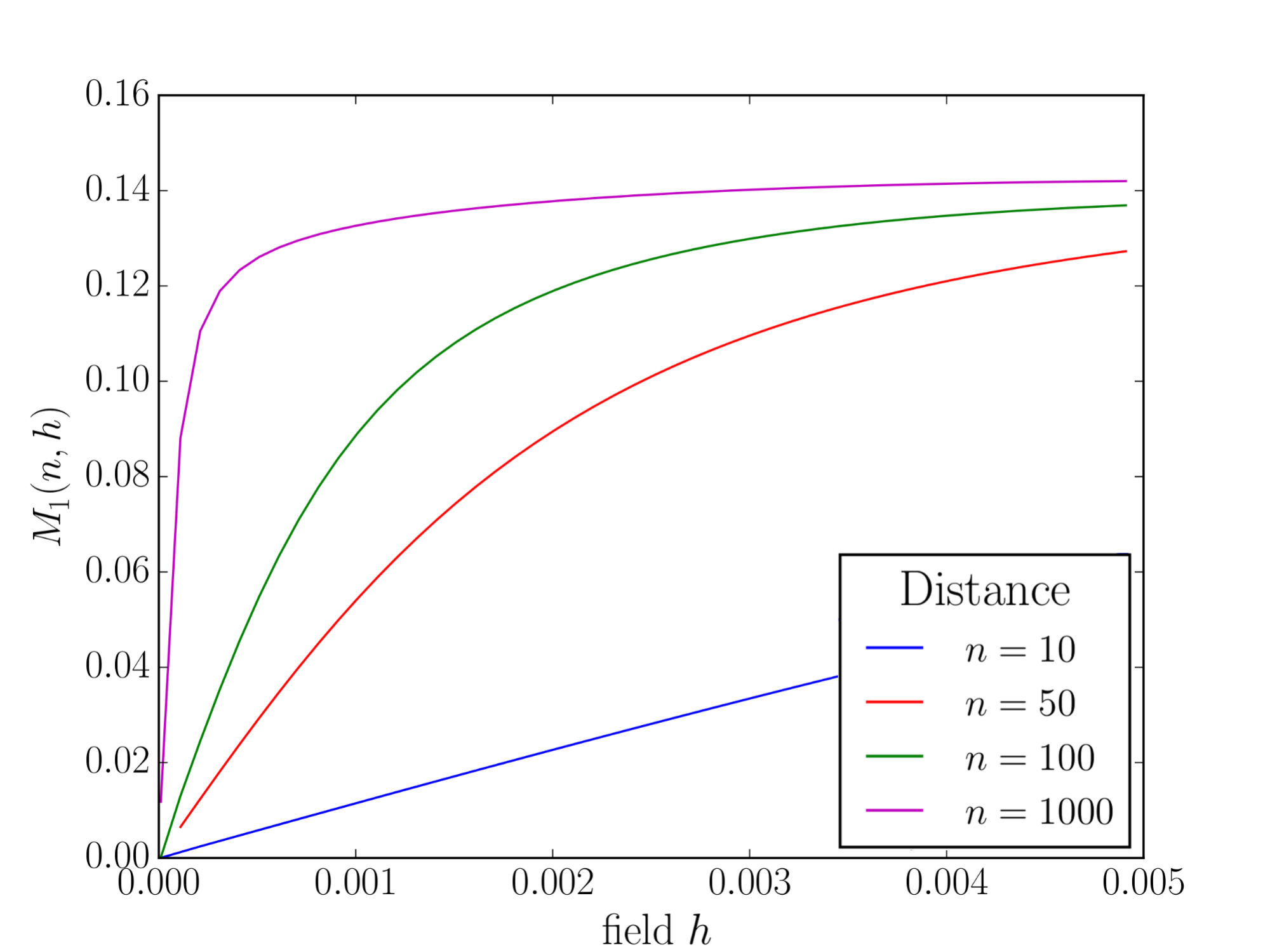}
\caption{Contribution to the magnetization from $D_1$ for the case of two vacancies on sublattices $1$ and $2$. (top) $M_1(n,h_z)/h_z$ vs distance $n$. At for small fields, the ratio is independent of $h_z$ and linear in $n$. For large separations, $M_1$ is independent of $n$ and approaches the isolated vacancy limit. (bottom) $M_1$ vs $h_z$ for different distances. At low fields, the magnetization is linear in $h_z$ with a slope that increases with distance. For large fields, $M_1$ approaches the isolated vacancy magnetization values.}
\label{oppsublatticeM1}
\end{figure}
\begin{figure}
\includegraphics[width=\columnwidth]{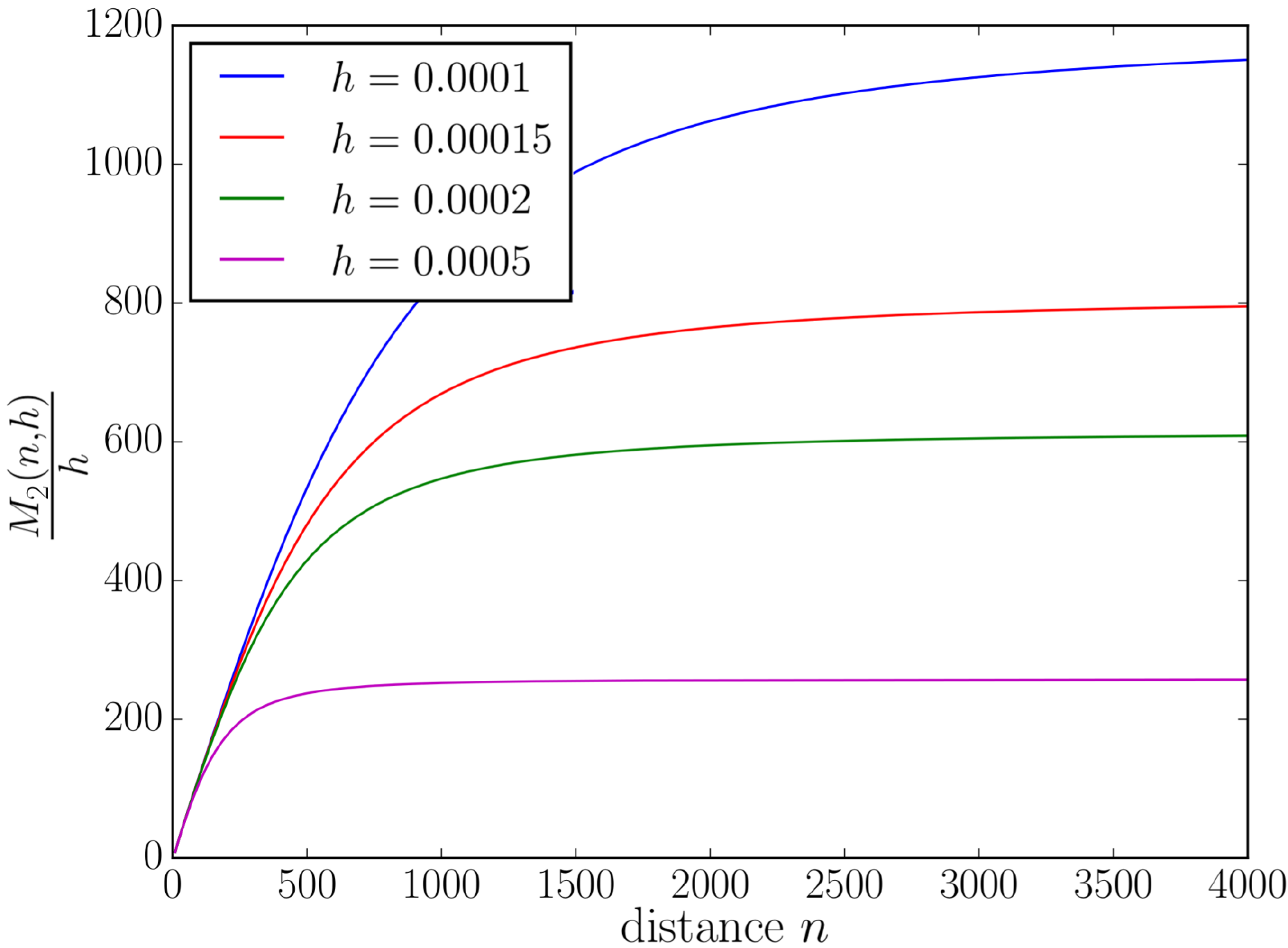}
\includegraphics[width=\columnwidth]{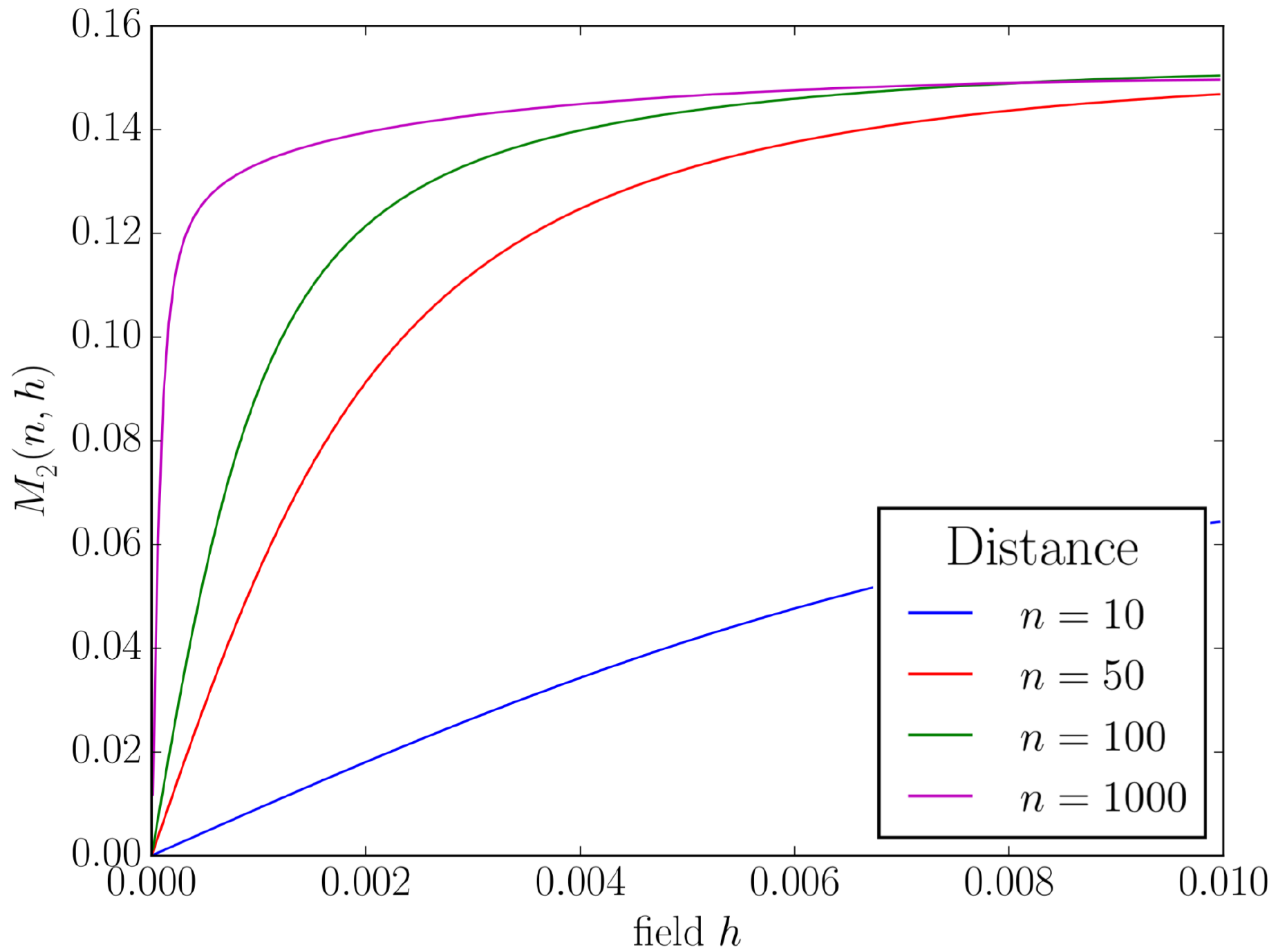}
\caption{Contribution to magnetization from $D_2$ for the case of two vacancies on sublattices $1$ and $2$. (top) $M_2(n,h_z)/h_z$ vs $n$ showing that the low-field magnetization is linear in $h_z$ and $n$. (bottom) $M_2(n,h_z)$ vs $h_z$ for different $n$.}
\label{oppsublatticeM2}
\end{figure}
In summary, the magnetization is given by
\begin{gather}
m_{\rm low}(h_z) \approx  \frac{4nC^2}{|\alpha|}  h_z \nonumber\\
m_{\rm high}(h_z) \approx \frac{2|C|}{\sqrt{2A\ln[1/h_z]}}
\end{gather}
The transition from low to high-field behavior occurs at an energy scale $\Omega$ where $g_{\mathbf{v,v}'}(\Omega)\approx g_{\mathbf{v,v}}(\Omega)$. This corresponds to $\Omega\sim \frac{1}{n\ln n}$

To understand the qualitative behavior of the magnetization for separations along directions away from the $z$-bonds, we again use the approximation of a simplified line-node (Appendix-\ref{app:circle-approx}). For a separation $n\mathcal{A}+\delta a_1$, $g_{\mathbf{vv'}}$ and therefore $\alpha$ are scaled down by a factor $J_0(R\delta)$ as shown in Eq-\ref{circ-approx:gvvprime}. Thus the magnetization along these directions are 
\begin{gather}
m_{\rm low}(h_z) \approx  \frac{4nC^2}{|\alpha J_0(R\delta)|} \times h_z \nonumber\\
m_{\rm high}(h_z) \approx \frac{2C}{\sqrt{2A\ln[1/h_z]}}
\end{gather}
This is accompanied by a corresponding decrease in the cross-over scale $\Omega\sim \frac{J_0(R\delta)}{n\ln n}$. It appears from the above expressions that, at those points where the Bessel function $J_0(R\delta)$ vanishes, the magnetization diverges. This however, is not true, as the cross-over scale also vanishes at these points. As a result, only the high-field behavior will be seen at these points.

As shown in Appendix-\ref{app:GFinGaplessPhase}, the qualitative behavior of $g((0,1),(0,4))$ for small $|Re[\omega]|$ is the same as that of $g((0,1),(0,2))$. As a result, the  magnetization of two vacancies on sublattices $1$ and $4$ is the same as that for the case of sublattice $1$ and $2$.

This completes our analysis of the one and two vacancy problems in the gapless spin liquid of the three dimensional Kitaev model on the hyper-honeycomb lattice.

\section{Summary and outlook}

We have analyzed the interplay of disorder and topology for isolated vacancies and pairs of vacancies in the gapped and gapless $\mathbb{Z}_2$ spin liquid phases of the Kitaev model on the hyper-honeycomb lattice. This presents an example of the effect of disorder on a three-dimensional QSL exhibiting the exotic Majorana excitations with different and unusual low-energy spectral properties. 

In the gapped phase, $n$ well-separated vacancies are associated with $2n-1$ low-energy Ising-like degrees of freedom. $n$ of these degrees of freedom correspond to spins localized around each vacancy. These can couple to an external magnetic field leading to an additional magnetization of the system at low fields. In this sense, properties of the low-energy vacancy degrees of freedom are similar to those in the two-dimensional honeycomb model, previously studied in Ref-[\onlinecite{Willans2010,*Willans2011},\onlinecite{Halasz2014}]. Crucially, however, unlike in two dimensions, the vacancies do not bind a $\mathbb{Z}_2$ flux  in three dimensions. \cite{Willans2010,*Willans2011}

In the gapped phase $J_z>J_x+J_y$, the vacancy-induced moments generate free magnetic moments, polarizing at arbitrarily small external magnetic fields. A pair of vacancies at finite separations interact through the gapped bulk modes only if its members are on opposite sublattices. The interaction is strongly anisotropic (the strongest interaction being in the direction of $z$-bonds), and decreases exponentially with distance. Such interactions are effectively anti-ferromagnetic, leading to a suppression of magnetization at low fields. This suggests that for small magnetic fields and a dilute concentration of vacancies, we should observe a finite magnetization whose strength is proportional to the vacancy concentration. Such low-field magnetization is absent in the pure system. In the gapped phase $J_z>J_x+J_y$, the magnetization of an isolated vacancy as well as the inter-vacancy interactions have analagous forms in the honeycomb and the hyper-honeycomb lattice. 

\begin{figure}
\includegraphics[width=\columnwidth]{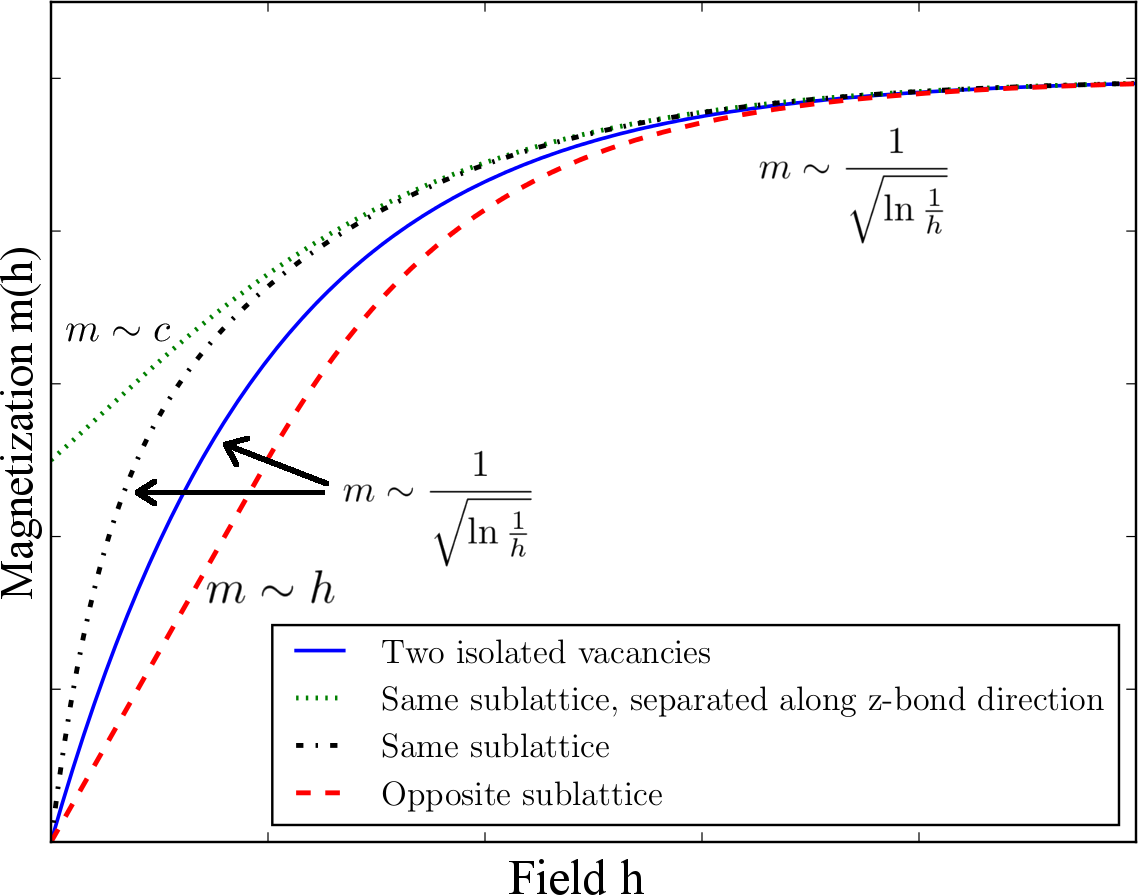}
\caption{Qualitative behavior of low-field magnetization of a pair of vacancy spins in the gapless phase.}
\label{interactionsummary}
\end{figure}

In the gapless phase, vacancy moments can hybridize with the gapless bulk modes of the surrounding spin liquid. Thus the local vacancy-induced mode, unlike in the gapped phase, can ``leak" into the bulk, leading to a suppression of the magnetization even for single vacancies. The magnetization however rises rapidly from $0$ as $m\propto\nicefrac{1}{\sqrt{-\ln [h_z]}}$ with an increasing external field. This magnetization is identical to what was estimated for the case of vacancies in the zero-flux sector of the honeycomb model in Ref-[\onlinecite{Willans2010,*Willans2011}]. Note, however, that vacancies in the ground state of the honeycomb lattice bind a flux and as a consequence, have a lower magnetization of $m\propto -h_z\ln h_z$.

In the gapless phase, the interaction between the vacancy-induced moments are much more prominent due to the presence of the gapless bulk modes through which the moments can effectively interact. The nature of the interaction depends qualitatively on the relative sublattice and the resultant net magnetization is shown schematically in Fig-\ref{interactionsummary}. The interaction between two vacancies on the same sublattice appears to be effectively ferromagnetic. At the isotropic point $J_{x,y,z}=1$, the interaction is strongest along the direction $\mathcal{A}$ of the $z$-bonds. For two vacancies on the same sublattice but separated along this direction, the interaction results in a finite magnetization at low fields. The interaction is weaker when the separation deviates from $\mathcal{A}$, and in this case, the magnetization is qualitatively the same as that of isolated vacancies, namely $m\propto\nicefrac{1}{\sqrt{-\ln [h_z]}}$. In comparison, two same-sublattice vacancies of the honeycomb system at the isotropic point were shown to have the same low-field magnetization as an isolated vacancy (both in the ground state flux sector and the zero-flux sector).\cite{Willans2010,*Willans2011} The anisotropic interaction in the hyper-honeycomb lattice is not surprising as the $z$-bonds in this lattice are distinct from the $x$ and $y$ bonds.

For two vacancies on opposite sublattices the interaction reduces the low-field magnetization to $m\propto h_z$. Unlike the case of the isotropic honeycomb lattice, there is a direction (along the  $z$-bonds) in which this interaction is stronger than in the other directions, while other qualititative features of the low-field magnetization behavior correspond to those of two opposite-sublattice vacancies in the honeycomb lattice (both in the ground state flux sector and the zero-flux sector).\cite{Willans2010,*Willans2011}

The interaction modifies the magnetization of a pair of vacancies only at small fields. At large fields above a crossover scale that depends on the separation between the vacancies, the magnetization approaches that of two isolated vacancies. Along the direction of strongest interaction, the crossover energy scale appears to decrease with distance as a power law, indicating a long-range interaction between the vacancy-induced spins.

The above results demonstrates very interesting emergent features resulting from vacancies in the Kitaev model in three dimensions. A particularly curious result is the absence of flux-binding to the vacancies in three dimensions, as opposed to two dimensions. Though this result is explicitly shown for the Kitaev model in this work, the fact that extended flux-loops in three dimensional $\mathbb{Z}_2$ QSLs are energetically expensive is likely to be a robust result valid under more general settings where exact analysis is not possible. Absence of flux-binding can have interesting implications for the statistics of holes doped into such systems. Such holes can be considered as vacancies which can now hop on the lattice sites. This forms a very interesting direction of future research, particularly in the context of $\beta$-${\rm Li}_2{\rm IrO}_3$. 

The sublattice-dependent sign of the interaction leads to the following fascinating question: if the concentration of the vacancies is increased, does this lead to a ``spin-glass" like transition for the vacancy-induced moments\cite{ASen2015} while the bulk still remains a QSL  ? We expect the features of such a ``frozen" spin-glass phase to be dependent on the gapped/gapless nature of the underlying QSL, as is evident from the nature of the interaction. 

More generally, an extension of this work to capture the many-body physics of defects in topological phases, and the response of the resulting disordered phases to external probes, present a rich and intriguing field for future studies. 

\begin{acknowledgments}
SGJ would like to thank M. Hermanns, O. Petrova and J. Rehn for useful discussions. RM thanks J. Chalker and A. Willans for collaboration on closely related work. SB thanks Y. B. Kim, E.-K. H. Lee, K. Hwang and R. Schaffer for past collaborations. This work was in part supported by DFG (SFB 1143).
\end{acknowledgments}

\appendix
\section{Fermion-parity constraint}
\label{App:FermionParity}

In this section, we sketch the proof of the fermion-parity constraint for the clean lattice as well as for a lattice with a vacancy. We assume that there are even number of unit-cells $L$ in each periodic direction. 

First, we consider the clean lattice without a vacancy. Consider the loops running around the periodic directions of the lattice given by vectors ${a}_1$ or ${a}_2$ (Fig-\ref{unitcell}). The loops along ${a}_1$ (${ a}_2$) contain sites of sublattices $1$ and $4$ ($2$ and $3$). Each $x$ or $y$ bond and each site occur in exactly one loop. The product of the all spin interaction terms  $\sigma_i^\alpha\sigma^\alpha_{i+1}$ (where $\alpha=x$ or $y$ depending on the bond-type) along all such loops can be written as
\begin{equation}
\prod_{\gamma\in xy-{\rm loops}}\prod_{i=1}^{L}\sigma_{\gamma_{i}}^{\alpha_{i}}\sigma_{\gamma_{i+1}}^{\alpha_{i}}=\prod_{\gamma\in xy-{\rm loops}}\prod_{i=1}^{L}\sigma_{\gamma_{i+1}}^{\alpha_{i}}\sigma_{\gamma_{i+1}}^{\alpha_{i+1}}
\label{loop-rearrangement}
\end{equation}
where, the loop-paths are represented as a sequence of sites $\gamma\equiv\{\gamma_i\}_{i=1\to L}$ with $L+i\equiv i$. The bond connecting $\gamma_i$ to $\gamma_{i+1}$ is represented by $\alpha_i$ ($=x$ or $y$). We take product of all such loops. The loops allow a rearrangement of the product over bonds (LHS) into a product over sites (RHS).

Using the spin-algebra $\imath\sigma^x\sigma^y\sigma^z+1=0$, the RHS of Eq-\ref{loop-rearrangement} can be reduced to a product over all $\sigma_z$ operators, i.e. $\prod_{\mathbf{r \in \mathcal{S}}}\sigma_\mathbf{r}^z$. Rearranging this product over all sites to product over $z$-bonds, we obtain:
\begin{equation}
\prod_{\gamma\in xy-{\rm loops}}\prod_{i=1}^{L}\sigma_{\gamma_{i}}^{\alpha_{i}}\sigma_{\gamma_{i+1}}^{\alpha_{i}}
=
\prod_{\langle ij\rangle \in z-{\rm bonds} } \sigma^z_i \sigma^z_j
\end{equation}
We now write this relation in terms of the Majorana operators in the extended Hilbert space to obtain:
\begin{equation}
\prod_{\left\langle \mathbf{r,s}\right\rangle \in x{\rm -bonds}}u_{\mathbf{rs}}\times\prod_{\left\langle \mathbf{r,s}\right\rangle \in y{\rm -bonds}}u_{\mathbf{rs}}=\prod_{\left\langle \mathbf{r,s}\right\rangle \in z{\rm -bonds}}u_{\mathbf{rs}}\imath c_{\mathbf{r}}c_{\mathbf{s}}\nonumber
\end{equation}
where $\mathbf{r}$ and $\mathbf{s}$ are on odd and even sublattices. Identifying the $u_{\bf rs}$ and $\imath c_\mathbf{r}c_\mathbf{s}$ to be the parities of bond and matter-fermions, we arrive at the relation
\begin{equation}
(-1)^{N_{\rm bond}+N_{\rm matter}}=1
\label{clean-system-fermion-parity}
\end{equation}
The LHS is an operator that commutes with all gauge-transformations $\mathfrak{D}$. This implies that a state in the extended space can have a physical projection iff it satisfies the constraint Eq-\ref{clean-system-fermion-parity}. 

This constraint is equivalent to the constraint $\mathcal{D}\equiv \prod_{\mathbf{r}\in\mathcal{S}}D_\mathbf{r}=1$ (See discussion below Eq-\ref{D=1constraint}) since $\mathcal{D}$ can be re-written after rearranging the terms of the product in a similar manner as 
\begin{equation}
\prod_{\langle\mathbf{rs}\rangle} u_{\mathbf{rs}}~~ \times \prod_{\langle\mathbf{rs}\rangle\in z-{\rm bonds}} \imath c_{\mathbf{r}} c_{\mathbf{s}}
\end{equation}

In the case of a single vacancy, one of the $xy$-loops becomes an open chain terminated at sites $\mathbf{1}$ and $\mathbf{2}$ of Fig-\ref{localdof}. But one can still construct a product over all the $x$ and $y$ bonds similar to the previous case. When written in terms of the operators of the extended space we obtain the following constraint on the fermion-numbers.
\begin{equation}
(-1)^{N_{\rm bond}+N_{\rm matter}} \times ib^z_\mathbf{3} c_\mathbf{3} \times ib^x_\mathbf{1} b^y_\mathbf{2} =1
\end{equation}
\section{Zero mode in the gapped phase}
\label{app:zeromode}

The single-mode Hamiltonian $\mathfrak{H}'$ for a gapped system with a vacancy is obtained from the Hamiltonian $\mathfrak{H}$ (Eq. \ref{cleanmatrixham}) of a clean system by removing the row and column corresponding the vacancy site. As discussed in Sec-\ref{sec:gapped-vac}, rank-deficiency of $\mathfrak{H}'$ implies that there exists a null vector $\psi$ for $\mathfrak{H}'$ with support only on the non-vacancy sites. Here, we show that the null vector is given by Eq-\ref{eq_majzero}, {\it i.e.}, 
\begin{align}
\psi \propto \mathfrak{H}^{-1}e_\mathbf{v}
\end{align}
where $e_\mathbf{v}$ is a vector in which all entries except the one at the vacancy site $\mathbf{v}$ are zero.

First note that, bipartiteness of $\mathfrak{H}$ and $\mathfrak{H}^{-1}$ implies that $\psi(\mathbf{v})=0$. Consider each row of the matrix equation $\mathfrak{H} \mathfrak{H}^{-1}e_\mathbf{v} = e_\mathbf{v}$. \begin{eqnarray}
\sum_{\mathbf{s}\in \mathcal{S}} \mathfrak{H}_\mathbf{rs}\psi(\mathbf{s})&=&1 \text{ if }\mathbf{r}=\mathbf{v}\nonumber\\
\sum_{\mathbf{s}\in \mathcal{S}} \mathfrak{H}_\mathbf{rs}\psi(\mathbf{s})&=&0\;\; \forall\;  {\mathbf{r}\neq \mathbf{v}}\nonumber
\end{eqnarray}
In both equations, the summations $\sum_{\mathbf{s}\in \mathcal{S}}$ can be replaced by $\sum_{\mathbf{s}\in \mathcal{S}-\{\mathbf{v}\}}$ as $\psi(\mathbf{v})=0$. With this, the second equation above is the same as $\mathfrak{H}'\psi_R=0$ where $\psi_R(\mathbf{s})=\psi(\mathbf{s})$ for $\mathbf{s}\neq \mathbf{v}$.

\section{Inverse of the Hamiltonian in the gapped phase}
\label{appendix-invert}
The single-mode Hamiltonian $\mathfrak{H}$ of the clean system  given in Eq-\ref{cleanmatrixham} can be inverted for parameters $J_{x,y,z}$ in the gapped phase, as it has no zero eigenvalues and therefore $\det[\mathfrak{H}]\neq 0$. We assume infinite system size and periodic boundary conditions. The Fourier transform $\mathfrak{h}$ of $\mathfrak{H}$ is defined such that
\begin{equation}
\mathfrak{H}(r,s)  = \int_{-\pi}^{\pi} \frac{d^3k}{8\pi^3} \mathfrak{h}(k) e^{\imath k.(s-r)}\nonumber
\end{equation}
where $k=k_1 d_1 + k_2 d_2 + k_3 d_3$, ($d_i$ are reciprocal vectors such that $a_i.d_j=\delta_{ij}$, and $k_i\in [-\pi,\pi]$) and $d^3 k\equiv dk_1 dk_2 dk_3$. For a  unit-cell location $r=\sum_{i=1}^3 n_i a_i$, $k.r$ is then $\sum_{i=1}^3 k_i n_i$. Domain of integration is the cube $[-\pi,\pi]^3$. $\mathfrak{H}(r,s)$ and $\mathfrak{h}(k)$ are $4\times 4$ matrices. 

The Fourier transform $\mathfrak{h}(k)$ is given by
\begin{gather}
\mathfrak{h}_{k}=\left[\begin{array}{cc}
0_{2\times2} & \mathfrak{M}(k)\\
\mathfrak{M}(k)^{\dagger} & 0_{2\times2}
\end{array}\right]\nonumber\\
\mathfrak{M}(k) = J_z \imath\left[\begin{array}{cc}
1 & \overline{\mathfrak{p}}\nonumber\\
\overline{\mathfrak{q}} & 1
\end{array}\right]\nonumber\\
\mathfrak{p}=e^{-\imath k_3} (j_x+j_y e^{\imath k_1})\nonumber\\
\mathfrak{q}=j_x+j_y e^{\imath k_2}\nonumber
\end{gather}
Inverse of the this matrix has the form 
\begin{equation}
\mathfrak{h}^{-1}=\left[\begin{array}{cc}
0 & [\mathfrak{M}^{-1}]^{\dagger}\\
\mathfrak{M}^{-1} & 0
\end{array}\right]\nonumber
\end{equation}
where $\mathfrak{M}^{-1}$ is given by 
\begin{equation}
\mathfrak{M}^{-1}=-\frac{1}{J_{z}}\frac{\imath}{1-\overline{\mathfrak{p}}\overline{\mathfrak{q}}}\left[\begin{array}{cc}
1 & -\overline{\mathfrak{p}}\\
-\overline{\mathfrak{q}} & 1
\end{array}\right]\nonumber
\end{equation}
The inverse of the real space single-mode Hamiltonian $\mathfrak{H}$ can be obtained as the inverse Fourier transform of $\mathfrak{h}^{-1}$ to get 
\begin{equation}
\mathfrak{H}^{-1}\left(r,s\right)=\left[\begin{array}{cc}
{{0_{2\times 2}}} & \left[g\left(r-s\right)\right]^{\dagger}\\
g\left(s-r\right) & {{0_{2\times2}}}
\end{array}\right].
\end{equation}
and $g(R)$ is given by 
\begin{gather}
g(R)=-\frac{\imath}{J_z}\left[\begin{array}{cc}
F(R) & F_{p}(R)\\
F_{q}(R) & F(R)
\end{array}\right]\nonumber
\end{gather}
where 
\begin{gather}
F_p(R)=-j_x F(R+a_3) -j_y F(R+a_3-a_1)\nonumber\\
F_q(R)=-j_x F(R) -j_y F(R-a_2).
\label{FqFp}
\end{gather}
The function $F(R_1 a_1 +R_2 a_2 + R_3 a_3)$ is 

\begin{equation}
F(R)=\begin{cases}
j_{x}^{2|R_{3}|}\left[\frac{j_y}{j_x}\right]^{R_{1}+R_{2}}\binom{|R_{3}|}{R_{1}}\binom{|R_{3}|}{R_{2}} & R_{3}<0\\                                          & \text{\& }R_{1},R_{2}\in[0,|R_{3}|]\\
0 & \text{otherwise}
\end{cases}
\label{F}
\end{equation}

\section{Calculation of t-matrix}
\label{binomialInverse}
The Green's function for a matrix $H$ is defined as $G(z)=(z\mathbb{I}-H)^{-1}$, where $z\in \mathbb{C}$.  The Green's function for a perturbed Hamiltonian $H+V$ can be calculated in terms of the Green's function of $H$ using a t-matrix approach. The new Green's function is given by 
\begin{equation}
G' = [z\mathbb{I}-(H+V)]^{-1}
\end{equation}
Perturbation matrix has a singular value decomposition $V=XDY$ such that $D$ is invertible. Using this decomposition and the Woodbury matrix identity \cite{woodbury1950} we get  
\begin{equation}
G'=G+G X [D^{-1} - YGX]^{-1} Y G
\end{equation}
The matrix $T=X [D^{-1} - YGX]^{-1} Y=V(1-GV)^{-1}$ is called the t-matrix. 

For example, for the matrix with a diagonal perturbation shown in Fig-\ref{pictureofHamiltonian},  $V = U\times (\frac{1}{\epsilon}\mathbb{I}_2) \times U^T$ where $U$ is a $2N\times 2$ matrix with all entries zero except $U_{\mathbf{v},1}=U_{\mathbf{v}',2}=1$; with $\mathbf{v}$ and $\mathbf{v}'$ being the vacancy locations. Plugging into the above expression, t-matrix for this particular perturbation is obtained as 
\begin{equation}
\left[\begin{array}{cc}
T_{\bf v,v} & T_{\bf v,v'}\\
T_{\bf  v',v} & T_{\bf v',v'}
\end{array}\right]=\left[\begin{array}{cc}
\epsilon-G_{\bf v,v} & -G_{\bf v,v'}\\
-G_{\bf v',v} & \epsilon-G_{\bf v',v'}
\end{array}\right]^{-1}\nonumber
\end{equation}
with all other matrix elements of $T$ being zero.
\section{Change in density of states, $\rho(\omega, h)$ in the gapless phase}
\label{appen_gapless_single}
Note that magnetizations and ground state energies depend only on the density of states of single-mode Hamiltonian $\mathfrak{H}_{h_z}^V$ (Eq-\ref{HamiltonianwithVacAndH}) which has the following form
\begin{equation}
\left[\begin{array}{cc}
0 & \imath X\\
-\imath X^T & 0
\end{array}\right]
\end{equation}
where $X$ is a real matrix. Its spectrum is invariant under a unitary transformation by $U=[(e^{i\pi/4}\mathbb{I},0),(0, e^{-i\pi/4}\mathbb{I})]$ which transform the above matrix to
\begin{equation}
\left[\begin{array}{cc}
0 & X\\
X^T & 0
\end{array}\right].
\label{effectiveRealHamiltonian}
\end{equation} 
We use the transformed form of the Hamiltonian for calculations of the density of states since in this new basis, all elements of the Hamiltonian matrix are real. The Green's function of this real matrix have simpler symmetries which simplify the calculations.

The Green's function $G_0$ for the single-mode Hamiltonian of a system with a vacancy and at zero field, can be obtained by introducing an infinite potential at the vacancy site $\mathbf{v}$. \cite{Willans2010,*Willans2011} Using t-matrix methods, this Green's function can be expressed in terms of the Green's function $g$ for the clean system as:
\begin{equation}
G_{h=0}(\omega,\mathbf{r},\mathbf{s})=\begin{cases}
g(\omega,\mathbf{r},\mathbf{s})-\frac{g(\omega,\mathbf{r},\mathbf{v})g(\omega,{\mathbf{v}},\mathbf{s})}{g(\omega,\mathbf{v},\mathbf{v})} & \mathbf{r},\mathbf{s}\neq \mathbf{r_{z}}\\
\frac{1}{\omega} & \mathbf{r}=\mathbf{s}=\mathbf{r_z}\\
0 & \text{otherwise}
\end{cases}
\end{equation}
where $\mathbf{r_z}$ is the row-index for the $b^z_\mathbf{3}$ mode in Fig-\ref{localdof}. Introduction of a magnetic field adds a perturbation $V_{ij}=h_z (\delta_{i,\mathbf{R_3}}\delta_{j ,\mathbf{r_z}}+\delta_{j ,\mathbf{R_3}}\delta_{i ,\mathbf{r_z}})$ to the single-mode Hamiltonian. Here $\mathbf{R_3}$ is the row-index for the $c_\mathbf{3}$ mode (Fig-\ref{localdof}). 
The Green's function $G_{h_z}$ in the presence of this  perturbation can be obtained in terms of $G_0$ using t-matrix methods (Appendix-\ref{binomialInverse}) as $G_{h_z}(\omega) = G_0(\omega) + G_0(\omega)  T G_0(\omega)$ where the non-zero entries of $T$ are
\begin{multline}
\left[\begin{array}{cc}
T_{\mathbf{r_z},\mathbf{r_z}} & T_{\mathbf{R_3},\mathbf{r_z}}\\
T_{\mathbf{r_z},\mathbf{R_3}} & T_{\mathbf{R_3},\mathbf{R_3}}
\end{array}\right]=\dots\nonumber\\ \frac{1}{1-\frac{h_z^{2}}{\omega}G_{0}\left(\omega,\mathbf{R_3},\mathbf{R_3}\right)}\left[\begin{array}{cc}
h_z^{2}G_{0}\left(\omega,\mathbf{R_3},\mathbf{R_3}\right) & h_z\\
h_z & \frac{h_z^{2}}{\omega}
\end{array}\right]
\end{multline}
Field-dependent change in density of single-modes can be calculated using Eq-\ref{densitychange}. The trace can be expressed as
\begin{multline}
{\rm Tr} [G_{h_z}-G_0] = {\rm Tr} [G_0 T G_0] = \frac{1}{1-\frac{h_z^2}{\omega}G_0(\mathbf{R_3},\mathbf{R_3})} \times \\
 \left[     \frac{h_z^2}{\omega^2} G_0(\mathbf{R_3},\mathbf{R_3}) -  \frac{h_z^2}{\omega}\sum_k G_0(\mathbf{R_3},k) G_0(k,\mathbf{R_3})\right]\nonumber
\end{multline}
Using the following identity for the Green's functions,
\begin{multline}
-\partial_{\omega} G^+_0(\omega, \mathbf{r},\mathbf{s}) = \left[ \frac{1}{(\omega+\imath 0^+-H^V)^2} \right]_{\mathbf{r,s}} \\
= \sum_{k} G^+_0(\omega, \mathbf{r},k)G^+_0(\omega, k,\mathbf{s})
\label{greenfuncidentity}
\end{multline}
the summation in the trace can be simplified to obtain
\begin{equation}
{\rm Tr} [G_{h_z}-G_0] = \frac{\partial_\omega \left[ 1-\frac{h_z^2}{\omega}G^+_0(\mathbf{R_3},\mathbf{R_3}) \right]}{1-\frac{h_z^2}{\omega}G^+_0(\mathbf{R_3},\mathbf{R_3})}
\end{equation}
Thus, the change in density upon introduction of a  field $h_z$ is given by Eq-\ref{singlevacdensity}.

\section{Green's functions in the gapless phase}
\label{app:GFinGaplessPhase}
In this section as well as the next, we present the details of the calculations and approximations involved in estimating  the Green's functions in the gapless phase. In this section, we discuss the calculations of $g(\omega,\mathbf{r},\mathbf{s})$, when the sites $\mathbf{r}$ and $\mathbf{s}$ are in the same unit-cell or separated in the direction of $z$-bonds. Estimation of the Green's functions along directions away from $z$-bond direction is described in the next section.

As noted in the Appendix-\ref{appen_gapless_single}  around Eq-\ref{effectiveRealHamiltonian}, in order to understand the low-energy spectrum, we work with an effective single-mode Hamiltonian with real entries that has the same spectrum as the Hamiltonian of interest. Such a Hamiltonian for the clean system has the following form
\begin{equation}
\mathcal{H}(r_1,r_2)=\left[\begin{array}{cc}
0_{2\times 2} & M(r_1,r_2)\\
M(r_2,r_1)^T & 0_{2\times 2}
\end{array}\right] 
\end{equation}
where $M(r,s)$ is 
\begin{equation}
\left[\begin{array}{cc}
J_{z}\delta_{r,s} & J_{x}\delta_{r,s+a_{3}}+J_{y}\delta_{r,s-a_{1}+a_{3}}\\
J_{x}\delta_{r,s}+J_{y}\delta_{r-a_{2},s} & J_{z}\delta_{r,s}
\end{array}\right]\nonumber
\end{equation}
where the indices $r_1,r_2,r_,s\in \mathcal{T}$ are unit-cell locations. Sublattices are indexed in the order $1,3,2,4$.

To study the gapless phase, we focus on the point $J_{x,y,z}=1$. The momentum space Green's function $g_0(\omega,k)$ at this point can be written as
\begin{equation}
g_0(\omega,k) = \left[\begin{array}{cc}
g_{d} & g_{oe}\\
g_{eo} & g_{d}
\end{array}\right].
\label{originalGF}
\end{equation}
Here $g_d$, $g_{eo}$ and $g_{oe}$ are given by 
\begin{gather}
g_d=\frac{\omega}{P(\omega)}\left[\begin{array}{cc}
\frac{2\omega^{2}-\Delta}{2} & \overline{\mathfrak{p}}+\mathfrak{q}\\
\mathfrak{p}+\overline{\mathfrak{q}} & \frac{2\omega^{2}-\Delta}{2}
\end{array}\right]\nonumber\\
g_{oe}=\frac{1}{P(\omega)}\left[\begin{array}{cc}
\omega^{2}+\overline{\mathfrak{p}}\overline{\mathfrak{q}}-1 & \mathfrak{q}(1-\overline{\mathfrak{p}}\overline{\mathfrak{q}})+\overline{\mathfrak{p}}\omega^{2}\\
\mathfrak{p}(1-\overline{\mathfrak{p}}\overline{\mathfrak{q}})+\overline{\mathfrak{q}}\omega^{2} & \omega^{2}+\overline{\mathfrak{p}}\overline{\mathfrak{q}}-1
\end{array}\right]\label{ExactGF}\\
g_{eo}=g_{oe}^\dagger(\bar{\omega})\nonumber
\end{gather}

where $\mathfrak{p}=e^{-\imath k_3}(1+e^{\imath k_1})$ and $\mathfrak{q}=1+e^{\imath k_2}$. $k_i=k.d_i\in[-\pi,\pi]$ where $d_i$ are the reciprocal vectors such that $d_i.a_j=\delta_{ij}$.The characteristic polynomial $P(\omega)$ for the $4\times 4$ matrix $g(\omega,k)$ is given by 
\begin{equation}
P(\omega,k)=\omega^4-\Delta \omega^2 +\delta \nonumber\\
\end{equation}
where 
\begin{gather}
\Delta = 2+4\cos^2{\frac{k_1}{2}}+4\cos^2{\frac{k_2}{2}};\;
\delta = (1-\mathfrak{pq})(1-\overline{\mathfrak{pq}})\nonumber
\label{definitionOfDeltas}
\end{gather}

Note that the momentum Green's functions satisfy $g(\omega,k)=g(\omega,-k)^T$, making the real space Green's functions inversion symmetric i.e. $g(\mathbf{r},\mathbf{s})=g(\mathbf{s},\mathbf{r})$.

The single-mode energies are given by zeros of $P(\omega)$ and occur at 
\begin{equation}
E_{\pm,\pm}(k)=\pm \left[ \frac{\Delta \pm \left[\Delta^2-4\delta \right]^{\frac{1}{2}}}{2}  \right]^{\frac{1}{2}}
\end{equation}
Two bands ($E_{\pm,-}(k)$) in the middle of the spectrum, shown in Fig-\ref{gaplessspectrum} intersect at $E_{\pm,-}=0$ along a closed contour in the Brillouin zone. The contour is at the intersection of the plane $k_3-k_2/2-k_1/2=0$ with the surface $4\cos \frac{k_1}{2} \cos \frac{k_2}{2}=1$. Close to this `line-node', the spectrum is linear along the directions normal to the line.  Energies close to the line-node have the form
\begin{equation}
E_{\pm}(k) = \pm \sqrt{\frac{\delta(k)}{\Delta (k)}}.
\end{equation}
$\sqrt{\delta(k)}$ near the line is a linear function of the displacement between the line-node and the point $k$ of the Brillouin zone. $\Delta$ is a smooth non-zero function of $k$ and determines the slope of the linear spectrum near the line-node. Different points on the line-node thus have different slopes depending on the local value of $\Delta$. 

Calculation of real space Green's function of the system by Fourier inverse transforming the above mentioned momentum space Green's function is difficult. We therefore look for approximations to the Green's function that preserves the qualitative aspects of the low-energy spectrum.  In this spirit, we make the following approximations

\begin{enumerate}
\item We assume that the linear spectrum has the same slope (with perpendicular distance from the line-node) everywhere along the line-node. The actual slope is determined by $\Delta(k)$ which has a range $[2,10]$ in the Brillouin zone. We chose this to be a constant $\Delta_0$. When numerical results are presented, $\Delta_0$ is chosen to be $7.5$.

\item For the purpose of studying the low-energy behavior, we need to consider only the two inner bands. Two outer bands can be discarded by removing the $\omega^4$ term in the characteristic polynomial i.e. $P(\omega)\approx-\Delta_0\omega^2+\delta$.
\end{enumerate}
\begin{figure}
\includegraphics[width=\columnwidth]{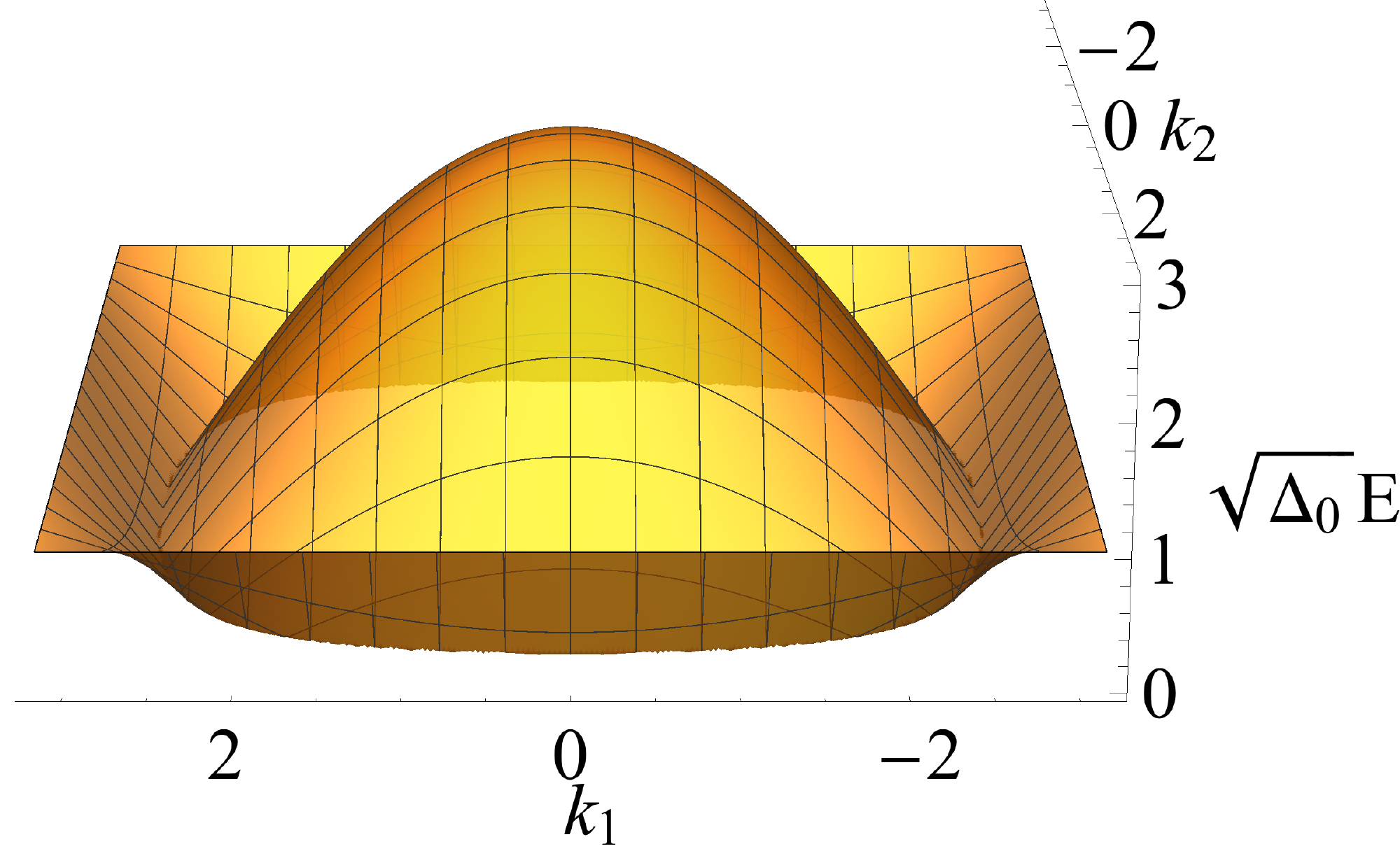}
\caption{Upper band approximated spectrum $E=\pm\sqrt{\frac{\delta}{\Delta_0}}$. Compared to the actual spectrum in Fig-\ref{gaplessspectrum}, the new spectrum preserves the linearity of the dispersion, and the location of the line-node. The slope of the spectrum is a constant ($\frac{1}{\sqrt{\Delta_0}}$) at all points along the line-node. In contrast, the actual spectrum has a dispersion whose slope near the line-node is $\frac{1}{\sqrt{10}}<\frac{1}{\sqrt{\Delta}}<\frac{1}{\sqrt{2}}$.}
\label{effectivespectrum}
\end{figure}
With these assumptions and approximations, the new spectrum (zeroes of the simplified characteristic polynomial) is shown in Fig-\ref{effectivespectrum}. 

It is useful to make the following change of coordinates from $(k_1,k_2,k_3)$ to $(u,v,\kappa)$. The transformations are made separately in each quadrant of $(k_1,k_2)$ plane.
\begin{gather}
u=4\cos \frac{k_1}{2} \cos \frac{k_2}{2}\nonumber  \;\; (u\in[0,4])\\
v=\frac{\sin \frac{k_1}{2}}{\sin \frac{k_2}{2}} \;\; (v\in[-\infty,\infty]) \nonumber\\
\kappa=k_3-\frac{k_1}{2}-\frac{k_2}{2}\nonumber
\end{gather}
Here $\kappa$ and $u$ are the coordinates along the two directions perpendicular to the line-node. $\kappa$ changes in the directions perpendicular to the plane of the line-node whereas $u$ is in the plane of the line-node. $v$ is the coordinate in the direction parallel to the node. In terms of the new coordinates, the line-node occurs at the intersection of  surfaces $u=1$ and $\kappa=0$. The function $\delta$ is given by
\begin{gather}
\delta = (u-1)^2+4u\sin^2{\frac{\kappa}{2}}\nonumber
\end{gather}
and can be approximated with $\delta = (u-1)^2 + \kappa^2$ near the line-node.
Jacobian of the transformation can be calculated to be
\begin{gather}
J(u,v)=\left|\frac{\partial (k_1,k_2,k_3)}{\partial (u,v,\kappa)}\right|=\left[{(v^2-1)^2+\frac{u^2v^2}{4}}\right]^{-\frac{1}{2}}
\end{gather}

Using these approximations and transformations, we are able to estimate the low-energy behavior of the Green's functions $g(\omega,(r,i)(r,j))$ and $g(\omega,(0,i)(n\mathcal{A},j))$ where $\mathcal{A}=a_3-a_1/2-a_2/2$ is the direction along the $z$-bonds. $\mathcal{A}$ points along the axis of the line-node. The symmetry of the separation relative to the line-node implies that all points on the line-node contribute in-phase to the Fourier integral, as will be seen below. As a result, the only $v$-dependence of the integrand is through the Jacobian. This dependence can be integrated out exactly using the following result
\begin{align}
\int_0^\infty  \frac{dv}{\sqrt{(v^2-1)^2+\frac{u^2v^2}{4}}} &= 
\frac{1}{2}\int_0^\pi \frac{d\theta}{\sqrt{\cos^2{\theta}+\frac{u^2}{16}\sin^2\theta}}\nonumber\\&=\frac{4}{u}K(1-\frac{16}{u^2})
\label{JacobianIntegral}
\end{align}
where $K$ is the complete Elliptic Integral of first kind.
The first equality follows from setting $v=\tan \frac{\theta}{2}$, and the second one from the results in Chapter-19.2 of Ref-\onlinecite{NIST:DLMF}.

In the remaining part of this section, we describe the calculation of matrix elements that are useful for the magnetization calculations presented in the main text.
\subsubsection*{Calculation of diagonal elements}
Leading contribution to the site diagonal Green's function $g(\omega,\mathbf{r},\mathbf{r})$, at small ${\rm Re}[\omega]$, is given by 
\begin{equation}
\frac{\omega}{2}\int_{-\pi}^{\pi} \frac{2\omega^2-\Delta}{P(\omega)} \frac{d^3 k}{8\pi^3} \approx \frac{\omega\Delta_{0}}{2}\int_{-\pi}^{\pi}\frac{1}{\Delta_{0}\omega^{2}-\delta}\frac{d^{3}k}{8\pi^{3}}
\end{equation}  Using the previously mentioned exact transformations from $(k_1,k_2,k_3)$ to $(u,v,\kappa)$, we obtain
\begin{multline}
\frac{\omega\Delta_{0}}{4\pi^{3}}\int_{0}^{\infty}dv\int_{0}^{4}du\int_{-\pi}^{\pi}d\kappa \dots\\ \dots\frac{J(u,v)}{\Delta_{0}\omega^{2}-(u-1)^{2}-4u\sin^{2}\frac{\kappa}{2}}
 \label{approximatediagonalGF}
\end{multline}
Integrating out the $v$-dependence using Eq-\ref{JacobianIntegral}, we obtain
\begin{equation}
\frac{\omega\Delta_0}{4\pi^3}\int_0^4 du\int_{-\pi}^{\pi}d\kappa \frac{\frac{4}{u}K(1-\frac{16}{u^2})}{\Delta_0 \omega^2-(u-1)^2-4u\sin^2 \frac{\kappa}{2}} 
\label{AfterIntegratingJ}
\end{equation}
Near the line-node at $u=1$, the numerator of integrand in  Eq-\ref{AfterIntegratingJ} is a smooth slowly varying function and can be treated as a constant $4K(-15)$. Also, the previously mentioned quadratic approximation to $\delta$ can be used here to obtain
\begin{equation}
\frac{\omega\Delta_0K(-15)}{\pi^3}\int d(u-1)\int d\kappa \frac{1}{\Delta_0 \omega^2-(u-1)^2-\kappa^2} \nonumber
\end{equation}
This gives the leading order behavior to be
\begin{equation}
g_0(\omega,{\bf r},{\bf r})\approx  \frac{\Delta_0K(-15)}{\pi^2}\omega\ln(-\frac{\Delta_0}{\Lambda^2}\omega^2)
\label{SiteDiagonalGF}
\end{equation}
where $\Lambda$ is an upper energy cut-off which, by comparing with the numerical estimates of the Green's function, we set to $\Lambda=4$. We find that this arbitrary constant does not appear in the leading order behavior of any of the magnetisations that we calculate. Fig-\ref{comparediagGF} shows the comparison between this estimate and numerical integration of the form of the Green's function in Eq-\ref{approximatediagonalGF}.
\begin{figure}
\includegraphics[width=\columnwidth]{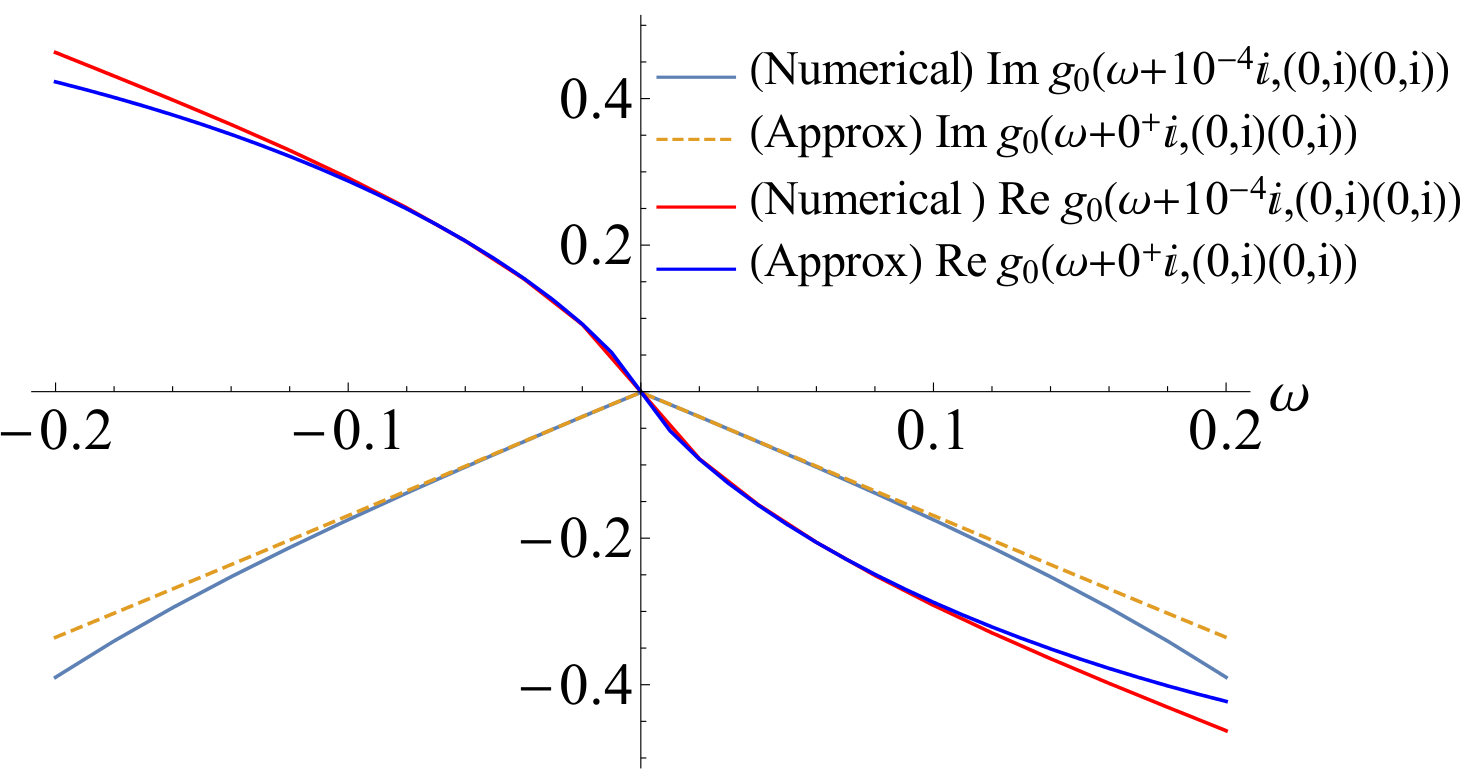}
\caption{Comparison between the low-energy approximation (Eq-\ref{SiteDiagonalGF}) to the diagonal elements of the Green's function and the Green's function numerically calculated from Eq-\ref{approximatediagonalGF}.}
\label{comparediagGF}
\end{figure}
\subsubsection*{Calculation of $g(\omega,(r,1)(r,2))$}
Nearest-neighbour sites $(r,1)$ and $(r,2)$ are connected by a single $z$-bond. The leading order behavior of the matrix element $g(\omega,(r,1)(r,2))$ is given by (from Eq-\ref{ExactGF})
\begin{equation}
\int_{-\pi}^{\pi} \frac{\overline{\mathfrak{p}\mathfrak{q}}-1}{P(\omega)} \frac{d^3k}{8\pi^3}
\end{equation}
Applying the changes of variables mentioned previously, we get 
\begin{multline}
-\frac{1}{2\pi^3} \int_{-\pi}^{\pi}d\kappa\int_0^\infty dv \int_0^4 du \dots \\ \dots \frac{(u\cos\kappa-1)J(u,v)}{\Delta_0 \omega^2-(u-1)^2-4u\sin^2\frac{\kappa}{2}} \label{approximate12GF}
\end{multline}
From numerical estimates we find that this integral is approximately a constant at small $\omega$. To estimate this, set $\omega=0$ and integrate out $\kappa$ exactly to get
\begin{equation}
g_0(\omega,(0,1),(0,2))=-\frac{4}{\pi^2}\int_0^{1}\frac{1}{u}K(1-\frac{16}{u^2})
\end{equation}
The last quantity can be numerically estimated to be $-0.4$. Fig-\ref{1020GF} shows the numerically obtained estimate of Eq-\ref{approximate12GF}.
\begin{figure}
\includegraphics[width=\columnwidth]{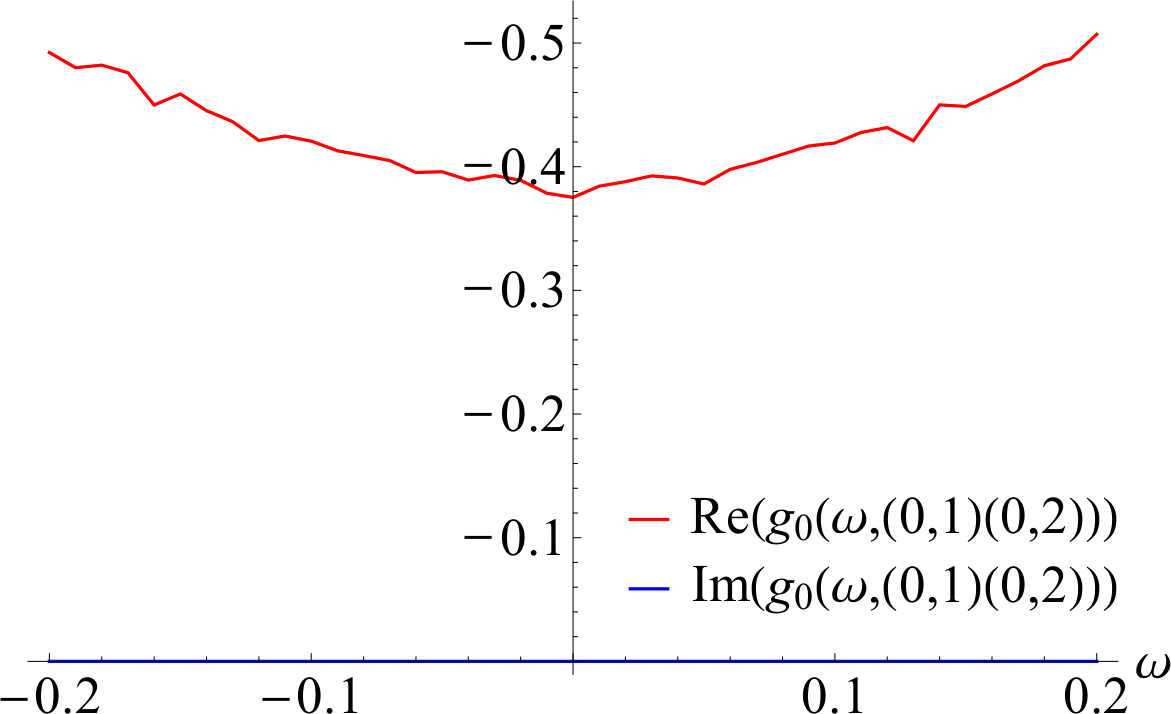}
\caption{Numerically calculated Green's function $g_0(\omega,(0,1)(0,2))$ using the form of  the Green's function in Eq-\ref{approximate12GF}. Compare with the low-energy approximation  $\sim-0.4$.}
\label{1020GF}
\end{figure}

\subsubsection*{Calculation of $g_0(\omega,(0,i)(n\mathcal{A},i))$}
The Green's function for the gapless system corresponding to $(J_i=1)$ decays the slowest along the direction $\mathcal{A}=a_3-a_2/2-a_1/2$. Greens function matrix elements between sites separated by $n\mathcal{A}$ (for large $n$) are easier to evaluate because $\mathcal{A}$ points along the axis of the line-node. The resulting symmetry simplifies the calculations.

We first consider the sublattice-diagonal terms $g_0(\omega,(0,i)(n\mathcal{A},i))$ of the Green's function matrix. The leading order behavior is given by (from Eq-\ref{ExactGF}):
\begin{equation}
\frac{\omega}{2}\int_{-\pi}^{\pi}\frac{\Delta_{0}}{\Delta_{0}\omega^{2}-\delta}e^{\imath\left(k_{3}-\frac{k_{1}}{2}-\frac{k_{2}}{2}\right)n}\frac{d^{3}k}{8\pi^{3}}
\end{equation}
After changing of variables from $(k_1,k_2,k_3)$ to $(u,v,\kappa)$ and integrating out the $v$-dependence in the Jacobian (Eq-\ref{JacobianIntegral}), we obtain 
\begin{equation}
\frac{\omega\Delta_{0}}{4\pi^{3}}\int_{-\pi}^{\pi}d\kappa\int_{0}^{4}du\frac{e^{\imath\kappa n}\frac{4}{u}K\left(1-\frac{16}{u^{2}}\right)}{\Delta_{0}\omega^{2}-(1-u)^{2}-4u\sin^{2}\frac{\kappa}{2}}
\end{equation}

The most significant contributions originate from near the line-node $(u,\kappa)=(1,0)$, where we use the quadratic approximation for $\delta$ and $\frac{4}{u}K(1-\frac{16}{u^2})\sim 4K(-15)$. With these approximations, the integral reduces to
\begin{equation}
\frac{\omega\Delta_{0}K\left(-15\right)}{\pi^{3}}\int d\left(u-1\right)d\kappa\frac{e^{\imath\kappa n}}{\Delta_{0}\omega^{2}-\left(u-1\right)^{2}-\kappa^{2}}
\end{equation}
Extending the integration limits to infinity, this can be evaluated to be
\begin{multline}
g(\omega\pm 0\imath,(0,i)(n\mathcal{A},i))\sim\frac{\Delta_{0}K\left(-15\right)}{\pi}\times \dots\\ \dots \left[\omega Y_{0}\left(\sqrt{\Delta_{0}}\left|n\omega\right|\right)\mp\imath \left|\omega\right|J_{0}\left(\sqrt{\Delta_{0}}\left|n\omega\right|\right)\right]
\end{multline}

\subsubsection*{Calculation of $g_0(\omega,(0,1)(n\mathcal{A},2))$}
Leading contribution to $g_0(\omega,(0,1)(n\mathcal{A},2))$ arises from 
\begin{equation}
-\int\frac{\mathfrak{\overline{pq}}-1}{\Delta_{0}\omega^{2}-\delta}e^{\imath(k_{3}-\frac{k_{1}}{2}-\frac{k_{2}}{2})n}\frac{d^{3}k}{8\pi^{3}}
\end{equation}
After changing variables from $(k_1,k_2,k_3)$ to $(u,v,\kappa)$, integrating out $v$-dependence, and using the previously mentioned approximations near the line-node, this reduces to 
\begin{equation}\frac{2K\left(-15\right)}{\pi^{3}}\int_0^4 du\int_{-\pi}^\pi d\kappa\frac{ue^{\imath\kappa}-1}{\Delta_{0}\omega^{2}-\left(u-1\right)^{2}-\kappa^{2}}e^{\imath\kappa n}
\end{equation}
Expanding the numerator to linear order around the line-node $(u,\kappa)=(1,0)$ this becomes
\begin{equation}
\frac{2K\left(-15\right)}{\pi^{3}}\int du\int d\kappa\frac{u-1+\imath\kappa}{\Delta_{0}\omega^{2}-\left(u-1\right)^{2}-\kappa^{2}}e^{\imath\kappa n}
\end{equation}
Extending the integration limits to infinity, this can be evaluated to be
\begin{multline}
g_0(\omega\pm \imath 0,(0,1)(n\mathcal{A},2))=\frac{2 K(-15) \sqrt{\Delta_0}}{\pi} \times \dots\\
\dots\times {\rm sign}(n)\left[ |\omega| Y_1(|n\omega|\sqrt{\Delta_0})\mp \imath \omega J_1(|n\omega|\sqrt{\Delta_0})\right]
\end{multline}

\subsubsection*{g((0,1)(r,3)) and g((0,1)(r,4))}
In this subsection, we argue that the Green's function matrix elements between two sites on sublattices $1$ and $3$ have a similar $\omega$-dependence at low-energy  as the Green's function between two sites on the sublattice $1$. Similarly, the Green's function matrix elements between two sites on sublattices $1$ and $4$ have a similar low-energy form as the Green's function between two sites on sublattices $1$ and $2$.

From Eq-\ref{ExactGF}, we see that the Green's function $g((0,1)(r,3))$ has the form 
\begin{equation}
-\omega\int \frac{\bar{\mathfrak{p}}+\mathfrak{q}}{\Delta_0\omega^2-\delta} e^{\imath k.r}\frac{d^3k}{8\pi^3}
\end{equation}
The integrand, apart from the numerator,  has the same form as in the case of $g((0,1),(r,1))$. The numerator is finite everywhere and non-zero along the line-node. As a result, $g((0,1),(r,3))$ at low energies is proportional to the Green's function $g((0,1)(r,1))$.

Similarly, the Green's function matrix element $g((0,1)(r,4))$, which has the form 
\begin{equation}
\int \frac{\mathfrak{q}(\overline{\mathfrak{p}\mathfrak{q}}-1)}{\Delta_0\omega^2-\delta} e^{\imath k.r}\frac{d^3 k}{8\pi^3},
\end{equation}
differs from $g((0,1)(r,2))$ only by the factor $\mathfrak{q}=1+e^{\imath k_2}$. We therefore expect this matrix element to differ from $g((0,1)(r,2))$ at low energies only by a multiplying form factor.

\section{Calculation of Green's function for a modified line-node}
\label{app:circle-approx}
\begin{figure}
\includegraphics[width=\columnwidth]{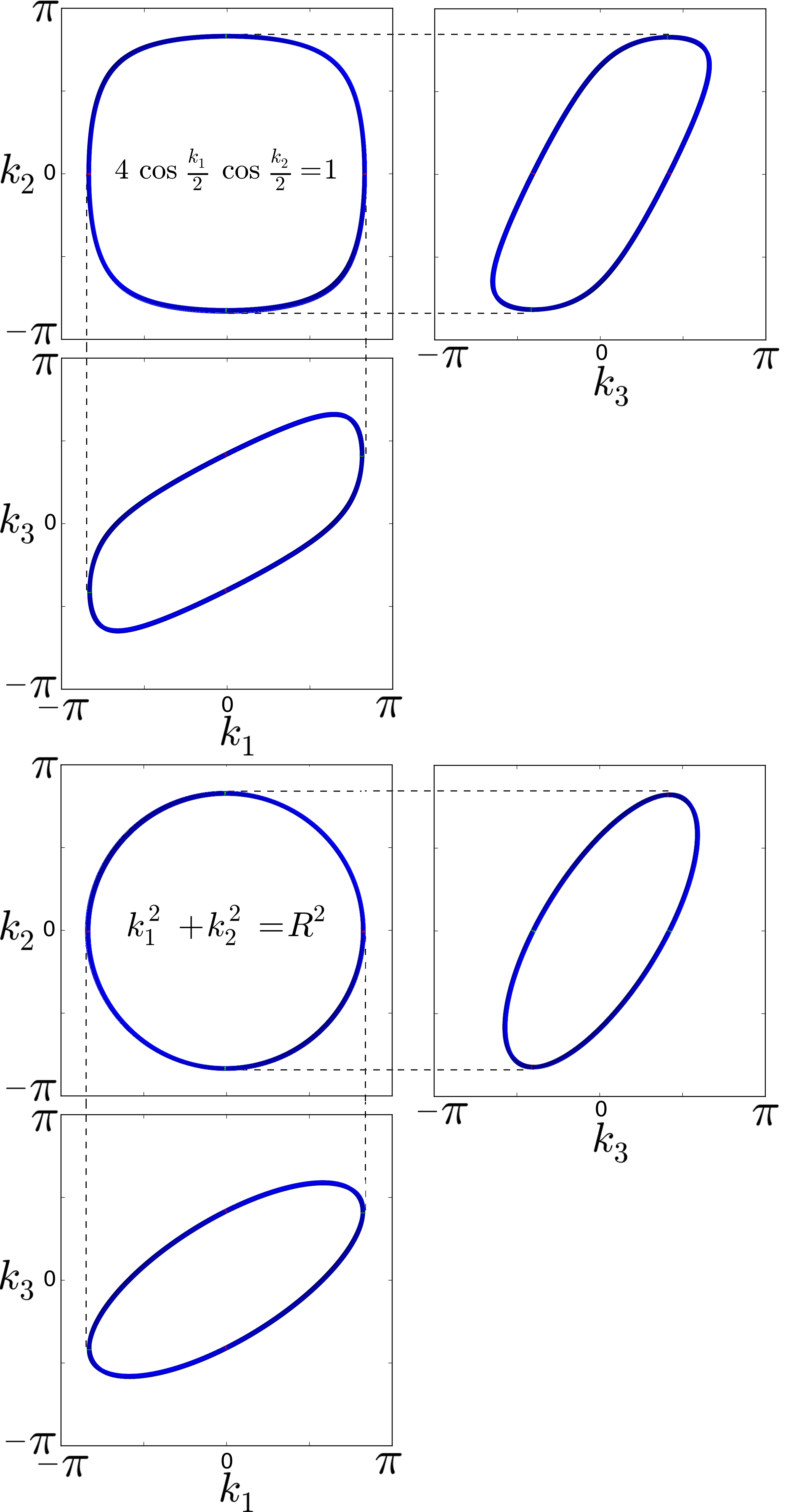}
\caption{(top) Orthogonal projection of the actual line-node onto the three coordinate planes. (bottom) Orthogonal projections of the modified line-node}
\label{orthprojlinenode}
\end{figure}
Calculation of the Green's function for directions away from the $\mathcal{A}$ axis is made difficult by the complex shape of the line-node. For the parameters we are studying $J_{x,y,z}=1$, the line node occurs along the contour formed when the plane $2k_3-k_1-k_2=0$ intersects the cylindrical surface $4\cos\frac{k_1}{2}\cos\frac{k_2}{2}=1$. The projection of the line-node on the $k_1-k_2$ plane (Fig-\ref{orthprojlinenode} (top)) is a four-fold symmetric closed contour. Radius of this projection varies between $\frac{2\sqrt{2}\pi}{3}\sim 3$ and $2{\cos^{-1}}\frac{1}{4}\sim 2.6$. The qualitative behavior of the Green's function can be extracted considering instead, a line-node whose projection on the $k_1-k_2$ plane is a circle of radius $R$, with linear dispersion along the two directions perpendicular to it ((Fig-\ref{orthprojlinenode} (bottom)). 

Form of the sublattice-diagonal elements $g(\omega,(0,i)(r,i))$ of the Green's function where $r=n_1 a_1 +n_2 a_2 +n_3 a_3$ of the Green's function matrix can be obtained from the following form
\begin{equation}
\frac{\omega\Delta_{0}}{2}\int\frac{e^{\imath k.r}}{\Delta_{0}\omega^{2}-\left(\sqrt{k_{1}^{2}+k_{2}^{2}}-R\right)^{2}-\left(k_{3}-\frac{k_{1}+k_{2}}{2}\right)^{2}}\frac{d^{3}k}{8\pi^{3}}
\end{equation}
This has a line-node along a circle of radius $R$ in the plane $2k_3=k_1+k_2$. The spectrum (zeros of the denominator) is linear along directions normal to the line-node. Changing the variables to $(k_1,k_2,\kappa)$, the integral becomes
\begin{equation}
\frac{\omega\Delta_{0}}{2}\int\frac{e^{\imath\kappa n_{3}+k_{1}\left(n_{1}+\frac{n_{3}}{2}\right)+k_{2}\left(n_{2}+\frac{n_{3}}{2}\right)}}{\Delta_{0}\omega^{2}-\left(\sqrt{k_{1}^{2}+k_{2}^{2}}-R\right)^{2}-\kappa^{2}}\frac{d^{2}k\, d\kappa}{8\pi^{3}}
\end{equation}
Using radial coordinates in the $(k_1,k_2)$ plane
\begin{equation}
\frac{\omega\Delta_{0}}{2}\int\frac{e^{\imath\kappa n_{3}+k_{r}n_{12}\cos\theta}}{\Delta_{0}\omega^{2}-\left(k_{r}-R\right)^{2}-\kappa^{2}}\frac{k_{r}dk_{r}d\kappa d\theta}{8\pi^{3}}
\end{equation}
where $n_{12}=\sqrt{\left(n_{1}+\frac{n_{3}}{2}\right)^{2}+\left(n_{2}+\frac{n_{3}}{2}\right)^{2}}$. Since the leading contribution arises from around the line-node $k_r=R$, we can expand $k_r$ around the line-node (i.e. $k_r=R+\delta k_r$) to obtain
\begin{equation}
\frac{\omega\Delta_{0}R}{2}\int\frac{e^{\imath\kappa n_{3}+\left(R+\delta k_{r}\right)n_{12}\cos\theta}}{\Delta_{0}\omega^{2}-\left(\delta k_{r}\right)^{2}-\kappa^{2}}\frac{d\delta k_{r}d\kappa d\theta}{8\pi^{3}}
\end{equation}
The leading contribution to the decay of the Green's function with $n_{12}$ can be obtained by retaining $R$ only in the exponent. The $\delta k_r$ has only a modulating effect. With this approximation, the integral can be evaluated to obtain
\begin{multline}
g(\omega \pm \imath 0^+,(0,i),(r,i)) = \frac{\Delta_{0}R}{8}J_{0}\left(Rn_{12}\right)\times \dots \\ \dots \left[\omega Y_{0}\left(\sqrt{\Delta_{0}}\left|n_3\omega\right|\right)\mp\imath\left|\omega\right|J_{0}\left(\sqrt{\Delta_{0}}\left|n_3\omega\right|\right)\right]
\end{multline}
where
\begin{gather}
r=\sum n_i a_i\nonumber\\
n_{12} = \sqrt{(n_1+n_3/2)^2 + (n_2+n_3/2)^2 }\nonumber
\end{gather}
Site diagonal element of the Green's function $g(\omega,(0,i),(0,i))$ can be obtained in a similar manner to be 
\begin{equation}
g(\omega,(r,i),(r,i)) = \frac{\omega R \Delta_0}{8\pi} \ln \left[ -\frac{\Delta_0\omega^2}{\Lambda^2} \right]
\end{equation}
where, as before, $\Lambda$ is an upper cutoff.

The Green's functions $g(\omega,(0,1),(r,2))$ can be estimated similarly. The leading contribution to the original form of this Greens function is (from Eq-\ref{originalGF}) is 
\begin{equation}
\int_{-\pi}^{\pi} \frac{\overline{\mathfrak{p}\mathfrak{q}}-1}{P(\omega)} e^{\imath k.r} \frac{d^3k}{8\pi^3}
\end{equation}
The numerator, in terms of coordinates $(u,\kappa)$ perpendicular to the line-node is $ue^{\imath\kappa}-1$, which can be approximated with $(u-1)+\imath \kappa$ very close to the line-node. Motivated by this form of the integrand close to the line-node, the form for Green's function for the case of a modified line-node can be written as 
\begin{equation}
-\int\frac{\left(k_{r}-R+\imath\kappa\right)e^{\imath\kappa n_{3}+\imath k_{r}n_{12}\cos\theta}}{\Delta_{0}\omega^{2}-\left(k_{r}-R\right)^{2}-\kappa^{2}}\frac{k_{r}dk_{r}d\theta}{8\pi^{3}}
\end{equation}
Using $k_r\sim R$ in the exponent, similar to the previous calculation, we can evaluate this integral to obtain
\begin{equation}
\frac{RJ_{0}\left(Rn_{12}\right){\rm sign}\left(n_{3}\right)}{4}[\left|\omega\right|Y_{1}(\left|n_{3}\omega\right|\sqrt{\Delta_{0}})\mp\omega J_{1}(\left|n_{3}\omega\right|\sqrt{\Delta_{0}})]
\label{circ-approx:gvvprime}
\end{equation}

\section{Leading terms in $X_1$ and $X_2$}
\label{app:X1X2}
The functions $X_1$ and $X_2$ defined in terms of $G_0$ in Eq-\ref{X1X2} can be expressed in terms of the Green's functions $g$ of the clean system by using the expressions for $G_0$ given in Eq-\ref{2vacGaplessGFnofield} and Eq-\ref{impurityTmatrx}. Inversion symmetry ($g_{\mathbf{a,b}}=g_{\mathbf{b,a}}$) and translation symmetry of the Green's function can be used to simplify the resultant expressions to obtain
\begin{equation}
X_i = \frac{N_i}{g_{\bf vv}^2-g_{\bf vv'}^2} \text{ for } i=1,2
\end{equation}
$N_1$ is given by
\begin{multline}
-2g_z^2 g_0 + 2g_z g_{\bf vv'}(g_{\bf R_3 v'} + g_{\bf R_3' v}) + \dots
\\ \dots - g_0 ( g_{\bf R_3 v'}^2 + g_{\bf  R_3' v}^2 + 2g_{\bf v v'}^2-2g_0^2)
\end{multline}
and $N_2$ is given by
\begin{multline}
g_z^4 + g_0^4 + 2g_z g_0 (g_{\bf R_3 v'} + g_{\bf R_3' v})(g_{\bf R_3 R_3'} + g_{\bf vv'})+\dots\\
\dots + (g_{\bf R_3 v'} g_{\bf R_3' v} - g_{\bf R_3 R_3'} g_{\bf vv'})^2 +\dots \\
\dots - 2g_z^2 (g_0^2 + g_{\bf R_3 v'} g_{\bf R_3' v} + g_{\bf R_3 R_3'} g_{\bf vv'})+\dots \\
\dots - g_0^2 (g_{\bf R_3 R_3'}^2+g_{\bf R_3 v'}^2 + g_{\bf R_3' v}^2 + g_{\bf vv'}^2)
\end{multline}
where $\mathbf{R_3}$ and $\mathbf{R_3}'$ are the row-indices corresponding to the modes of type $c_\mathbf{3}$  (Fig-\ref{localdof}) located next to the vacancies $\bf{v}$ and $\bf{v}'$. The symbols $g_0$ represents the site-diagonal Green's function $g_{\bf a,a}$; and $g_{z}=g_{\mathbf{v,R_3}}=g_{\mathbf{v',R'_3}}$ represents the Green's function between two sites separated by a $z$-bond. 

For sufficiently far-separated vacancies, and small $|{\rm Re}[\omega]|$ close to the real axis, the only relevant terms in the expression are 
\begin{gather}
N_1 \approx -2 g_z^2 g_0 \nonumber\\
N_2 \approx g_z^4
\end{gather}
All other terms are sub-leading because they contain either powers of $g_0 \sim \omega \ln \omega$ or powers of matrix elements of $g$ connecting far separated sites. With this approximation, we obtain
\begin{align}
X_1 \approx -\frac{2g_{\bf vR_3}^2g_{\bf vv}}{g_{\bf vv}^2-g_{\bf vv'}^2}\\
X_2 \approx \frac{g_{\bf vR_3}^4}{g_{\bf vv}^2-g_{\bf vv'}^2}\nonumber
\end{align}

\bibliography{biblio_kitaev.bib}
\end{document}